\documentclass[aps,pre,final,twocolumn,10pt,superscriptaddress,nofootinbib,nobibnotes]{revtex4-1}
\usepackage{amsmath,amssymb}
\usepackage[squaren]{SIunits}
\usepackage{graphicx,color,colortbl}
\usepackage{rotating,array,tabularx,booktabs}
\usepackage{braket}
\usepackage{appendix}

\newlength{\myl}%
\newcolumntype{Y}{>{\centering\arraybackslash}X}

\usepackage[capitalize]{cleveref}

\newcommand{\INT}[3]{\settowidth{\myl}{$\displaystyle\int_{#1}^{#2}$}{\int_{#1}^{#2}\;\;\;\hspace{-\the\myl}\dif #3}\,}
\newcommand{\TINT}[3]{\settowidth{\myl}{$\int_{#1}^{#2}$}{\int_{#1}^{#2}\!\ifthenelse{\equal{#1#2}{}}{}{\;\;\;\;\hspace{-\the\myl}}\dif #3}\,}%
\newcommand{\EINT}[3]{\settowidth{\myl}{$\int_{#1}^{#2}$}{\int_{#1}^{#2}\;\;\;\,\hspace{-\the\myl}\dif #3}\,}

\allowdisplaybreaks

\newcommand{\dif}{\mathrm{d}}%
\newcommand{\rt}{(\vec{r},t)}%
\newcommand{\Eins}{\mathbf{1}}%
\newcommand{\fdif}{\operatorname{\delta}}
\newcommand{\Fdif}[2]{\frac{\fdif\!#1}{\fdif\!#2}}
\newcommand{\ii}{\mathrm{i}}%
\newcommand{\Nabla}{\vec{\nabla}}%
\newcommand{\Laplace}{\Nabla^2}%
\newcommand{\tdif}[2]{\frac{\dif#1}{\dif#2}}%
\newcommand{\pdif}[2]{\frac{\partial#1}{\partial#2}}%
\newcommand{\R}{\mathbb{R}}%
\newcommand{\C}{\mathbb{C}}%
\newcommand{\Tr}{\operatorname{Tr}}%
\newcommand{\rhom}{\rho_{\mathrm{m}}}
\newcommand{\rhomo}{\rho_{\mathrm{m},0}}
\newcommand{\norm}[1]{\lVert#1\rVert}%
\newcommand{\ZT}[1]{\textquotedblleft#1\textquotedblright}%
\newcommand{\rs}{\vec{r}\hskip1pt'}%
\newcommand{\rss}{\vec{r}\hskip1pt''}%
\newcommand{\rsss}{\vec{r}\hskip1pt'''}%
\newcommand{\HO}{\hat{H}}%
\newcommand{\Jw}{\hat{\vec{\mathcal{J}}}}%
\newcommand{\DT}{\mathcal{D}}%
\newcommand{\CF}{C}%
\newcommand{\CV}{\vec{C}}%

\begin{document}
\title{Extended dynamical density functional theory for nonisothermal binary systems including momentum density}

\author{Michael te Vrugt}
\affiliation{Institut f\"ur Physik, Johannes Gutenberg-Universit\"at Mainz, 55128 Mainz, Germany}

\author{Hartmut L{\"o}wen}
\affiliation{Institut f{\"u}r Theoretische Physik II: Weiche Materie, Heinrich-Heine-Universit{\"a}t D{\"u}sseldorf, 40225 D{\"u}sseldorf, Germany}

\author{Helmut R. Brand}
\affiliation{Department of Physics, University of Bayreuth, 95540 Bayreuth, Germany}

\author{Raphael Wittkowski}
\email[Corresponding author: ]{rgwitt25@dwi.rwth-aachen.de}
\affiliation{Department of Physics, RWTH Aachen University, 52074 Aachen, Germany}
\affiliation{DWI -- Leibniz Institute for Interactive Materials, 52074 Aachen, Germany}
\affiliation{Institute of Theoretical Physics, Center for Soft Nanoscience, University of M\"unster, 48149 M\"unster, Germany}


\begin{abstract}
In order to describe the nonisothermal dynamics of two-phase flows or binary mixtures such as colloidal suspensions consisting of colloidal particles and solvent on a microscopic level, we derive a new extended dynamical density functional theory (EDDFT) that includes the total mass density, the local concentration of one species, the total momentum density, and the energy density as variables using the Mori-Zwanzig-Forster projection operator technique. Through the incorporation of the momentum density into EDDFT, not only the diffusive but also the convective dynamics is taken into account. We derive an exact entropy and free-energy functional for the case of hard spheres. The hydrodynamic limit of our new EDDFT and its relation to the mode-coupling theory of the glass transition are discussed. It is shown that EDDFT allows to obtain the correct value for the speed of sound. 
\end{abstract}



\maketitle


\section{\label{sec:introduction}Introduction}
Standard dynamical density functional theory (DDFT) \cite{MarconiT1999,MarconiT2000,ArcherE2004,EspanolL2009,Evans1979,Munakata1989,Fraaije1993,Kawasaki1994}, reviewed by \citet{teVrugtLW2020}, provides a dynamical equation for the local concentration of colloidal particles in a colloidal suspension.
This theory has successfully been applied to the diffusive dynamics of colloidal suspensions. It is also used in a wide variety of other fields, such as biology \cite{AngiolettiBD2014,AngiolettiBD2018,AlSaediHAW2018,ChauviereLC2012}, chemistry \cite{WerkhovenSvR2019,WerkhovenESvR2018,GaoX2018,QingLZTQMXZ2020}, disease spreading \cite{teVrugtBW2020,teVrugtBW2020b}, plasma physics \cite{DiawM2015,DiawM2016,DiawM2017}, and polymer physics \cite{FraaijevVMPEHAGW1997,ManthaQS2020,KnollHLKSZM2002,KnollLHKSZM2004,LudwigsBVRMK2003}. However, it fails to describe the convective dynamics of such systems that is especially relevant in case of a pronounced flow field of the solvent. 
To improve DDFT in this respect, besides the local concentration of colloidal particles a further variable 
-- the momentum density or equivalently the velocity field -- has to be taken into account. Extensions towards inertial or nonisothermal flow are central for industrial applications, e.g., to describe metal alloys \cite{OforiFGEP2013} or cement flow \cite{RousselFC2010,MonlouisVP2004}. More recently, fluid mechanics with inertia and temperature gradients has gained importance in the description of airborne transmission of infectious diseases, in particular COVID-19 \cite{BhagatWDL2020,AbkarianMXYS2020}.

Due to the successes of standard DDFT, a significant amount of work has been done on the development of extensions towards additional order parameters (see our review article \cite{teVrugtLW2020} for an overview). Of particular interest here is the study of \textit{inertial} dynamics. This significantly extends the applicability of DDFT and allows to connect it to existing work in kinetic theory \cite{MarconiM2014} and fluid mechanics \cite{Archer2009,StierleG2021}. This is particularly beneficial for industrial applications, where convective flow plays an important role. Work in this direction has been done by a variety of authors \cite{Archer2006,Archer2009,MarconiT2006,MarconiM2007,ChanF2005,QiaoZYQBZL2021,Schmidt2018}. In inertial extensions of DDFT, the \textit{momentum density} is the main order parameter that is required in addition to the local concentration.

In the derivation of DDFTs with momentum density, two main approaches can be distinguished. The first and more popular one is to start from dynamical theories for the one-body phase-space distribution, which have been studied in statistical mechanics since the pioneering work of \citet{Boltzmann1872} (see \citet{BrownUM2009} for historical remarks). Dynamic equations for mass and momentum (and sometimes energy) density, which correspond to moments of the phase-space distribution, can be found by integrating out the phase space dynamics \cite{Marconi2011,MarconiM2009,GoddardHO2020,ZhaoW2011,GoddardNSYK2013}. One then obtains an infinite hierarchy of equations, the so-called Bogoliubov-Born-Green-Kirkwood-Yvon (BBGKY) hierarchy, which can be closed using DDFT \cite{DiawM2015}. Approaches of this form, which are also applicable in multicomponent systems \cite{MarconiM2011,MarconiM2011b,Marconi2011,MarconiM2014,MonteferranteMM2014,MarconiM2012}, have been used for simple fluids \cite{MarconiM2009}, colloidal fluids \cite{Marconi2011,MillsGA2024}, plasmas \cite{DiawM2015}, and granular media \cite{GoddardHO2020}. 

The second route, which has been employed in more recent work, is the Mori-Zwanzig-Forster technique (MZFT) \cite{Nakajima1958,Zwanzig1960,Mori1965,Grabert1982,teVrugtW2019,teVrugtW2019d,Netz2024,MeyerVS2017,JungJ2023}, often also referred to as Mori-Zwanzig formalism. It allows to project the microscopic dynamics of a many-particle system onto an arbitrary set of relevant variables, for which closed equations of motion are thereby obtained. For example, one can derive DDFT by projecting the microscopic dynamics onto the one-particle density \cite{Yoshimori1999,Munakata2003,Yoshimori2005,EspanolL2009,EspanolV2002,Kawasaki1994}. Extensions are obtained by introducing additional order parameters, such as the energy density for nonisothermal systems \cite{WittkowskiLB2012,AneroET2013,JiaK2021,KaufmantV2026}, or the momentum density for systems in which inertia is relevant \cite{CamargodlTDBCE2019,DuqueZumajoCdlTCE2019,DuqueZumajodlTCE2019,CamargodlTDZEDBC2018,DuranYGK2017,DuranYGK2017,Haussmann2016,Haussmann2022}. A general framework based on this idea is \textit{extended dynamical density functional theory} (EDDFT), developed by \citet{WittkowskiLB2012,WittkowskiLB2013}, which allows to extend DDFT to a set of arbitrary order parameters. Approaches based on the MZFT have a variety of advantages: They are more general since the MZFT is also applicable to quantum systems \cite{teVrugtW2019}, they give a natural connection to the equations of hydrodynamics which can also be derived from the MZFT \cite{Piccirelli1968,Grabert1982}, they allow to relate the resulting DDFT to the mode coupling theory (MCT) of the glass transition \cite{Goetze2009}, which is a further important application of the MZFT \cite{Janssen2018,teVrugtLW2020,WittkowskiLB2012}, and they provide a natural connection to thermodynamics since they are based on thermodynamic functionals and describe the approach to thermodynamic equilibrium \cite{teVrugtW2019d,WittkowskiLB2013,teVrugt2020}.

In this article, we use the general framework of EDDFT \cite{WittkowskiLB2012} in order to 
derive an EDDFT for the following variables: 
(a) the total mass density, 
(b) the concentration field of one species, 
(c) the total momentum density, and 
(d) the energy density.
This new EDDFT generalizes the standard DDFT \cite{MarconiT1999,MarconiT2000,ArcherE2004,EspanolL2009} and is appropriate to describe both the diffusive and the convective dynamics of, e.g., two-phase flow and colloidal suspensions, where two particle species are present (species 1 for fluid 1 or the solvent, species 2 for fluid 2 or the colloidal particles). Note that we do not assume the two species to have a very different mass, such that the EDDFT is applicable to arbitrary mixtures. Furthermore, it describes the flow field of the solvent directly and takes also hydrodynamic interactions between the particles into account. Our EDDFT thus constitutes a set of generalized Navier-Stokes equations for two-component systems -- such as colloidal suspensions consisting of colloids and solvent -- that contains 
the Navier-Stokes equations for simple liquids \cite{LandauL1987} as well as the standard DDFT as special cases. In contrast to the Navier-Stokes equations, it is capable of describing phase transitions and can thereby, e.g., be used to simulate the formation and dynamics of air bubbles in a fluid.
Owing to the consideration of the energy density as a relevant variable, this EDDFT can also be applied to describe heat transport in two-phase flow and colloidal suspensions as well as the coupling of heat transport to the concentration of the second species (Ludwig-Soret effect and Dufour effect) and to flow (thermal convection). We explicitly demonstrate that, due to the incorporation of thermal fluctuations, EDDFT gives the correct speed of sound and thereby improves previous DDFT-based approaches \cite{Archer2006}.
Since EDDFT is not restricted to large wavelengths and small frequencies, we also present its general 
hydrodynamic limit. Moreover, we improve its practical applicability by deriving the explicit form of the entropy functional of EDDFT for the case of hard spheres. We also discuss its relation to MCT. 

This article is organized as follows: In \cref{sec:I} we present the derivation of the new EDDFT in detail. 
Afterwards, special cases of this theory and its relation to MCT are discussed in \cref{sec:II}. Applications are presented in \cref{applications}. Finally, we conclude in \cref{sec:conclusions}.

\section{\label{sec:I}EDDFT including momentum density}
The Mori-Zwanzig-Forster projection operator technique (MZFT) 
\cite{Mori1965,ZwanzigM1965,Forster1974,Grabert1982,Forster1990,Zwanzig2001,teVrugtW2019d,teVrugtW2019} is applied 
in order to derive an EDDFT, which includes the momentum density field as a variable,  
from the classical microscopic Hamiltonian equations 
for the phase-space variables of a binary mixture. Thereby, we extend earlier work \cite{WittkowskiLB2012,WittkowskiLB2013} on the derivation of EDDFT using the MZFT.  

\subsection{Microscopic dynamics}
We consider a mixture of $N_{\mathrm{c}}$ colloidal (c) particles that are suspended in a solvent (s) 
consisting of $N_{\mathrm{s}}$ atomic or molecular isotropic particles. In principle, our derivation does not only apply to colloidal suspensions, but to any classical many-particle system consisting of two types of isotropic particles (or of one species if the concentration of the other one is set to zero). For instance, oil-water demixing would also be a possible application of our model. For most of this work, we will nevertheless often refer to one particle species as \ZT{colloidal particles} and to the other one as \ZT{solvent particles} (motivated by the fact that previous work on EDDFT \cite{WittkowskiLB2012} focused on colloidal suspensions). Clearly our analysis is straightforwardly generalizable but more tedious to three and more component.

The phase-space variables of this system are the positions $\vec{r}^{\mathrm{c}}_{i}(t)$ and translational momenta 
$\vec{p}^{\mathrm{c}}_{i}(t)$ with $i=1,\dotsc,N_{\mathrm{c}}$ of the colloidal particles as well as the 
positions $\vec{r}^{\mathrm{s}}_{i}(t)$ and momenta $\vec{p}^{\mathrm{s}}_{i}(t)$ with $i=1,\dotsc,N_{\mathrm{s}}$ 
of the solvent particles at time $t$.
In general, these particles are subjected to external potentials $U^{\mathrm{c}}_{1}(\vec{r})$ 
and $U^{\mathrm{s}}_{1}(\vec{r})$ that act on the colloidal and solvent particles, respectively. Here, we assume these potentials to be time-independent.
Interactions between the particles are taken into account by the pair-interaction potentials 
$U^{(\mathrm{cc})}_{2}(\norm{\vec{r}-\rs})$, $U^{(\mathrm{cs})}_{2}(\norm{\vec{r}-\rs})$, and 
$U^{(\mathrm{ss})}_{2}(\norm{\vec{r}-\rs})$ (with the Euclidean norm $\norm{\cdot}$) for colloid-colloid, colloid-solvent, and solvent-solvent particle interactions, respectively. We thereby assume the interaction potential to depend only on distances between particles. Examples for this would be hard spheres or point particles with rigid isotropic interactions, counterexamples would be rods or other particles with orientational degrees of freedom.

If $m_{\mathrm{c}}$ denotes the mass of a colloidal particle and $m_{\mathrm{s}}$ the mass of a solvent particle, 
the system's dynamics is completely determined by its Hamiltonian 
\begin{equation}
\begin{split}
\HO&=\!\sum_{\mu\in\{\mathrm{c},\mathrm{s}\}}\sum^{N_{\mu}}_{i=1}\HO^{\mu}_{i},\\
\HO^{\mu}_{i}&=\frac{(\vec{p}^{\mu}_{i}(t))^{2}}{2m_{\mu}}+\frac{1}{2}\!\!\underset{(\mu,i)\neq(\nu,j)}{\sum_{\nu\in\{\mathrm{c},\mathrm{s}\}}\sum^{N_{\nu}}_{j=1}}\,
U^{(\mathrm{\mu\nu})}_{2}(\norm{\vec{r}^{\mu}_{i}(t)\!-\vec{r}^{\nu}_{j}(t)})\\
&\quad\:\! + U_1^\mu(\vec{r}_i^\mu),
\end{split}\label{eq:H}%
\end{equation}
or equivalently by the Liouvillian 
\begin{equation}
\mathcal{L}=\!\sum_{\mu\in\{\mathrm{c},\mathrm{s}\}}\sum^{N_{\mu}}_{i=1} 
\Big(\Nabla_{\vec{p}^{\mu}_{i}}\HO\Big)\!\cdot\!\Nabla_{\vec{r}^{\mu}_{i}}
-\!\!\!\sum_{\mu\in\{\mathrm{c},\mathrm{s}\}}\sum^{N_{\mu}}_{i=1}
\Big(\Nabla_{\vec{r}^{\mu}_{i}}\HO\Big)\!\cdot\!\Nabla_{\vec{p}^{\mu}_{i}} .
\label{liouvillian}
\end{equation}
Here we have assumed $U_2^{\mu\nu} \equiv U_2^{\nu\mu}$. On the basis of the huge number of phase-space variables, we define a few slow relevant variables $\hat{a}_{i}\rt$ 
with $i=1,\dotsc,n$ and $n\ll N_{\mathrm{c}},N_{\mathrm{s}}$ that are appropriate to describe the considered system.
These relevant variables are assumed to be classical, real-valued, and linearly independent, so that the dynamics of a 
relevant variable $\hat{a}_{i}\rt$ is given by 
\begin{equation}
\dot{\hat{a}}_{i}\rt+\Nabla_{\vec{r}}\!\cdot\!\hat{\vec{J}}^{(i)}\rt=\hat{Q}_i\rt
\label{eq:a_i}%
\end{equation}
with the corresponding current $\hat{\vec{J}}^{(i)}\rt$ and source term $\hat{Q}_i\rt$. The source term vanishes for a conserved variable.
In particular, we choose the following relevant variables:
the total mass density $\hat{\rho}_{\mathrm{m}}\rt$ of the colloidal suspension, the local concentration of the colloidal particles $\hat{c}\rt$,  the total momentum density $\hat{\vec{g}}\rt$, and the energy density $\hat{\varepsilon}\rt$.
This leads to the six independent relevant variables $\hat{a}_{1}\rt\equiv\hat{\rho}_{\mathrm{m}}\rt$, 
$\hat{a}_{2}\rt\equiv\hat{c}\rt$, $(\hat{a}_{3}\rt,\hat{a}_{4}\rt,\hat{a}_{5}\rt)\equiv\hat{\vec{g}}\rt$, and $\hat{a}_{6}\rt\equiv\hat{\varepsilon}\rt$. Alternatively, one can use the solvent density $\hat{s}\rt$ or the total number density $\hat{n}\rt$ instead of the total mass density. Which choice is the most useful one depends on the context. Later, we will switch between different options making use of the fact that $\hat{s}\rt$ and $\hat{n}\rt$ can be expressed as linear combinations of $\hat{\rho}_{\mathrm{m}}\rt$ and $\hat{c}\rt$.

These relevant variables have different symmetry properties. While the mass density $\hat{\rho}_{\mathrm{m}}\rt$, 
the concentration field $\hat{c}\rt$, and the energy density $\hat{\varepsilon}\rt$ are even under parity (spatial inversion) and time reversal, the momentum density $\hat{\vec{g}}\rt$ is odd under parity or time reversal.
Explicit expressions for the chosen relevant variables are
{\allowdisplaybreaks%
\begin{align}%
\begin{split}%
\hat{\rho}_{\mathrm{m}}\rt&=\sum_{\mu\in\{\mathrm{c},\mathrm{s}\}}\sum^{N_{\mu}}_{i=1} 
m_{\mu} \delta\big(\vec{r}-\vec{r}^{\mu}_{i}(t)\big) ,
\end{split}\label{eq:rhom_operator}\\%
\begin{split}%
\hat{c}\rt&=\sum^{N_{\mathrm{c}}}_{i=1} \delta\big(\vec{r}-\vec{r}^{\mathrm{c}}_{i}(t)\big) ,
\end{split}\label{eq:c_operator}\\%
\begin{split}%
\hat{\vec{g}}\rt&=\sum_{\mu\in\{\mathrm{c},\mathrm{s}\}}\sum^{N_{\mu}}_{i=1} 
\vec{p}^{\mu}_{i}(t)\delta\big(\vec{r}-\vec{r}^{\mu}_{i}(t)\big) ,
\end{split}\label{eq:g_operator}\\%
\begin{split}%
\hat{\varepsilon}\rt&=\sum_{\mu\in\{\mathrm{c},\mathrm{s}\}}\sum^{N_{\mu}}_{i=1} 
\HO^{\mu}_{i}(t) \delta\big(\vec{r}-\vec{r}^{\mu}_{i}(t)\big) .
\end{split}\label{eq:e_operator}%
\end{align}}%
The dynamics of these relevant variables can be derived from a time derivative of Eqs.\ \eqref{eq:rhom_operator}-\eqref{eq:e_operator}, the Hamiltonian equations, and the Hamiltonian \eqref{eq:H} and is given by 
{\allowdisplaybreaks%
\begin{gather}%
\begin{split}%
\dot{\hat{\rho}}_{\mathrm{m}}\rt+\Nabla_{\vec{r}}\!\cdot\!\hat{\vec{J}}^{\rhom }\rt=0 ,
\end{split}\label{eq:rhom_hat_dyn}\\%
\begin{split}%
\dot{\hat{c}}\rt+\Nabla_{\vec{r}}\!\cdot\!\hat{\vec{J}}^{c}\rt=0 ,
\end{split}\label{eq:c_hat_dyn}\\%
\begin{split}%
\dot{\hat{\vec{g}}}\rt+\Nabla_{\vec{r}}\!\cdot\!\hat{\mathrm{J}}^{\vec{g}}\rt=\hat{\vec{F}}\rt,
\end{split}\label{eq:g_hat_dyn}\\%
\begin{split}%
\dot{\hat{\varepsilon}}\rt+\Nabla_{\vec{r}}\!\cdot\!\hat{\vec{J}}^{\varepsilon}\rt=0 
\end{split}\label{eq:e_hat_dyn}%
\end{gather}}%
with the local currents $\hat{\vec{J}}^{\rhom }\rt$, $\hat{\vec{J}}^{c}\rt$, $\hat{\mathrm{J}}^{\vec{g}}\rt$, and $\hat{\vec{J}}^{\varepsilon}\rt$ and the external force density $\hat{\vec{F}}\rt$. (The typesetting for $\hat{\mathrm{J}}^{\vec{g}}\rt$ indicates that it is a tensor.) Equation \eqref{eq:g_hat_dyn} involves, in addition to the conserved part of the dynamics, also a non-conserved part since the momentum is not conserved in the presence of external forces. All other relevant variables are strictly conserved.
For the conserved variables, explicit expressions for the local currents follow from\footnote{For the non-conserved variable $\vec{\hat{g}}\rt$, we have to set the external potentials to zero before this relation can be applied.}  
$\Nabla_{\vec{r}}\!\cdot\!\hat{\vec{J}}^{a_{i}}\rt=-\mathcal{L}\hat{a}_{i}\rt$ 
\cite{WittkowskiLB2012} and are -- for the present set of variables -- given 
by\footnote{The divergence $\Nabla_{\vec{r}}\!\cdot\!\mathrm{M}$ of a tensor $\mathrm{M}(\vec{r})$ is defined as 
$\partial_{j}M_{ij}$ with $\Nabla_{\vec{r}}=(\partial_{1},\partial_{2},\partial_{3})$ and $\mathrm{M}=(M_{ij})_{i,j=1,2,3}$.}
{\allowdisplaybreaks%
\begin{align}%
\begin{split}%
\hat{\vec{J}}^{\rhom }\rt
&=\sum_{\mu\in\{\mathrm{c},\mathrm{s}\}}\sum^{N_{\mu}}_{i=1}\vec{p}^{\mu}_{i}(t)
\delta\big(\vec{r}-\vec{r}^{\mu}_{i}(t)\big) ,
\end{split}\label{eq:J_rhom}\\%
\begin{split}%
\hat{\vec{J}}^{c}\rt 
&=\sum^{N_{\mathrm{c}}}_{i=1}\frac{\vec{p}^{\mathrm{c}}_{i}(t)}{m_{\mathrm{c}}}\:\!
\delta\big(\vec{r}-\vec{r}^{\mathrm{c}}_{i}(t)\big) ,
\end{split}\label{eq:J_c}\\%
\begin{split}%
\hat{\mathrm{J}}^{\vec{g}}\rt
&=\sum_{\mu\in\{\mathrm{c},\mathrm{s}\}}\sum^{N_{\mu}}_{i=1} 
\frac{\vec{p}^{\mu}_{i}(t)\!\:\!\otimes\!\:\!\vec{p}^{\mu}_{i}(t)}{m_{\mu}}\:\! 
\delta\big(\vec{r}-\vec{r}^{\mu}_{i}(t)\big) \\
&\quad\:\! -\frac{1}{2}\!\!\!\!\!\sum_{\mu,\nu\in\{\mathrm{c},\mathrm{s}\}}\!\!\!\!\!\!\!\!
\sum^{N_{\mu}}_{\begin{subarray}{c}i=1\\(\mu,i)\neq(\nu,j)\end{subarray}}\!\!\!\sum^{N_{\nu}}_{j=1}
\frac{\vec{r}^{\mu\nu}_{ij}(t)\!\:\!\otimes\!\:\!\vec{r}^{\mu\nu}_{ij}(t)}{\norm{\vec{r}^{\mu\nu}_{ij}(t)}}
\tdif{U^{(\mu\nu)}_{2}(r)}{r} 
\!\:\!\\
&\quad\:\! \int^{1}_{0}\!\!\!\!\dif\lambda\,\:\!\delta\big(\vec{r}-\vec{r}^{\mu}_{i}(t)\!\:\!
+\lambda\:\!\vec{r}^{\mu\nu}_{ij}(t)\big) , 
\end{split}\label{eq:J_g}\raisetag{3em}\\%
\begin{split}%
\hat{\vec{J}}^{\varepsilon}\rt 
&=\sum_{\mu\in\{\mathrm{c},\mathrm{s}\}}\sum^{N_{\mu}}_{i=1} 
\frac{\vec{p}^{\mu}_{i}(t)}{m_{\mu}}\:\! \HO^{\mu}_{i}(\vec{r}_i^\mu(t))\:\! 
\delta\big(\vec{r}-\vec{r}^{\mu}_{i}(t)\big) \\
&\quad\:\! -\frac{1}{4}\!\!\!\!\!\sum_{\mu,\nu\in\{\mathrm{c},\mathrm{s}\}}\!\!\!\!\!\!\!\!
\sum^{N_{\mu}}_{\begin{subarray}{c}i=1\\(\mu,i)\neq(\nu,j)\end{subarray}}\!\!\!\sum^{N_{\nu}}_{j=1}
\frac{\vec{r}^{\mu\nu}_{ij}(t)\!\:\!\otimes\!\:\!\vec{r}^{\mu\nu}_{ij}(t)}{\norm{\vec{r}^{\mu\nu}_{ij}(t)}} \\
&\quad\:\! \Big(\frac{\vec{p}^{\mu}_{i}(t)}{m_{\mu}}+\frac{\vec{p}^{\nu}_{j}(t)}{m_{\nu}}\Big)\tdif{U^{(\mu\nu)}_{2}(r)}{r} \,
\!\!\:\!\\
&\quad\:\! \int^{1}_{0}\!\!\!\!\dif\lambda\,\:\!
\delta\big(\vec{r}-\vec{r}^{\mu}_{i}(t)\!\:\!+\lambda\:\!\vec{r}^{\mu\nu}_{ij}(t)\big) 
\end{split}\label{eq:J_e}%
\end{align}}%
with $\vec{r}^{\mu\nu}_{ij}(t)=\vec{r}^{\mu}_{i}(t)-\vec{r}^{\nu}_{j}(t)$ and the dyadic product $\otimes$. Note that the current $\hat{\vec{J}}^{\rhom }\rt$ corresponding to the 
total mass density $\hat{\rho}_{\mathrm{m}}\rt$ is again a relevant variable, 
the total momentum density $\hat{\vec{g}}\rt$:
$\hat{\vec{J}}^{\rhom}\rt=\hat{\vec{g}}\rt$. 
This is a unique feature of the total mass density. The external force density is given by
\begin{equation}
\hat{\vec{F}}\rt = - \sum_{\mu\in\{\mathrm{c},\mathrm{s}\}}\sum^{N_{\mu}}_{i=1}\delta(\vec{r} - \vec{r}_i^\mu(t)) \Nabla_{\vec{r}_i^\mu}U_1^\mu(\vec{r}_i^\mu(t)).  
\end{equation}
The form of the interaction potential contributions in \cref{eq:J_g,eq:J_e} is very complicated and deserves some comment: From the Liouvillian \eqref{liouvillian}, one obtains the dynamic equations
\begin{align}
\begin{split}
\dot{\hat{\vec{g}}}\rt &=\sum_{\mu\in\{\mathrm{c},\mathrm{s}\}}\sum^{N_{\mu}}_{i=1} \Nabla_{\vec{r}_i^\mu}\cdot
\frac{\vec{p}^{\mu}_{i}(t)\!\:\!\otimes\!\:\!\vec{p}^{\mu}_{i}(t)}{m_{\mu}}\:\!\delta(\vec{r} - \vec{r}_i^\mu(t))\\
&\quad\:\!-\underset{(\mu,i)\neq(\nu,j)}{\sum^{N_{\mu}}_{i=1}\sum^{N_{\nu}}_{j=1}} \delta(\vec{r} - \vec{r}_i^\mu(t))\Nabla_{\vec{r}_i^\mu} U_2^{\mu\nu}(\norm{\vec{r}_{ij}^{\mu\nu}(t)})\\
&\quad\:\!- \sum_{\mu\in\{\mathrm{c},\mathrm{s}\}}\sum^{N_{\mu}}_{i=1} \delta(\vec{r} - \vec{r}_i^\mu(t))\Nabla_{\vec{r}_i^\mu} U_1^{\mu}(\vec{r}_i^\mu(t))
\label{momentumsimple}
\end{split}\\
\begin{split}
\dot{\hat{\varepsilon}}\rt &=- \Nabla_{\vec{r}}\cdot\!\!\!\!\sum_{\mu\in\{\mathrm{c},\mathrm{s}\}}\sum^{N_{\mu}}_{i=1}\frac{\vec{p}^{\mu}_{i}(t)}{m_{\mu}}\hat{H}_i^\mu \delta(\vec{r}- \vec{r}_i^\mu(t))\\
&\quad\:\! -\frac{1}{2}\underset{(\mu,i)\neq(\nu,j)}{\sum^{N_{\mu}}_{i=1}\sum^{N_{\nu}}_{j=1}} \delta(\vec{r} - \vec{r}_i^\mu(t)) \Big(\frac{\vec{p}^{\mu}_{i}(t)}{m_{\mu}} + \frac{\vec{p}^{\nu}_{j}(t)}{m_{\nu}}\Big)\\
&\qquad\:\! \cdot\Nabla_{\vec{r}_i^\mu} U_2^{\mu\nu}(\norm{\vec{r}_{ij}^{\mu\nu}(t)}).
\label{energysimple}
\end{split}
\end{align}
Equations \eqref{eq:J_g} and \eqref{eq:J_e} are then obtained using \cite{CamargodlTDZEDBC2018}
\begin{equation}
\begin{split}
&\underset{(\mu,i)\neq(\nu,j)}{\sum^{N_{\mu}}_{i=1}\sum^{N_{\nu}}_{j=1}}\delta(\vec{r} - \vec{r}_i^\mu(t))\Nabla_{\vec{r}_i^\mu} U_2^{\mu\nu}(\norm{\vec{r}^{\mu\nu}_{ij}})\\
&=\frac{1}{2}\underset{(\mu,i)\neq(\nu,j)}{\sum^{N_{\mu}}_{i=1}\sum^{N_{\nu}}_{j=1}}(\delta(\vec{r} - \vec{r}_i^\mu(t))-\delta(\vec{r} - \vec{r}_j^\nu(t)))\Nabla_{\vec{r}_i^\mu} U_2^{\mu\nu}(\norm{\vec{r}^{\mu\nu}_{ij}})
\end{split}
\end{equation}
and the identity \cite{Grabert1982}
\begin{equation}
\begin{split}
&\delta(\vec{r}-\vec{r}_1) - \delta(\vec{r}- \vec{r}_2) \\
&= - \Nabla_{\vec{r}}\cdot \frac{\vec{r}_1 \otimes \vec{r}_2}{\norm{\vec{r}_{12}}} 
\INT{0}{\norm{\vec{r}_{12}}}{r}\delta\Big(\vec{r}-\vec{r}_1+ r\frac{\vec{r}_{12}}{\norm{\vec{r}_{12}}}\Big).
\end{split}
\end{equation}
The advantage of \cref{eq:J_g,eq:J_e} compared to \cref{momentumsimple,energysimple} is that the conservation law structure is more obvious. Note that the definitions \eqref{eq:J_g} and \eqref{eq:J_e} are not unique since any tensor or vector whose divergence in combination with Eqs.\ \eqref{eq:g_hat_dyn} and \eqref{eq:e_hat_dyn} gives \cref{momentumsimple,energysimple} can also be used. Some alternative equations for the momentum current can be found in the literature \cite{WajnrybAD1995,delasHerasS2018}.

\subsection{Mori-Zwanzig-Forster projection operator technique}
Corresponding to every microscopic variable $\hat{a}_{i}\rt$ there is an averaged variable\footnote{We write $\hat{a}_{i}(\vec{r})$ for $\hat{a}_{i}(\vec{r},0)$.} 
$a_{i}\rt=\Tr\!\big(\rho(t)\hat{a}_{i}(\vec{r})\big)$, where $\Tr$ denotes the grand-canonical trace and $\rho(t)$ is a relevant probability density that can be defined in different ways. A possible definition of $\rho(t)$ is given further below. 
Notice that the averaged momentum density $\vec{g}\rt$ is directly related to the velocity field $\vec{v}\rt$ 
of the colloidal suspension: $\vec{g}\rt=\rhom \rt\vec{v}\rt$ (see \cref{momentum}).

The MZFT allows for the systematic derivation of field theories for an arbitrarily chosen set of relevant variables. It was derived by \citet{Nakajima1958}, \citet{Zwanzig1960}, and \citet{Mori1965}, and later extended to allow for an improved description of, e.g., far-from-equilibrium systems \cite{Grabert1978} or time-dependent Hamiltonians \cite{teVrugtW2019,MeyerVS2019,Netz2024}. An accessible introduction was given by \citet{teVrugtW2019d}, a more recent review by \citet{Schilling2022}. A general overview of the form used here is the textbook by \citet{Grabert1982}. The essential step in the MZFT is to approximate the unknown microscopic phase-space distribution by a \textit{relevant density} that is solely a functional of the relevant macroscopic variables \cite{Grabert1982}. In particular, in so-called \textit{canonical projection operator formalisms} \cite{Kawasaki2000,Kawasaki2006b}, the relevant density is constructed in such a way that it solely depends on the average values of the macroscopic variables \cite{EspanolL2009}, here given by $\rhom \rt$, $c\rt$, 
$\vec{g}\rt$, and $\varepsilon\rt$. This is useful since the macroscopic information that is available to an experimentalist is typically limited to these average values.

Since the energy density is among the relevant variables, the most appropriate choice for the relevant density is 
\begin{equation}
\begin{split}
\rho(t) &= \frac{1}{Z(t)}\exp\!\bigg(-\int_{\R^{3}}\!\!\!\!\!\:\!\dif^{3}r\,\varepsilon^{\flat}\rt\hat{\varepsilon}(\vec{r}) \\
&\quad\:\! -\int_{\R^{3}}\!\!\!\!\!\:\!\dif^{3}r\,
\rho^{\flat}_{\mathrm{m}}\rt\hat{\rho}_{\mathrm{m}}(\vec{r})
-\int_{\R^{3}}\!\!\!\!\!\:\!\dif^{3}r\,c^{\flat}\rt\hat{c}(\vec{r}) \\
&\quad\:\! -\sum^{3}_{i=1}\int_{\R^{3}}\!\!\!\!\!\:\!\dif^{3}r\,
g^{\flat}_{i}\rt\hat{g}_{i}(\vec{r}) \;\bigg).
\end{split}\label{relevantdensity}
\end{equation}
In \cref{relevantdensity}, which is an extension of the form used by \citet{Grabert1982}, we have introduced the normalization $Z(t)$ (partition function) which ensures
\begin{equation}
\Tr(\rho(t)) = 1,    
\end{equation}
as well as the conjugate variables -- denoted by a superscript $\flat$ -- which ensure that the macroequivalence condition
\begin{equation}
\Tr(\rho(t)\hat{a}(\vec{r}))=a\rt 
\end{equation}
holds. The trace $\Tr$ corresponds, in the classical case, to a phase-space integral. If we introduce the dimensionless entropy functional \cite{Grabert1982}
\begin{equation}
\mathcal{S}(t)= - \Tr(\rho(t)\ln(\rho(t))), 
\label{entropyfunctional}
\end{equation}
the conjugate variables are given by
\begin{equation}
\begin{split}
\rho^{\flat}_{\mathrm{m}}\rt=\Fdif{\mathcal{S}}{\rhom\rt} ,&\quad
c^{\flat}\rt=\Fdif{\mathcal{S}}{c\rt} ,\quad\\
\vec{g}^{\flat}\rt=\Fdif{\mathcal{S}}{\vec{g}\rt} ,&\quad
\varepsilon^{\flat}\rt=\Fdif{\mathcal{S}}{\varepsilon\rt}.
\end{split}
\label{eq:TC}%
\end{equation}
The conjugate variable $\varepsilon^\flat$ can be identified with the thermodynamic $\beta$ (which is a function of space and time in the nonisothermal system), that is inversely proportional to the temperature $T$. We thus have
\begin{equation}
\varepsilon^\flat \rt = \beta \rt = \frac{1}{k_{\mathrm{B}}T\rt},   
\end{equation}
where $k_{\mathrm{B}}$ is the Boltzmann constant. This allows to introduce the thermodynamic conjugates
\begin{equation}
\begin{split}
\rho^{\natural}_{\mathrm{m}}\rt=&-k_{\mathrm{B}}T \rt\rho^\flat \rt,\\
c^{\natural}\rt=& - k_{\mathrm{B}}T\rt c^\flat\rt,\\
\vec{g}^{\natural}\rt=& - k_{\mathrm{B}}T\rt \vec{g}^\flat\rt.
\end{split}
\label{eq:TC2}%
\end{equation}

A comment on the notation is in order: In the literature, the superscript $\natural$ usually denotes a thermodynamic conjugate, irrespective of the specific thermodynamic functional that is employed. Consequently, one should in principle write $a^\natural = \delta \mathcal{S}/ \delta a$ for the conjugate to the variable $a$ that is based on the entropy functional $\mathcal{S}$, in analogy to the notation $a^\natural = \delta \mathcal{F}_{\mathrm{dim}}/ \delta a$ with the dimensional free-energy functional\footnote{We use a subscript dim to distinguish the standard free-energy functional with dimension \ZT{energy} from the dimensionless functional used in this work.} $\mathcal{F}_{\mathrm{dim}}$ used in previous work on EDDFT \cite{WittkowskiLB2012,WittkowskiLB2013}. In this work, we require \textit{both} $\delta \mathcal{S}/ \delta a$ and $-k_{\mathrm{B}}T\delta \mathcal{S}/ \delta a$. Since $-k_{\mathrm{B}} T\delta \mathcal{S}/ \delta a$ reduces to $\delta \mathcal{F}_{\mathrm{dim}}/ \delta a$ for isothermal systems if $\mathcal{F}$ is defined in the usual way, we write $a^\flat = \delta \mathcal{S}/ \delta a$ and $a^\natural = - k_{\mathrm{B}}T a^{\flat}$ such that $a^\natural$ gets its usual meaning in the isothermal limit.

The MZFT consists in the application of a projection operator $\hat{\mathcal{P}}_{t}=1-\hat{\mathcal{Q}}_{t}$ with 
\begin{equation}
\hat{\mathcal{Q}}_{t}X=X-\Tr(\rho(t) X)-\int_{\R^{3}}\!\!\!\!\!\:\!\dif^{3}r\,\Delta\hat{a}_{i}\rt
\Tr\!\Big(\Fdif{\rho(t)}{a_{i}\rt}X\Big) 
\label{eq:Q_Operator}
\end{equation}
and $\Delta\hat{a}_{i}\rt=\hat{a}_{i}(\vec{r})-a_{i}\rt$ that projects from the huge phase space onto a space that is spanned only by the chosen relevant variables: 
$\hat{\mathcal{P}}_{t}\hat{a}_{i}\rt=\hat{a}_{i}\rt$.
In case of exclusively conserved\footnote{At this point, we make the approximation that the contributions of the external force to the memory kernel are negligible, such that it only contributes to the organized drift and can thus simply be added on the level of the macroscopic transport equation for $\vec{g}$. This assumption is very common and allows for a significant simplification of the resulting transport equations. A more general model can also be derived in the MZFT framework.} relevant variables, the exact transport equations read
\begin{equation}
\begin{split}
\dot{a}_{i}\rt &= - \Nabla_{\vec{r}}\!\cdot\!\Tr\!\big(\rho(t)\hat{\vec{J}}^{(i)}(\vec{r})\big) \\
&\quad\, - \sum^{n}_{j=1}\Nabla_{\vec{r}}\!\cdot\!\!\int_{\R^{3}}\!\!\!\!\!\:\!\dif^{3}r' \INT{0}{t}{t'} R^{(ij)}(\vec{r},\vec{r}'\!,t,t')
\Nabla_{\rs}a^{\flat}_{j}(\rs\!,t')
\end{split}\raisetag{4em}\label{eq:EDDFT_exact}%
\end{equation}
with the memory kernel 
\begin{equation}
R^{(ij)}_{kl}(\vec{r},\vec{r}'\!,t,t') = \Tr\!\Big(\rho(t')\big(\hat{\mathcal{Q}}_{t'} J^{(j)}_l(\vec{r}')\big)G(t',t)\big(\hat{\mathcal{Q}}_t J^{(i)}_k(\vec{r})\big)\Big)
\label{rij}
\end{equation}
and the operator
\begin{equation}
G(t',t)=\exp_R \!\bigg(\INT{t'}{t}{t''} \mathcal{L} \mathcal{Q}_{t''}\bigg), 
\end{equation}
where $\exp_R$ is a right-time-ordered exponential. 

\subsection{\label{eddfteq}Extended dynamical density functional theory}
If we moreover assume slow\footnote{One makes here the approximation that the memory term appearing in the exact transport equations \eqref{eq:EDDFT_exact} can be approximated by a dissipative contribution that is local in time. Moreover, we have in \cref{eq:EDDFT_exact} assumed that the initial phase space density is given by the relevant density, otherwise \cref{eq:EDDFT_allg} would include a noise term. A more detailed discussion of the necessary approximations can be found in the literature \cite{teVrugtW2019,Grabert1982,teVrugt2022}.} relevant variables, this projection leads to the \textit{general EDDFT equations} 
\begin{equation}
\begin{split}
\dot{a}_{i}\rt &= - \Nabla_{\vec{r}}\!\cdot\!\Tr\!\big(\rho(t)\hat{\vec{J}}^{(i)}(\vec{r})\big) + \Tr(\rho(t)\hat{Q}_i(\vec{r}))\\
&\quad\:\! - \sum^{n}_{j=1}\Nabla_{\vec{r}}\!\cdot\!\!\int_{\R^{3}}\!\!\!\!\!\:\!\dif^{3}r'\, \DT^{(ij)}(\vec{r},\rs\!,t)
\Nabla_{\rs}a^{\flat}_{j}(\rs\!,t)
\end{split}\label{eq:EDDFT_allg}%
\end{equation}
with the reversible contributions $\Tr\!\big(\rho(t)\hat{\vec{J}}^{(i)}(\vec{r})\big)$ and the diffusion tensors
\begin{equation}
\DT^{(ij)}_{kl}(\vec{r},\rs\!,t)=
\!\int^{\infty}_{0}\!\!\!\!\!\!\dif t'\,\Tr\!\Big(\rho(t)\big(\hat{\mathcal{Q}}_{t}\hat{J}^{(j)}_{l}(\rs)\big)
e^{\mathcal{L}t'}\!\big(\hat{\mathcal{Q}}_{t}\hat{J}^{(i)}_{k}(\vec{r})\big)\!\Big) .
\label{eq:Dijkl_allg}%
\end{equation}
The general EDDFT equations \eqref{eq:EDDFT_allg} and \eqref{eq:Dijkl_allg} correspond to Eqs.\ (58) and (59) in the first EDDFT article by \citet{WittkowskiLB2012}, with differences arising from the different definition of the thermodynamic conjugates (we introduce them here based on the entropy and not based on the free energy as done previously \cite{WittkowskiLB2012}) and from the inclusion of a source term $\Tr(\rho(t)\hat{Q}_i\rt)$ in the organized drift. 
 
Equations \eqref{eq:EDDFT_allg} and \eqref{eq:Dijkl_allg} are now specified to our particular set of relevant variables.
For the reversible contributions in Eq.\ \eqref{eq:EDDFT_allg} we obtain\footnote{See the textbook by \citet{Grabert1982} for some useful mathematical manipulations. In Eq.\ \eqref{eq:Tr_e}, we assumed that the pair-interaction described by the potential $U^{(\mu\nu)}_{2}(r)$ is short-range so that the velocity field $\vec{g}^{\natural}\rt$ (see Eq.\ \eqref{gnatural} below) is approximately constant on the length scale of the particle-particle interaction.} 
{\allowdisplaybreaks%
\begin{gather}%
\begin{split}%
\Tr\!\big(\rho(t)\hat{\vec{J}}^{\rhom }(\vec{r})\big)=\vec{g}\rt ,
\end{split}\label{eq:Tr_rhom}\\%
\begin{split}%
\Tr\!\big(\rho(t)\hat{\vec{J}}^{c}(\vec{r})\big)=c\rt\vec{g}^{\natural}\rt ,
\end{split}\label{eq:Tr_c}\\%
\begin{split}%
\Tr\!\big(\rho(t)\hat{\mathrm{J}}^{\vec{g}}(\vec{r})\big)
&=\rhom \rt\vec{g}^{\natural}\rt\!\:\!\otimes\!\:\!\vec{g}^{\natural}\rt 
+\Pi\rt, \\
\end{split}\label{eq:Tr_g}\\%
\begin{split}%
\Tr\!\big(\rho(t)\hat{\vec{J}}^{\varepsilon}(\vec{r})\big)= 
\big(\varepsilon\rt\:\!\Eins+\Pi\rt\big)\vec{g}^{\natural}\rt
\end{split}\label{eq:Tr_e}%
\end{gather}}%
with the identity matrix $\Eins$ and the pressure tensor \cite{KadanoffB1989}
{\allowdisplaybreaks%
\begin{align}%
\Pi\rt &= \Pi_{0}\rt+\Pi_{1}\rt,\label{eq:PI}\\
\Pi_{0}\rt &= \frac{\Eins}{\beta\rt\:\! m_{\mathrm{s}}}
\big(\rhom \rt -(m_{\mathrm{c}}-m_{\mathrm{s}})c\rt\big),
\label{eq:PIN}\\%
\begin{split}%
\Pi_{1}\rt &=
-\frac{1}{2}\!\!\!\!\!\sum_{\mu,\nu\in\{\mathrm{c},\mathrm{s}\}}\!\!\!\!\!\int_{\R^{3}}\!\!\!\!\!\:\!\dif^{3}r'
\frac{\rs\!\!\:\!\otimes\!\:\!\rs}{\norm{\rs}} U^{(\mu\nu)\prime}_{2}(\norm{\rs}) \\
&\qquad \int^{1}_{0}\!\!\!\!\dif\lambda\,\:\! \rho^{(2)}_{\mu\nu}(\vec{r}+\lambda\rs\!,\vec{r}+(\lambda-1)\rs\!,t) ,
\end{split}\label{eq:PII}%
\end{align}}%
the derivative $U^{(\mu\nu)\prime}_{2}(r)=\dif U^{(\mu\nu)}_{2}(r)/\dif r$, and the two-particle densities \cite{WittkowskiLB2012}
\begin{equation}
\begin{split}%
\rho^{(2)}_{\mu\nu}(\vec{r},\rs\!,t)=\!\!\!\underset{(\mu,i)\neq(\nu,j)}{\sum^{N_{\mu}}_{i=1}\sum^{N_{\nu}}_{j=1}}\!
\Tr\!\big(\tilde{\rho}(t)\delta\big(\vec{r}-\vec{r}^{\mu}_{i}\big)\delta\big(\rs\!-\vec{r}^{\nu}_{j}\big)\big) , 
\end{split}\label{eq:rhoII}\raisetag{2em}%
\end{equation}
with the Galileian-transformed relevant density $\tilde{\rho}$ given by \cref{eq:rho_GT}.
If the pair-interaction potentials $U^{(\mu\nu)}_{2}(r)$ are short-ranged and the system is close to local thermodynamic equilibrium, the two-particle densities $\rho^{(2)}_{\mu\nu}(\vec{r},\rs\!,t)$ are locally translationally and rotationally invariant and one can approximate 
$\rho^{(2)}_{\mu\nu}(\vec{r}+\lambda\rs\!,\vec{r}+(\lambda-1)\rs\!,t)\approx \rho^{(\mu\nu)}_{\mathrm{LTE}}(\norm{\rs},t)$ in Eq.\ \eqref{eq:PII}. 
The pressure tensor $\Pi\rt$ then reduces to 
\begin{equation}
\Pi\rt=p\rt \Eins 
\label{eq:pkI}%
\end{equation}
with the scalar hydrostatic pressure $p\rt$ and its virial expression (the special case of a one-component system is discussed in the literature \cite{Grabert1982,HansenMD2009}) 
\begin{equation}
\begin{split}
p\rt &= \frac{1}{\beta\rt\:\! m_{\mathrm{s}}}\big(\rhom \rt
-(m_{\mathrm{c}}-m_{\mathrm{s}})c\rt\big) \\
&\quad\:\! -\frac{2\pi}{3} \!\!\!\!\!\sum_{\mu,\nu\in\{\mathrm{c},\mathrm{s}\}}\INT{0}{\infty}{r'} r'^3 U^{(\mu\nu)\prime}_{2}(r')\rho^{(\mu\nu)}_{\mathrm{LTE}}(r',t)
\end{split}\label{eq:pkII}\raisetag{2.5em}%
\end{equation}
with $r'=\norm{\rs}$.
Notice that the pressure tensor $\Pi\rt$ is a functional of the total mass density $\rhom \rt$, of the local concentration $c\rt$, and of the energy density $\varepsilon\rt$, but that due to Galilean invariance $\Pi\rt$ is independent of the momentum density $\vec{g}\rt$.

With Eqs.\ \eqref{eq:J_rhom}-\eqref{eq:J_e} and \eqref{eq:Q_Operator}, Eqs.\ \eqref{eq:Tr_rhom}-\eqref{eq:Tr_e} 
directly lead to the projected currents 
{\allowdisplaybreaks%
\begin{align}%
\begin{split}%
\hat{\mathcal{Q}}_{t}\hat{\vec{J}}^{\rhom }(\vec{r},\tau) &= \vec{0} ,
\end{split}\label{eq:QJ_rhom}\\%
\begin{split}%
\hat{\mathcal{Q}}_{t}\hat{\vec{J}}^{c}(\vec{r},\tau) &= \hat{\vec{J}}^{c}(\vec{r},\tau)
+\hat{\rho}_{\mathrm{m}}(\vec{r},\tau)\frac{c\rt\vec{g}\rt}{\rho^{2}_{\mathrm{m}}\rt} \\
&\quad\:\! -\hat{c}(\vec{r},\tau)\frac{\vec{g}\rt}{\rhom \rt}
-\hat{\vec{g}}(\vec{r},\tau)\frac{c\rt}{\rhom \rt} ,
\end{split}\label{eq:QJ_c}\\%
\begin{split}%
\hat{\mathcal{Q}}_{t}\hat{\mathrm{J}}^{\vec{g}}(\vec{r},\tau) &= \hat{\mathrm{J}}^{\vec{g}}(\vec{r},\tau) \\
&\quad\:\! +\hat{\rho}_{\mathrm{m}}(\vec{r},\tau)\Big(\frac{\vec{g}\rt\!\:\!\otimes\!\:\!
\vec{g}\rt}{\rho^{2}_{\mathrm{m}}\rt}-\frac{\Eins}{\beta\rt\:\! m_{\mathrm{s}}}\Big) \\
&\quad\:\! +\hat{c}(\vec{r},\tau)\frac{m_{\mathrm{c}}-m_{\mathrm{s}}}{\beta\rt\:\! m_{\mathrm{s}}}\Eins \\
&\quad\:\!-\frac{\hat{\vec{g}}(\vec{r},\tau)\!\:\!\otimes\!\:\!\vec{g}\rt
+\vec{g}\rt\!\:\!\otimes\!\:\!\hat{\vec{g}}(\vec{r},\tau)}{\rhom \rt}  \\
&\quad\:\! -\Delta\hat{\Upsilon}_{t}(\vec{r},\tau)-\Pi_{1}\rt ,
\end{split}\label{eq:QJ_g}\raisetag{6em}\\%
\begin{split}%
\hat{\mathcal{Q}}_{t}\hat{\vec{J}}^{\varepsilon}(\vec{r},\tau) &= 
\hat{\vec{J}}^{\varepsilon}(\vec{r},\tau) 
+\hat{\rho}_{\mathrm{m}}(\vec{r},\tau)\Big(\frac{\varepsilon\rt\:\!\Eins+\Pi\rt}{\rhom \rt} \\
&\quad\:\! -\frac{\Eins}{\beta\rt\:\! m_{\mathrm{s}}}\Big)\frac{\vec{g}\rt}{\rhom \rt} \\
&\quad\:\! +\hat{c}(\vec{r},\tau)\frac{m_{\mathrm{c}}-m_{\mathrm{s}}}{\beta\rt\:\! m_{\mathrm{s}}}\frac{\vec{g}\rt}{\rhom \rt} \\
&\quad\:\! -\frac{\hat{\vec{g}}(\vec{r},\tau)}{\rhom \rt}\big(\varepsilon\rt\:\!\Eins+\Pi\rt\big) \\
&\quad\:\! -\hat{\varepsilon}(\vec{r},\tau)\frac{\vec{g}\rt}{\rhom \rt} \\
&\quad\:\! -\big(\Delta\hat{\Upsilon}_{t}(\vec{r},\tau)+\Pi_{1}\rt\big)\frac{\vec{g}\rt}{\rhom \rt}
\end{split}\label{eq:QJ_e}\raisetag{9em}%
\end{align}}%
with the generalized microscopic bulk modulus 
\begin{equation}
\begin{split}
\hat{\Upsilon}_{t}(\vec{r},\tau)&=\int_{\R^{3}}\!\!\!\!\:\!\dif^{3}r'\Big(\hat{\rho}_{\mathrm{m}}(\rs\!,\tau)
\Fdif{\Pi_{1}\rt}{\rhom (\rs\!,t)}
+\hat{c}(\rs\!,\tau)\Fdif{\Pi_{1}\rt}{c(\rs\!,t)} \\
&\quad\:\! +\hat{\varepsilon}(\rs\!,\tau)\Fdif{\Pi_{1}\rt}{\varepsilon(\rs\!,t)}\Big) ,
\end{split}\raisetag{2em}%
\end{equation}
its average $\Upsilon_{t}(\vec{r},\tau)=\Tr(\rho(\tau)\hat{\Upsilon}_{t}(\vec{r}))$, and the reduced microscopic bulk modulus $\Delta\hat{\Upsilon}_{t}(\vec{r},\tau)=\hat{\Upsilon}_{t}(\vec{r},\tau)-\Upsilon_{t}(\vec{r},\tau)$. 
Hence, the \textit{EDDFT equations for colloidal suspensions} are 
{\allowdisplaybreaks%
\begin{align}%
\begin{split}%
\dot{\rho}_{\mathrm{m}}\rt &= -\Nabla_{\vec{r}}\!\cdot\!\vec{g}\rt ,
\end{split}\label{eq:EDDFT_rhom}\\%
\begin{split}%
\dot{c}\rt &= -\Nabla_{\vec{r}}\!\cdot\!\big(c\rt\vec{g}^{\natural}\rt\big) \\
&\quad\:\! -\Nabla_{\vec{r}}\!\cdot\!\!\int_{\R^{3}}\!\!\!\!\!\:\!\dif^{3}r'\, \DT^{(cc)}(\vec{r},\rs\!,t) 
\Nabla_{\rs}c^{\flat}(\rs\!,t) \\
&\quad\:\! -\Nabla_{\vec{r}}\!\cdot\!\!\int_{\R^{3}}\!\!\!\!\!\:\!\dif^{3}r'\, \DT^{(c\varepsilon)}(\vec{r},\rs\!,t) \Nabla_{\rs}\varepsilon^{\flat}(\rs\!,t) ,
\end{split}\label{eq:EDDFT_c}\\%
\begin{split}%
\dot{g}_{i}\rt &= -\big(\Nabla_{\vec{r}}\!\cdot\!\big(\rhom \rt\vec{g}^{\natural}\rt
\!\:\!\otimes\!\:\!\vec{g}^{\natural}\rt+\Pi\rt \big)\big)_{i} \\
&\quad\:\! -\sum^{3}_{j=1}\Nabla_{\vec{r}}\!\cdot\!\!\int_{\R^{3}}\!\!\!\!\!\:\!\dif^{3}r'\, \DT^{(g_{i}g_{j})}(\vec{r},\rs\!,t)
\Nabla_{\rs}g^{\flat}_{j}(\rs\!,t) \\
&\quad\:\! + F_i\rt,
\end{split}\label{eq:EDDFT_g}\raisetag{2.5em}\\%
\begin{split}%
\dot{\varepsilon}\rt &= -\Nabla_{\vec{r}}\!\cdot\!\big((\varepsilon\rt\Eins+\Pi\rt)\vec{g}^{\natural}\rt\big) \\
&\quad\:\! -\Nabla_{\vec{r}}\!\cdot\!\!\int_{\R^{3}}\!\!\!\!\!\:\!\dif^{3}r'\, \DT^{(\varepsilon c)}(\vec{r},\rs\!,t) \Nabla_{\rs}c^{\flat}(\rs\!,t) \\
&\quad\:\! -\Nabla_{\vec{r}}\!\cdot\!\!\int_{\R^{3}}\!\!\!\!\!\:\!\dif^{3}r'\, \DT^{(\varepsilon\varepsilon)}(\vec{r},\rs\!,t) \Nabla_{\rs}\varepsilon^{\flat}(\rs\!,t) .
\end{split}\label{eq:EDDFT_e}%
\end{align}}%
The EDDFT equations \eqref{eq:EDDFT_rhom}-\eqref{eq:EDDFT_e} are an important and far-reaching extension of standard DDFT \cite{MarconiT1999,MarconiT2000,ArcherE2004,EspanolL2009} and therefore constitute the main result of this article. Explicit expressions for the diffusion tensors are given in \cref{diffusiontensors}.

\subsection{\label{momentum}Momentum density}
One can show for fluids with only one velocity the relation $\vec{g}\rt=\rhom \rt\vec{g}^{\natural}\rt$
between the total momentum density $\vec{g}\rt$ and its thermodynamic conjugate $\vec{g}^{\natural}\rt$ 
\cite{HohenbergM1965,Khalatnikov1989}. To see this, note that \cite{Grabert1982}
\begin{equation}
\begin{split}
\Tr(\tilde{\rho}(t)\hat{\vec{g}}\rt) &= \Tr(\rho(t)(\hat{\vec{g}}\rt - g^\natural\rt \hat{\rho}_\mathrm{m}\rt) \\
&= \vec{g}\rt - g^\natural\rt \rhom \rt = 0.
\end{split}
\end{equation}
In the second step, we have performed an inverse Galilei transformation, and in the last step we have used the fact that the Galilean-transformed density $\tilde{\rho}$ given by \cref{eq:rho_GT} is even in the momenta. The relation $\vec{g}\rt=\rhom \rt\vec{g}^{\natural}\rt$ allows to identify the function $\vec{g}^{\natural}\rt$ with the velocity field $\vec{v}\rt\equiv\vec{g}^{\natural}\rt$ and thus provides an explicit expression for it:
\begin{equation}
\vec{g}^{\natural}\rt= \frac{\vec{g}\rt}{\rhom \rt} .
\label{gnatural}
\end{equation}
The free-energy functional $\mathcal{F}[\rhom ,c,\vec{g},T]$ of a hard-sphere system (derived in \cref{twoparticle}) can be written as
\begin{equation}
\begin{split}
\mathcal{F}[\rhom ,c,\vec{g},T] &= \mathcal{F}_{1}[\rhom ,\vec{g},T]
+\mathcal{F}_{2}[\rhom ,c,T] ,\\
\mathcal{F}_{1}[\rhom ,\vec{g},T] &=  \int_{\R^{3}}\!\!\!\!\!\:\!\dif^{3}r\:\!\frac{\beta\rt(\vec{g}\rt)^{2}}{2\rhom \rt} 
\end{split}
\label{f1f2}
\end{equation}
with the contribution $\mathcal{F}_{2}[\rhom ,c,T]$ that is independent of the momentum density. This leads to the relation 
\begin{equation}
\vec{g}^\natural\rt = \frac{1}{\beta\rt}\Fdif{\mathcal{F}}{\vec{g}\rt}. 
\label{gnaturalf}
\end{equation}
Similarly, from the entropy functional $\mathcal{S}[\rhom,c,\vec{g},\varepsilon]$, we can obtain the velocity field as
\begin{equation}
\vec{g}^\natural\rt = - \frac{1}{\beta\rt}\Fdif{\mathcal{S}}{\vec{g}\rt}.    
\end{equation}
as shown in \cref{consistencycheck} below. For an isothermal system with $\beta\rt = \beta_0=1/(k_\mathrm{B}T_0)$ (where $\beta_0$ is a constant inverse energy and $T_0$ a constant reference temperature), a dimensional free-energy functional can be defined as 
\begin{equation}
\mathcal{F}_{\mathrm{dim}} = k_{\mathrm{B}}T_0 \mathcal{F},
\label{dimensionalization}
\end{equation} 
\cref{gnaturalf} simplifies to
\begin{equation}
\vec{g}^\natural\rt =\Fdif{\mathcal{F}_{\mathrm{dim}}}{\vec{g}\rt},
\end{equation}
showing that $\vec{g}^\natural = \vec{v}$ is indeed the thermodynamic conjugate of $\vec{g}$.

\section{\label{ddft}Entropy, free energy, and adiabatic approximation}
For the general EDDFT equations \eqref{eq:EDDFT_rhom}-\eqref{eq:EDDFT_e} to be useful in practice, one requires explicit expressions for the entropy functional $\mathcal{S}$ and for the pressure term $\Nabla \cdot\Pi$. In fact, these problems are closely related, since extensions of DDFT that include the momentum density typically approximate the pressure term using a free-energy functional \cite{Archer2009,BurghardtB2006}. Hence, we will address the problem of deriving explicit expressions in three steps: First (\cref{hard}), we construct the entropy functional (which is done for the important case of hard spheres). Second (\cref{twoparticle}), we use this result to obtain the free-energy functional. Third (\cref{adiabatic}), the free-energy functional is used to approximate the organized drift of momentum and energy density (\ZT{adiabatic approximation}).

\subsection{\label{hard}Entropy functional}
In this subsection, we will consider the special case of a system of hard spheres, for which we will derive the explicit form of the entropy functional $\mathcal{S}$. This has two main reasons: First, hard spheres are arguably among the most important systems DFT and DDFT are applied to, such that by deriving an EDDFT functional for hard spheres, we significantly improve the applicability of EDDFT in practice. Second, hard spheres are \textit{athermal}, which, as demonstrated by \citet{AneroET2013}, allows to construct the entropy functional based on a known free-energy functional. Therefore, free-energy functionals such as that of fundamental measure theory (FMT) \cite{Roth2010,Rosenfeld1989}, which provides an accurate description of hard spheres in (D)DFT, can, with little modification, also be used in EDDFT. Here, we will generalize the derivation by \citet{AneroET2013} (which was done for one particle species and which did not take into account the momentum density) to a mixture of particles where convection is relevant. To simplify the notation, we will drop the dependence on $\vec{r}$ and $t$ for most of this section unless this dependence is not clear. 

We start by introducing the dimensionless grand-canonical potential
\begin{equation}
\Omega[\rhom ^\flat,c^\flat,\vec{g}^\flat,\beta] = - \ln(Z),  
\label{omega}
\end{equation}
which is a functional of the conjugate variables. It is related to the entropy functional, which depends on the relevant variables, by the Legendre transformation
\begin{equation}
\begin{split}
\mathcal{S}[\rhom ,c,\vec{g},\varepsilon] &= - \Omega[\rhom ^\flat,c^\flat,\vec{g}^\flat,\beta] \\
&\quad\:\! + \int_{\R^{3}}\!\!\!\!\:\!\dif^{3}r \Big(\rhom ^\flat \rhom  + c^\flat c  + \sum_{i=1}^{3}g_i^\flat g_i + \beta \varepsilon\Big).
\end{split}\label{legend}\raisetag{4em}%
\end{equation}
Using the potential \eqref{omega}, we can compute the relevant variables by
\begin{equation}
\rhom =\Fdif{\Omega}{\rhom ^{\flat}} ,\quad
c=\Fdif{\Omega}{c^{\flat}} ,\quad
g_{i}=\Fdif{\Omega}{g_{i}^{\flat}} ,\quad
\varepsilon=\Fdif{\Omega}{\beta}.
\label{eq:TC4}%
\end{equation}
For a general set of $n$ relevant variables $\{\hat{a}_i\rt\}$ with conjugates $\{a^\flat_i\rt\}$, the partition function is given by
\begin{equation}
Z(t)=\Tr\!\Big(\exp\!\Big(-\sum_{i=1}^{n}\INT{}{}{^3r}a^\flat_i\hat{a}_i\Big)\Big).
\label{zvont}
\end{equation}
We start by considering the isothermal limit, where the functional dependence of $Z$ and $\Omega$ on $\beta$ becomes a parametric dependence on the constant $\beta_0$. Moreover, we first ignore the momentum density. In this case, the partition function \eqref{zvont} is given by
\begin{widetext}
\begin{equation}
\begin{split}
Z_\mathrm{iso}[\rhom ^\flat,c^\flat,(\beta_0)] &=  \sum^{\infty}_{N_{\mathrm{c}},N_{\mathrm{s}}=0}\frac{1}{N_\mathrm{c}! N_\mathrm{s}!}\frac{1}{h^{3 N_\mathrm{c}}}\frac{1}{h^{3 N_\mathrm{s}}}\prod_{\mu \in \{\mathrm{c},\mathrm{s}\}}\prod_{i=1}^{N_\mathrm{\mu}}\int_{\R^{3}}\!\!\!\!\!\:\!\dif^{3}r_i^\mu \int_{\R^{3}}\!\!\!\!\!\:\!\dif^{3}p_i^\mu\\ 
&\quad\:\! \exp\!\bigg(-\beta_0\hat{H} 
- \int_{\R^{3}}\!\!\!\!\!\:\!\dif^{3}r \,\Big(\rho^{\flat}_{\mathrm{m}}(\vec{r})\hat{\rho}_{\mathrm{m}}(\vec{r})
+c^{\flat}(\vec{r})\hat{c}(\vec{r})\Big)\bigg) \\
&= \sum^{\infty}_{N_{\mathrm{c}},N_{\mathrm{s}}=0}\frac{1}{N_\mathrm{c}! N_\mathrm{s}!}\prod_{\mu \in \{\mathrm{c},\mathrm{s}\}}\prod_{i=1}^{N_\mathrm{\mu}}\int_{\R^{3}}\!\!\!\!\!\:\!\dif^{3}r_i^\mu \Big(\frac{2 m_\mathrm{\mu} \pi h^2}{\beta_0}\Big)^{\frac{3}{2}}\\ 
&\quad\:\! \exp\!\bigg(-\sum_{\mu \in \{\mathrm{c},\mathrm{s}\}}\sum_{i=1}^{N_\nu}\Big(\beta_0U_1^\nu(\vec{r}_i^\mu)
+ m_\mu\rho^{\flat}_{\mathrm{m}}(\vec{r}_i^\mu)\Big)
-\sum_{i=1}^{N_c}c^{\flat}(\vec{r}^\mathrm{c}_i)\bigg) f_{\mathrm{iso}}(\{\vec{r}_i^\mu \})
\end{split}    
\label{ziso}
\end{equation}
with the Planck constant $h$ and the function 
\begin{equation}
f_{\mathrm{iso}}(\{\vec{r}_i^\mu \}) = \exp\!\bigg(-\frac{1}{2}\!\!\underset{(\mu,i)\neq(\nu,j)}{\sum_{\nu\in\{\mathrm{c},\mathrm{s}\}}\sum^{N_{\nu}}_{j=1}}\,
\beta_0 U^{(\mathrm{\mu\nu})}_{2}(\norm{\vec{r}^{\mu}_{i}\!-\vec{r}^{\nu}_{j}})\Big)\bigg).
\label{function1}
\end{equation}
We denote the partition function in the isothermal case by $Z_\mathrm{iso}$ to distinguish it from the nonisothermal partition function $Z$. In the second step, we have performed the spatial integral in the exponent as well as the integrals over the momenta. For the nonisothermal case with momentum density, we apply to \cref{zvont} the Galilei transformation $\vec{p}_i^\mu \to \vec{p}_i^\mu - \beta^{-1}(\vec{r}_i^\mu)g^\flat(\vec{r}_i^\mu)m_\mu$, which is possible by the Galilei invariance of the trace. We find
\begin{equation}
\begin{split}
Z[\rhom ^\flat,c^\flat,\vec{g}^\flat,\beta] &=  \sum^{\infty}_{N_{\mathrm{c}},N_{\mathrm{s}}=0}\frac{1}{N_\mathrm{c}! N_\mathrm{s}!}\frac{1}{h^{3 N_\mathrm{c}}}\frac{1}{h^{3 N_\mathrm{s}}}\prod_{\mu \in \{\mathrm{c},\mathrm{s}\}}\prod_{i=1}^{N_\mathrm{\mu}}\int_{\R^{3}}\!\!\!\!\!\:\!\dif^{3}r_i^\mu \int_{\R^{3}}\!\!\!\!\!\:\!\dif^{3}p_i^\mu\\ 
&\quad\:\! \exp\!\Big(-\int_{\R^{3}}\!\!\!\!\!\:\!\dif^{3}r \,\Big(\beta (\vec{r}) \hat{\varepsilon} 
+ \Big(\rho^{\flat}_{\mathrm{m}}(\vec{r})
-\text{\footnotesize$\frac{1}{2\beta (\vec{r})}$}\big(\vec{g}^{\flat}(\vec{r})\big)^{2}\Big)\hat{\rho}_{\mathrm{m}}(\vec{r})
+c^{\flat}(\vec{r})\hat{c}(\vec{r})\Big)\Big) \\
&=  \sum^{\infty}_{N_{\mathrm{c}},N_{\mathrm{s}}=0}\frac{1}{N_\mathrm{c}! N_\mathrm{s}!}\prod_{\mu \in \{\mathrm{c},\mathrm{s}\}}\prod_{i=1}^{N_\mathrm{\mu}}\int_{\R^{3}}\!\!\!\!\!\:\!\dif^{3}r_i^\mu \Big(\frac{2 m_\mathrm{\mu} \pi h^2}{\beta (\vec{r}_i^\mu)}\Big)^{\frac{3}{2}}\\
&\quad\:\! \exp\!\Big(-\sum_{\mu \in \{\mathrm{c},\mathrm{s}\}}\sum_{i=1}^{N_\nu}\Big(\beta(\vec{r}_i^\mu) U_1^\nu(\vec{r}_i^\mu)
+ m_\mu\Big(\rho^{\flat}_{\mathrm{m}}(\vec{r}_i^\mu)
-\text{\footnotesize$\frac{1}{2\beta (\vec{r}_i^\mu)}$}\big(\vec{g}^{\flat}(\vec{r}_i^\mu)\big)^{2}\Big)\Big)
-\sum_{i=1}^{N_c}c^{\flat}(\vec{r}^\mathrm{c}_i)\Big) f(\{\vec{r}_i^\mu \})\\
&= \sum^{\infty}_{N_{\mathrm{c}},N_{\mathrm{s}}=0}\frac{1}{N_\mathrm{c}! N_\mathrm{s}!}\prod_{\mu \in \{\mathrm{c},\mathrm{s}\}}\prod_{i=1}^{N_\mathrm{\mu}}\int_{\R^{3}}\!\!\!\!\!\:\!\dif^{3}r_i^\mu \Big(\frac{2 m_\mathrm{\mu} \pi h^2}{\beta_0}\Big)^{\frac{3}{2}}\\
&\quad\:\! \exp\!\Big(-\sum_{\mu \in \{\mathrm{c},\mathrm{s}\}}\sum_{i=1}^{N_\nu}\Big(\beta_0 U_1^\nu(\vec{r}_i^\mu)
+ m_\mu\Big(\rho^{\flat}_{\mathrm{m}}(\vec{r}_i^\mu)
-\text{\footnotesize$\frac{1}{2\beta (\vec{r}_i^\mu)}$}\big(\vec{g}^{\flat}(\vec{r}_i^\mu)\big)^{2} + \frac{\beta(\vec{r}_i^\mu) - \beta_0}{m_\mathrm{s}}U_1^\mathrm{s}(\vec{r}_i^\mu) + \frac{3}{2m_\mathrm{s}}\ln\Big(\frac{\beta (\vec{r}_i^\mu)}{\beta_0}\Big)\Big)\Big) \\
&\quad\:\! -\sum_{i=1}^{N_c}\Big(c^{\flat}(\vec{r}_c) + (\beta(\vec{r}_i^c) - \beta_0)\Big(U_1^\mathrm{c} (\vec{r}_i^\mathrm{c}) - \frac{m_\mathrm{c}}{m_\mathrm{s}}U_1^\mathrm{s}(\vec{r}_i^\mathrm{s})\Big) + \frac{3}{2}\Big(1-\frac{m_\mathrm{c}}{m_\mathrm{s}}\Big)\ln\!\Big(\frac{\beta(\vec{r}_i^\mathrm{c})}{\beta_0}\Big)\Big) f(\{\vec{r}_i^\mu \})
\end{split}    
\label{znoniso}\raisetag{14em}%
\end{equation}
\end{widetext}
with the function
\begin{equation}
f(\{\vec{r}_i^\mu \}) = \exp\!\Big(-\frac{1}{2}\!\!\underset{(\mu,i)\neq(\nu,j)}{\sum_{\nu\in\{\mathrm{c},\mathrm{s}\}}\sum^{N_{\nu}}_{j=1}}\,
\beta(\vec{r}^{\mu}_{i}) U^{(\mathrm{\mu\nu})}_{2}(\norm{\vec{r}^{\mu}_{i}\!-\vec{r}^{\nu}_{j}})\Big)\Big).
\label{function2}
\end{equation}
For hard spheres, the function $f$ is zero if any two spheres overlap and one otherwise, such that $f$ does not depend on the temperature (athermal system) \cite{AneroET2013}. Consequently, the functions $f_{\mathrm{iso}}$ given by \cref{function1}, and the function $f$ given by \cref{function2} are identical, which is why we can replace $\beta(\vec{r}_i^\mu)$ by $\beta_0$ in the last step of Eq.\ \eqref{function2}. Comparing \cref{ziso,znoniso} shows that
\begin{equation}
\begin{split}
& Z[\rhom ^\flat,c^\flat,\vec{g}^\flat,\beta] \\
&= Z_\mathrm{iso}\Big[\rhom ^\flat + \frac{\beta - \beta_0}{m_\mathrm{s}}U_1^\mathrm{s} - \frac{(\vec{g}^\flat)^2}{2\beta} + \frac{3}{2m_\mathrm{s}}\ln\!\Big(\frac{\beta}{\beta_0}\Big),\\
&\quad\;\:\! c^\flat + (\beta-\beta_0)\Big(U_1^\mathrm{c} - \frac{m_\mathrm{c}}{m_\mathrm{s}}U_1^\mathrm{s}\Big) \\
&\quad\:\! + \frac{3}{2}\Big(1-\frac{m_\mathrm{c}}{m_\mathrm{s}}\Big)\ln\!\Big(\frac{\beta}{\beta_0}\Big),(\beta_0)\Big].
\end{split}\label{zcomparison}
\end{equation} 
Using \cref{omega,zcomparison} then gives
\begin{equation}
\begin{split}
&\Omega[\rhom ^\flat,c^\flat,\vec{g}^\flat,\beta] \\
&= \Omega_\mathrm{iso}\Big[\rhom ^\flat + \frac{\beta - \beta_0}{m_\mathrm{s}}U_1^\mathrm{s} - \frac{(\vec{g}^\flat)^2}{2\beta} 
+ \frac{3}{2m_\mathrm{s}}\ln\!\Big(\frac{\beta}{\beta_0}\Big),\\
&\quad\;\:\! c^\flat + (\beta-\beta_0)\Big(U_1^\mathrm{c} - \frac{m_\mathrm{c}}{m_\mathrm{s}}U_1^\mathrm{s}\Big) \\
&\quad\:\! + \frac{3}{2}\Big(1-\frac{m_\mathrm{c}}{m_\mathrm{s}}\Big)\ln\!\Big(\frac{\beta}{\beta_0}\Big),(\beta_0)\Big].
\end{split}\label{omegacomparison}
\end{equation} 
From \cref{eq:TC4,omegacomparison}, we find
\begin{align}
\rhom  &= \frac{\delta \Omega_\mathrm{iso}}{\delta \rhom ^\flat},\\
c &= \frac{\delta \Omega_\mathrm{iso}}{\delta c^\flat},\\
\vec{g} &= -\rhom \frac{\vec{g}^\flat}{\beta},\label{gflat}\\ 
\begin{split}
\varepsilon &= \frac{\rhom }{m_\mathrm{s}}U_1^\mathrm{s} 
+ \Big(U_1^\mathrm{c} - \frac{m_\mathrm{c}}{m_\mathrm{s}}U_1^\mathrm{s}\Big) c + \rhom\frac{(\vec{g}^\flat)^2}{2\beta^2} 
+ \frac{3}{2}\frac{\rhom }{m_\mathrm{s}\beta} \\
&\quad\:\! + \frac{3}{2}\Big(1 - \frac{m_\mathrm{c}}{m_\mathrm{s}}\Big)\frac{c}{\beta}.\label{varepsilon}\raisetag{2em}
\end{split}
\end{align}
The result \eqref{gflat} could also have been obtained from \cref{eq:TC2,gnatural}, thus confirming the consistency of our approach. Equation \eqref{varepsilon} looks very complicated, but this is mostly a consequence of our choice of relevant variables. We can introduce the total number density
\begin{equation}
\hat{n}(\vec{r}) = \sum_{\mu \in \{ \mathrm{c},\mathrm{s}\}}\sum_{i=1}^{N_\mu}\delta(\vec{r}-\vec{r}_i^\mu),
\end{equation}
with average value $n\rt = \Tr(\rho(t)\hat{n})$, which can be written as a linear combination of the relevant variables $\rhom $ and $c$ in the form
\begin{equation}
\hat{n} = \frac{1}{m_\mathrm{s}}\hat{\rho}_m + \Big(1 - \frac{m_\mathrm{c}}{m_\mathrm{s}}\Big) \hat{c}.\label{linearcombination}
\end{equation}
Moreover, we can introduce the number density of the solvent as
\begin{equation}
\hat{s} = \hat{n} - \hat{c}, 
\label{solvent}
\end{equation}
which has the average value $s\rt = \Tr(\rho(t)\hat{s})$. Using \cref{gflat,linearcombination,solvent}, \cref{varepsilon} takes the much simpler form
\begin{equation}
\varepsilon = \frac{3n}{2\beta} + c U_1^\mathrm{c} + s U_1^\mathrm{s} + \frac{\vec{g}^2}{2\rhom }.
\label{varepsilon2}%
\end{equation}
Equation \eqref{varepsilon2} has a clear physical interpretation: The energy density has a thermal contribution\footnote{For the isothermal case with $\vec{g} = \vec{0}$ and $U_1^\mathrm{c} + U_1^\mathrm{s} = 0$, \cref{varepsilon2} gives the standard relation $E = \frac{3}{2} N k_{\mathrm{B}}T$ with particle number $N$, known from elementary statistical mechanics.} (first term on the right-hand side), a contribution from the potential energy (second and third term) and a contribution that can be interpreted as a kinetic energy density arising from the momentum (fourth term). In particular, we can invert \cref{varepsilon2} to express $\beta$ (or, equivalently, the temperature $T =1/(k_{\mathrm{B}}\beta)$) to get
\begin{equation}
\beta = \frac{3}{2} \frac{n}{\varepsilon -c U_1^\mathrm{c} - s U_1^\mathrm{s} - \frac{\vec{g}^2}{2\rhom }}.  
\label{beta}
\end{equation}
In the isothermal overdamped case, we can construct the nondimensional isothermal intrinsic free energy $\mathcal{F}_\mathrm{iso}$ as
\begin{equation}
\begin{split}
&\mathcal{F}^0_\mathrm{iso}[\rhom ,c,(\beta_0)] \\
&= \Omega_{\mathrm{iso}}[\rho^\flat,c^\flat,(\beta_0)] \\
&\quad\:\! 
- \int_{\R^{3}}\!\!\!\!\:\!\dif^{3}r \big(\rhom ^\flat \rhom  + c^\flat c  
- \beta_0( s U_1^\mathrm{s} + c U_1^\mathrm{c})\big).
\end{split}\label{freeenergync}%
\end{equation}
For hard spheres, very good approximations for this free-energy functional have been developed in the framework of FMT \cite{Roth2010,Rosenfeld1989}. Following the procedure discussed by \citet{AneroET2013}, we can now use our knowledge of $\mathcal{F}^0_\mathrm{iso}$ together with \cref{omegacomparison,freeenergync} to obtain the nonisothermal grand potential
\begin{equation}
\begin{split}
&\Omega[\rhom ^\flat,c^\flat,\vec{g}^\flat,\beta] \\
&= \Omega_\mathrm{iso}\Big[\rhom ^\flat + \frac{\beta - \beta_0}{m_\mathrm{s}}U_1^\mathrm{s} - \frac{(\vec{g}^\flat)^2}{2\beta}  
+ \frac{3}{2m_\mathrm{s}}\ln\!\Big(\frac{\beta}{\beta_0}\Big),\\
&\quad\;\:\! c^\flat + (\beta-\beta_0)\Big(U_1^\mathrm{c} - \frac{m_\mathrm{c}}{m_\mathrm{s}}U_1^\mathrm{s}\Big) 
+ \frac{3}{2}\Big(1-\frac{m_\mathrm{c}}{m_\mathrm{s}}\Big),(\beta_0)\Big]\\
&= \mathcal{F}^0_\mathrm{iso}[\rhom,c,(\beta_0)] + \int_{\R^{3}}\!\!\!\!\:\!\dif^{3}r\, \beta_0( s U_1^\mathrm{s} + c U_1^\mathrm{c}) \\
&\quad\:\! + \Big(\rhom ^\flat + \frac{\beta - \beta_0}{m_\mathrm{s}}U_1^\mathrm{s} - \frac{(\vec{g}^\flat)^2}{2\beta} 
+ \frac{3}{2m_\mathrm{s}}\ln\!\Big(\frac{\beta}{\beta_0}\Big)\Big)\rhom  \\
&\quad\:\! + \Big(c^\flat + (\beta-\beta_0)(U_1^\mathrm{c} - \frac{m_\mathrm{c}}{m_\mathrm{s}}U_1^\mathrm{s}) \\
&\quad\:\! + \frac{3}{2}\Big(1-\frac{m_\mathrm{c}}{m_\mathrm{s}}\Big)\ln\!\Big(\frac{\beta}{\beta_0}\Big)\Big)c.
\end{split}
\label{omegalong}\raisetag{3em}
\end{equation}
Finally, we can combine \cref{legend,omegalong,gflat,varepsilon,linearcombination} to get the \textit{EDDFT entropy functional for nonisothermal underdamped hard spheres}, given by
\begin{equation}
\begin{split}
&\mathcal{S}[\rhom ,c,\vec{g},\varepsilon] \\
&= - \Omega[\rhom ^\flat,c^\flat,\vec{g}^\flat,\beta] \\
&\quad\:\! + \int_{\R^{3}}\!\!\!\!\:\!\dif^{3}r \Big(\rhom ^\flat \rhom  + c^\flat c  + \sum_{i=1}^{3}g_i^\flat g_i + \beta \varepsilon\Big)\\
&= -\Omega_\mathrm{iso}\Big[\rhom ^\flat + \frac{\beta - \beta_0}{m_\mathrm{s}}U_1^\mathrm{s} - \frac{(\vec{g}^\flat)^2}{2\beta} 
+ \frac{3}{2m_\mathrm{s}}\ln\!\Big(\frac{\beta}{\beta_0}\Big),\\
&\quad\;\:\! c^\flat + (\beta-\beta_0)\Big(U_1^\mathrm{c} - \frac{m_\mathrm{c}}{m_\mathrm{s}}U_1^\mathrm{s}\Big) 
+ \frac{3}{2}\Big(1-\frac{m_\mathrm{c}}{m_\mathrm{s}}\Big),(\beta_0)\Big]\\
&\quad\:\! + \int_{\R^{3}}\!\!\!\!\:\!\dif^{3}r \Big(\rhom ^\flat \rhom  + c^\flat c  + \sum_{i=1}^{3}g_i^\flat g_i + \beta \varepsilon\Big)\\
&= -\mathcal{F}^0_\mathrm{iso}[\rhom ,c,(\beta_0)] + \int_{\R^{3}}\!\!\!\!\:\!\dif^{3}r \bigg(\beta_0(- s U_1^\mathrm{s} - c U_1^\mathrm{c}) \\
&\quad\:\! - \Big(\rhom ^\flat + \frac{\beta - \beta_0}{m_\mathrm{s}}U_1^\mathrm{s} - \frac{(\vec{g}^\flat)^2}{2\beta}  
+ \frac{3}{2m_\mathrm{s}}\ln\!\Big(\frac{\beta}{\beta_0}\Big)\Big)\rhom  \\
&\quad\:\! - \Big(c^\flat + (\beta-\beta_0)\Big(U_1^\mathrm{c} - \frac{m_\mathrm{c}}{m_\mathrm{s}}U_1^\mathrm{s}\Big) \\
&\quad\:\! + \frac{3}{2}\Big(1-\frac{m_\mathrm{c}}{m_\mathrm{s}}\Big)\ln\!\Big(\frac{\beta}{\beta_0}\Big)\Big)c \\
&\quad\:\! + \Big(\rhom ^\flat \rhom  + c^\flat c  -\frac{\beta \vec{g}^2}{\rhom} + \beta \varepsilon\Big)\bigg)\\
&= -\mathcal{F}^0_\mathrm{iso}[\rhom ,c,(\beta_0)] 
+ \int_{\R^{3}}\!\!\!\!\:\!\dif^{3}r \bigg( \beta_0 (-s U_1^\mathrm{s} - c U_1^\mathrm{c}) \\
&\quad\:\! - (\beta - \beta_0) (s U_1^\mathrm{s} + c U_1^\mathrm{c}) +\frac{\beta \vec{g}^2}{2\rhom}  
+ \frac{3}{2}n\ln\!\Big(\frac{\beta}{\beta_0}\Big)-\frac{\beta \vec{g}^2}{\rhom} \\
&\quad\:\! + \frac{3}{2}n + \beta (c U_1^c + s U_1^s) + \frac{\beta \vec{g}^2}{2\rhom}\bigg)\\
&= - \mathcal{F}^0_\mathrm{iso}[\rhom ,c,(\beta_0)] \\
& \quad- \int_{\R^{3}}\!\!\!\!\!\:\!\dif^{3}r \frac{3}{2}n(\rhom,c)\Big(\ln\!\Big(\frac{\beta(\rhom,c,\vec{g},\varepsilon)}{\beta_0}\Big) -1\Big),
\end{split}\label{eddftfunctional}\raisetag{18em}%
\end{equation}
where $\beta$ is given by \cref{beta}, $n$ by \cref{linearcombination} and $\beta_0$ is an arbitrary constant. Despite its simplicity, the result \eqref{eddftfunctional} is exact (although approximations will be required for the excess free energy that contributes to $\mathcal{F}^0_\mathrm{iso}$). Interestingly, the form of \cref{eddftfunctional} is identical to the one derived by \citet{AneroET2013} for a one-component overdamped system. A difference, however, arises from the form of $\beta$, which now also depends on concentration and momentum density, and from the fact that $n$ is now the total number density of a two-component system.

As a consistency check\footnote{Note that this is a consistency check for \cref{eddftfunctional} and not a derivation of \cref{gflat}, since \cref{gflat} has been used in the derivation of \cref{eddftfunctional}.}, we demonstrate that \cref{eddftfunctional} gives the correct result for the conjugate to the momentum density (\cref{gflat}), namely $\vec{g}^\flat = -\beta \vec{g}/\rhom$. From \cref{eq:TC,beta,eddftfunctional}, we find
\begin{equation}
\begin{split}
\vec{g}^\flat &= \Fdif{S}{\vec{g}} = \Fdif{S}{\beta} \pdif{\beta}{\vec{g}} \\
&= -\frac{3}{2}\frac{n}{\beta}\frac{3}{2}\frac{n}{(\varepsilon - c U_1^\mathrm{s} - s U_1^\mathrm{s} - \frac{\vec{g}^2}{2\rhom})^2}\frac{\vec{g}}{\rhom}\\
&=- \frac{3}{2} \frac{2}{3}\frac{n (\varepsilon - c U_1^\mathrm{s} - s U_1^\mathrm{s} - \frac{\vec{g}^2}{2\rhom})}{n}\\
&\quad\, \frac{3}{2}\frac{n}{(\varepsilon - c U_1^\mathrm{s} - s U_1^\mathrm{s} + \frac{\vec{g}^2}{2\rhom})^2}\frac{\vec{g}}{\rhom}\\
&= -\frac{3}{2}\frac{n}{\varepsilon - c U_1^\mathrm{s} - s U_1^\mathrm{s} - \frac{\vec{g}^2}{2\rhom}}\frac{\vec{g}}{\rhom}\\
&= -\frac{\beta \vec{g}}{\rhom}.
\end{split} 
\label{consistencycheck}
\end{equation}

\subsection{\label{twoparticle}Free-energy functional}
Based on the entropy functional $\mathcal{S}$, we can introduce a non-dimensional free-energy functional $\mathcal{F}$ using the Legendre transformation
\begin{equation}
\mathcal{F}[\rho,c,\vec{g},T] = - \mathcal{S}[\rho,c,\vec{g},\varepsilon] 
+ \int_{\R^{3}}\!\!\!\!\!\:\!\dif^{3}r\, \beta \rt \varepsilon\rt.
\label{freeenergylegend}
\end{equation}
In the isothermal case $\beta\rt = \beta_0$, we can obtain the standard free energy $k_{\mathrm{B}}T_0 \mathcal{F}$. For a nonisothermal system as considered here, however, $T$ is a function of space and time, such that the choice between a dimensional and a non-dimensional free-energy functional is not merely a matter of choosing a prefactor since the former would be proportional to a space-dependent function.\footnote{Obviously, we can in principle also use an arbitrary constant $\beta_0$ to get a dimensional free energy.} Therefore, the nondimensional functional should be preferred. What is also important is that $\mathcal{F}$ is a functional of $T\rt$, whereas the isothermal functional $\mathcal{F}_{\mathrm{iso}}$, whose intrinsic part is introduced by \cref{freeenergync}, is a function of $T_0$.

In the isothermal case, the intrinsic free energy can be written as
\begin{equation}
\mathcal{F}^0_{\mathrm{iso}}[\{c_\mu\}] = \sum_{\mu\in\{\mathrm{c},\mathrm{s}\}} \mathcal{F}_\mathrm{id}[c_\mu] + \mathcal{F}_{\mathrm{exc}}[\{c_\mu\}]  + \sum_{\mu\in\{\mathrm{c},\mathrm{s}\}} \mathcal{F}_{\mathrm{ext}}[c_\mu]   
\label{freeenergyeq}
\end{equation}
with the ideal gas free energy
\begin{equation}
\mathcal{F}_\mathrm{id}[c_\mu] = \int_{\R^{3}}\!\!\!\!\!\:\!\dif^{3}r\, c_\mu(\vec{r})(\ln(c_\mu(\vec{r})\Lambda_\mu^3) -1)
\label{eq:IdealGasFreeEnergy}
\end{equation}
with the thermal de Broglie wavelength
\begin{equation}
\Lambda_\mu = \sqrt{\frac{h^2\beta}{2m_\mu\pi}},
\label{thermaldeBroglie}
\end{equation}
and the excess part $\mathcal{F}_{\mathrm{exc}}$ that depends on particle interactions. 

In the nonisothermal case, however, we first need to define what we mean by \ZT{excess free energy} for the nonequilibrium functional defined by \cref{freeenergylegend}, since we are by no means guaranteed that equilibrium relations such as \cref{freeenergyeq} carry over to the nonequilibrium case. Moreover, we require a way to calculate this excess free energy. Here, we address this problem for the case of hard spheres, for which we have derived the explicit form of the entropy functional in \cref{hard}. Using \cref{eddftfunctional,freeenergylegend,varepsilon2}, we find the \textit{free-energy functional for nonisothermal hard spheres}, given by
\begin{equation}
\begin{split}
\mathcal{F}[\{c_\mu\},\vec{g},T] &= \mathcal{F}^0 + \int_{\R^{3}}\!\!\!\!\!\:\!\dif^{3}r \bigg( \frac{\beta \vec{g}^2}{2\rhom } + \frac{3}{2}n\ln\!\Big(\frac{\beta}{\beta_0}\Big) + \beta \varepsilon\\
&= \mathcal{F}_\mathrm{exc}^0 + \sum_{\mu\in\{\mathrm{c},\mathrm{s}\}} \int_{\R^{3}}\!\!\!\!\!\:\!\dif^{3} c_\mu (\ln(c_\mu\Lambda_\mu^3 (T))-1)\\
&\quad\:\! +\beta c_\mu U_1^\mu +\frac{\beta\vec{g}^2}{2\rhom}\bigg).
\end{split}
\label{freeenergynoneq}\raisetag{3em}%
\end{equation}
The functional \eqref{freeenergynoneq} has a structure that is similar to the equilibrium form \eqref{freeenergyeq}. It consists of four terms:
\begin{enumerate}
    \item An excess free energy, which, for hard spheres, is identical to the expression from the isothermal case. This makes sense as the direct correlation functions obtained from FMT in the isothermal case do not depend on the temperature. Consequently, the excess free energy can be defined in the usual way and the direct correlation functions take the same form as in the isothermal case.
    \item An ideal gas free energy. It has the same form as in the isothermal case. However, the thermal de Broglie wavelength given by \cref{thermaldeBroglie} is now a function of space and time through its temperature-dependence. 
    \item A contribution from the external potential that also has a similar form as in equilibrium. Note, however, that there is now a prefactor $\beta$, which is a function of space and time.
    \item A term that is quadratic in the momentum density, which is the only term where $\vec{g}$ appears and which is not present in equilibrium DFT or in (E)DDFT without momentum density. Such a nonequilibrium contribution has also been used in underdamped PFC models \cite{GalenkoSE2015}.
\end{enumerate}
For a system that does not consist of hard spheres, we may still employ the parametrization \eqref{freeenergynoneq} and use it as a definition of the nondimensional excess free energy, although in this case it will not generally have the same form as in equilibrium.

\subsection{\label{adiabatic}Adiabatic approximation for the pressure}
For practical applications of EDDFT with momentum density, it is important to have an expression for the pressure tensor $\Pi\rt$, which is difficult due to the presence of interactions. In the derivation of DDFT, interaction terms are dealt with using the \ZT{adiabatic approximation} \cite{MarconiT1999,ArcherE2004}, which corresponds to assuming that the unknown two-body distribution function is identical to that of an equilibrium system with the same one-body density. This assumption allows to approximate the interaction terms via the functional derivative of the excess free-energy functional.

In extensions of DDFT that include the momentum density, the organized drift (in a one-component system of particles with mass $m$) is typically assumed to have the form \cite{Archer2009}
\begin{equation}
\rhom\rt(\dot{\vec{v}}\rt + (\vec{v}\rt\cdot\Nabla)\vec{v}\rt) = - \rhom\rt\Nabla \Fdif{\mathcal{F}_{\mathrm{dim}}}{\rhom\rt}. \label{generalizedeuler}
\end{equation}
Note that \cref{generalizedeuler} conserves the total momentum despite not having the form of a continuity equation, this is shown in \cref{momentumconservation}. When comparing \cref{generalizedeuler} to the organized momentum drift obtained here (corresponding to the standard Euler equation), which reads
\begin{equation}
\rhom\rt(\dot{\vec{v}}\rt + (\vec{v}\rt\cdot\Nabla)\vec{v}\rt) = -\Nabla\cdot \Pi\rt,
\label{generalizedeuler2}
\end{equation}
we can note that the pressure is approximated via the functional derivative of a free-energy functional $\mathcal{F}_{\mathrm{dim}}$ that has a dimension of energy (see \cref{dimensionalization}). The advantage of \cref{generalizedeuler} is that it uses the free-energy functional, which is known from DFT, and therefore provides a closed expression for the momentum drift. This corresponds to an adiabatic approximation. However, it is not immediately clear how one can read off a pressure $\Pi$ from \cref{generalizedeuler}. It is tempting to define the pressure tensor as
\begin{equation}
\Pi\rt = \Nabla^{-1}\rhom\rt\Nabla\Fdif{\mathcal{F}_{\mathrm{dim}}}{\rhom\rt},    
\label{adiabaticpressure}
\end{equation}
where $\Nabla^{-1}$ is an \ZT{inverse divergence}. 
This idea has actually been used by \citet{delasHerasS2018}, who introduce an adiabatic pressure based on \cref{adiabaticpressure}. They define the operator $\Nabla^{-1}$ as
\begin{equation}
\Nabla^{-1}f(\vec{r})=\INT{}{}{^3r'}\frac{\vec{r}-\vec{r}'}{4\pi\norm{\vec{r}-\vec{r}'}^3}f(\vec{r}'),  
\label{nablaminuseins}
\end{equation}
where $f$ is a test function. The identity
\begin{equation}
\delta(\vec{r})=\Nabla\cdot\bigg(\frac{\vec{r}}{4\pi\norm{\vec{r}}^3}\bigg)   
\label{deltaidentity}
\end{equation}
allows to show \cite{delasHerasS2018}
\begin{equation}
\Nabla\cdot\Nabla^{-1}f(\vec{r}) = f(\vec{r}).   
\label{nablainversion}
\end{equation}
Note that using $\Nabla^{-1}$, it is easily possible to write down a continuity equation for \textit{any} field, irrespective of whether it is actually conserved (see \cref{inversenabla} for a discussion of this point).

In our case, we face a further difficulty, namely that $T$ is not a constant but a function of space and time. In this case, the \ZT{dimensionalization} \eqref{dimensionalization} is, as discussed in  \cref{twoparticle}, not trivial, and the nondimensional functional $\mathcal{F}$ should be preferred. On the other hand, the adiabatic approximation requires the factor $k_{\mathrm{B}}T$ for dimensional reasons, such that we have to find an appropriate way of incorporating it. Moreover, our expression for the pressure tensor should recover the expression $\Pi_0$ given by \cref{eq:PIN} in the ideal gas limit. A final constraint is provided by the mean-field limit, in which we should have
\begin{equation}
\begin{split}
&\rhom\rt(\dot{\vec{v}}\rt + (\vec{v}\rt\cdot\Nabla)\vec{v}\rt) 
\\&= - \rhom\rt\Nabla U_1\rt 
\\&\quad\,- \frac{1}{m}\rhom\rt\Nabla\INT{}{}{^3r'}\rhom(\vec{r}',t)U_2(\norm{\vec{r}-\vec{r}'}).  
\end{split}
\label{generalizedeuler3}
\end{equation}
To see this, take Eq.\ (22) in Ref.\ \cite{teVrugt2026}, ignore all dependencies on particle orientations (passive spherical particles) and set $g=1$ in Eq.\ (23) of the same work (mean-field approximation). The mean-field form of $\mathcal{F}$ is obtained by replacing $U_1(\vec{r},t)$ by $U_1(\vec{r},t)+(1/m)\TINT{}{}{^3r'}\rhom(\vec{r}',t)U_2(\norm{\vec{r}-\vec{r}'})$ and ignoring everything else that comes from interactions. From \cref{freeenergynoneq}, we get with this approximation (which amounts to setting $\mathcal{F}_\mathrm{exc}^0=0$ and adjusting the external potential in the described way) the form
\begin{equation}
\begin{split}
\mathcal{F}[c_\mu,\vec{g},T] &=\sum_{\mu\in\{\mathrm{c},\mathrm{s}\}} \int_{\R^{3}}\!\!\!\!\!\:\!\dif^{3} \bigg(c_\mu(\vec{r}) (\ln(c_\mu(\vec{r})\Lambda_\mu^3 (T(\vec{r})))-1)\\
&\quad\:\! +\beta(\vec{r}) c_\mu(\vec{r}) U_1^\mu(\vec{r}) +\frac{\beta\vec{g}(\vec{r})^2}{2\rhom(\vec{r})}\\
&\quad\:\!  + \sum_{\nu\in\{\mathrm{c},\mathrm{s}\}}\beta(\vec{r})\int_{\R^{3}}\!\!\!\!\!\:\!\dif^{3}r'c_\mu(\vec{r}')c_\nu(\vec{r})U_2^{\mu\nu}(\norm{\vec{r}-\vec{r}'})\bigg).
\end{split}
\label{freeenergynoneq2}\raisetag{5em}%
\end{equation}
The requirements \ZT{correct ideal gas limit}, \ZT{correct isothermal limit}, and \ZT{inserting \cref{freeenergynoneq2} gives \cref{generalizedeuler3} (correct isothermal limit)} are satisfied by the approximation
\begin{equation}
\begin{split}
\Pi\rt &= k_{\mathrm{B}}\!\! \sum_{\mu\in\{\mathrm{c},\mathrm{s}\}}\bigg(c_\mu\rt T\rt\Eins \\
&\quad\:\!+\Nabla^{-1} c_\mu\rt\Nabla \bigg( T\rt \Fdif{\mathcal{F}_{\mathrm{exc}}}{c_\mu\rt}\bigg)\!\bigg).
\end{split}
\label{adiabaticapproximation}\raisetag{5em}%
\end{equation}
An alternative approach would be to, in the spirit of force DFT \cite{TschoppSHSB2022}, use \cref{eq:PII} combined with the fact that DFT allows to express the two-particle density appearing there as a functional of the one-particle density.

\section{\label{sec:II}Special cases and relation to MCT}
Here, we discuss isothermal systems, the hydrodynamic limit of the new EDDFT equations \eqref{eq:EDDFT_rhom}-\eqref{eq:EDDFT_e}, 
the special case of a one-component atomic or molecular system without a solvent (as, e.g., a metal), and the relation of the EDDFT equations to MCT.
Obvious special cases that do not need to be discussed in detail in the following are the standard DDFT \cite{MarconiT1999,MarconiT2000,ArcherE2004,EspanolL2009} and the DDFT approach of \citet{RauscherDKP2007} with a coupling to a flow field. 

For ease of notation, we will from now on write $\Nabla$ for $\Nabla_{\vec{r}}$ and drop the dependence on space and time if there is no danger of ambiguity (e.g., we will often write $\rhom$ instead of $\rhom \rt$).

\subsection{Isothermal limit}
We can recover standard DDFT and extensions that involve solely the momentum density by considering the isothermal limit where $\beta$ is a constant $\beta_0 = 1/(k_{\mathrm{B}}T_0)$ with a constant temperature $T_0$. From \cref{eq:TC2}, we can infer that the relevant density \eqref{relevantdensity} can, in this case, be written as
\begin{equation}
\rho(t)= \frac{1}{Z_\mathrm{iso}(t)}\exp(-\beta_0 \hat{H}_\mathrm{eff}(t))    
\end{equation}
with the effective Hamiltonian
\begin{equation}
\begin{split}
\hat{H}_\mathrm{eff}(t) &= \hat{H} - \int_{\R^{3}}\!\!\!\!\:\!\dif^{3}r\,
\rho^{\natural}_{\mathrm{m}}\rt\hat{\rho}_{\mathrm{m}}(\vec{r}) -
\int_{\R^{3}}\!\!\!\!\:\!\dif^{3}r\,c^{\natural}\rt\hat{c}(\vec{r}) \\
&\quad\:\! -\sum^{3}_{i=1}\int_{\R^{3}}\!\!\!\!\!\:\!\dif^{3}r\,
g^{\natural}_{i}\rt\hat{g}_{i}(\vec{r}) ,
\end{split}\raisetag{3em}%
\end{equation}
where we have used $\hat{H}= \int_{\R^3}\!\dif^3 r\, \hat{\varepsilon}$ and $\beta = \varepsilon^\flat$. Based on the free-energy functional \cite{Grabert1982}
\begin{equation}
\mathcal{F}_{\mathrm{dim}} = \braket{H} - k_{\mathrm{B}}T_0 \mathcal{S} 
\label{standardfreeenergy}
\end{equation}
with $\braket{H} =  \Tr(\rho(t)\hat{H})$, we can write the conjugate variables, given by \cref{eq:TC2}, as
\begin{equation}
\rho^{\natural}_{\mathrm{m}}=\Fdif{\mathcal{F}_{\mathrm{dim}}}{\rhom } ,\qquad
c^{\natural}=\Fdif{\mathcal{F}_{\mathrm{dim}}}{c} ,\qquad
g^{\natural}_{i}=\Fdif{\mathcal{F}_{\mathrm{dim}}}{g_{i}}.
\label{eq:TC3}%
\end{equation}
Note that this is now a complete set of thermodynamic conjugates, since $\varepsilon$ is no longer a relevant variable. The general EDDFT equations \eqref{eq:EDDFT_allg} then read
\begin{equation}
\begin{split}
\dot{a}_{i}\rt &= - \Nabla_{\vec{r}}\!\cdot\!\Tr\!\big(\rho(t)\hat{\vec{J}}^{(i)}(\vec{r})\big) \\
&\quad\:\! + \sum^{n}_{j=1}\Nabla_{\vec{r}}\!\cdot\!\!\int_{\R^{3}}\!\!\!\!\!\:\!\dif^{3}r'\, \beta_0\DT^{(ij)}(\vec{r},\rs\!,t)
\Nabla_{\rs}a^{\natural}_{j}(\rs\!,t).
\end{split}\label{eq:EDDFT_iso}\raisetag{4em}%
\end{equation}

\subsection{\label{hl}Hydrodynamic limit}
In general, the expressions for the diffusion tensors $D^{(ij)}(\vec{r},\vec{r}',t)$ are quite complicated. However, they drastically simplify in the \textit{hydrodynamic limit} characterized by small wavenumbers $k$ and small frequencies $z$. For this purpose, we go back to the non-Markovian transport equations with the kernel $R^{(ij)}(\vec{r},\vec{r}',t,t')$. Assuming translational invariance in space and time, this can be written as $R^{(ij)}(\vec{r}-\vec{r}',t-t')$. The assumption that the memory kernel depends on $t-t'$ is, as can be seen from \cref{rij}, justified if we approximate $\rho(t)$ by the equilibrium expression $\hat{\rho} = \exp(-\beta_0 \hat{H})/Z$, in which case the projection operator $\mathcal{P}$ is time-independent \cite{teVrugtW2019}. Translational invariance in space is given if we ignore the external potential. Then, we perform a Fourier-Laplace transformation (see \cref{fourierlap}) to get
\begin{equation}
\begin{split}
&R^{(ij)}(\vec{r}-\vec{r}',t-t') \\
&= \frac{1}{(2\pi )^4\ii } \int_{\R^{3}}\!\!\!\!\:\!\dif^{3}k\int_{z_0-\ii \infty}^{z_0 + \ii \infty}\!\!\!\!\!\!\!\!\!\!\:\!\dif z \, \tilde{R}^{(ij)}(\vec{k},z) e^{\ii \vec{k}\cdot(\vec{r}-\vec{r}')}e^{z (t-t')},
\end{split}\raisetag{3.5em}%
\end{equation}
where $z_0>z$ is a constant (see \cref{fourierlap}) and the transformed kernel $\tilde{R}^{(ij)}$ is given by
\begin{equation}
\tilde{R}^{(ij)}(\vec{k},z) = \int_{\R^{3}}\!\!\!\!\!\:\!\dif^{3}r\int_{0}^{\infty}\!\!\!\!\!\!\:\!\dif t\, R^{(ij)}(\vec{r},t) e^{-\ii\vec{k}\cdot \vec{r}}e^{-zt}.   
\end{equation}
Next, we take the hydrodynamic limit $\vec{k} = \vec{0}$ and $z = 0$. This gives
\begin{equation}
\begin{split}
&R^{(ij)}(\vec{r}-\vec{r}',t-t') \\
&= \frac{1}{(2\pi)^4 \ii } \int_{\R^{3}}\!\!\!\!\!\:\!\dif^{3}k\int_{z_0-\ii \infty}^{z_0 + \ii \infty}\!\!\!\!\!\!\!\!\!\!\:\!\dif z\,  D^{(ij)}_{\mathrm{HL}} e^{\ii \vec{k}\cdot(\vec{r}-\vec{r}')}e^{z (t-t')},
\end{split}\label{rijtrans}
\end{equation}
with the constant
\begin{equation}
D^{(ij)}_{\mathrm{HL}} =  \int_{\R^{3}}\!\!\!\!\!\:\!\dif^{3}r\int_{0}^{\infty}\!\!\!\!\!\!\:\!\dif t\, R^{(ij)}(\vec{r},t), 
\end{equation}
that is, using \cref{rij}, translationally invariant and the approximation $\rho(t) \approx \hat{\rho}$ (leading to a time-independent projection operator), given by
\begin{equation}
D^{(ij)}_{\mathrm{HL},kl} =\int_{\R^{3}}\!\!\!\!\!\:\!\dif^{3}r\int_{0}^{\infty}\!\!\!\!\!\!\:\!\dif t\, \Tr\!\Big(\hat{\rho}  \big(\hat{\mathcal{Q}}J^j_l(\vec{0})\big) e^{\mathcal{L} \hat{\mathcal{Q}}t} \big(\hat{\mathcal{Q}} J^i_k (\vec{r})\big)\Big). 
\end{equation}
Evaluating the integrals in \cref{rijtrans} gives
\begin{equation}
R^{(ij)}(\vec{r}-\vec{r}',t-t') =  D^{(ij)}_{\mathrm{HL}} \delta(\vec{r}-\vec{r}')\delta(t-t').  
\label{rijdelta}
\end{equation}
Using \cref{rijdelta}, the memory term in \cref{eq:EDDFT_exact} is found to be\footnote{We have here used the convention $\TINT{0}{t}{t'} f(t')\delta(t-t') = f(t)$ rather than $\TINT{0}{t}{t'} f(t')\delta(t-t') = \frac{1}{2}f(t)$, since the latter convention does not give the correct result for the diffusion tensor. The physical reason is that the convention $\TINT{0}{t}{t'} f(t')\delta(t-t') = \frac{1}{2}f(t)$ is motivated by the fact that the Delta distribution $\delta(t)$ can be approximated by a function $d(t)$ with small width and large amplitude and the properties $\TINT{-\infty}{\infty}{t}d(t) = 1$ and $d(t)=d(-t)$, where the latter property implies $\TINT{-\infty}{t}{t'}d(t-t') = \frac{1}{2}$. However, for reasons of causality, $R^{(ij)}(t-t')$ cannot have contributions from $t'>t$, such that the distribution $\delta(t-t')$ in \cref{rijdelta} should be approximated by a function $d(t-t')$ that vanishes for $t'>t$. This implies $\TINT{-\infty}{t}{t'}d(t-t') =1$.}
\begin{equation}
\int_{\R^{3}}\!\!\!\!\!\:\!\dif^3 r \INT{0}{t}{t'} D^{(ij)}_{\mathrm{HL}} \delta(\vec{r}-\vec{r}')\delta(t-t')a_j^\beta(\vec{r}',t') = D^{(ij)}_{\mathrm{HL}}a_j^\beta \rt.
\end{equation}
In the hydrodynamic limit, i.e., in the special case of small wave vectors $\vec{k}$ and frequencies $z$ ($\vec{k}\approx\vec{0}$, $z\approx 0$)\footnote{More precisely, the hydrodynamic limit can be characterized by the conditions $\norm{\vec{k}}l_{\mathrm{c}}\ll 1$ and $z t_{\mathrm{c}}\ll 1$, where $l_{\mathrm{c}}$ and $t_{\mathrm{c}}$ are characteristic microscopic length and time scales, respectively.}, the diffusion tensors \eqref{eq:D_cc}-\eqref{eq:D_ee} thus reduce to 
{\allowdisplaybreaks%
\begin{align}%
\begin{split}%
\DT^{(cc)}_{\mathrm{HL}}(\vec{r},\rs) &= D^{(cc)}_{\mathrm{HL}}\Eins \delta(\vec{r}-\rs) ,
\end{split}\label{eq:Dcc_HL}\\%
\begin{split}%
\DT^{(c\varepsilon)}_{\mathrm{HL}}(\vec{r},\rs) &= D^{(c\varepsilon)}_{\mathrm{HL}}\Eins \delta(\vec{r}-\rs) ,
\end{split}\label{eq:Dce_HL}\\%
\begin{split}%
\DT^{(g_{i}g_{j})}_{\mathrm{HL}}(\vec{r},\rs) &= \mathrm{D}^{(g_{i}g_{j})}_{\mathrm{HL}}\delta(\vec{r}-\rs) , 
\end{split}\label{eq:Dgg_HL}\\%
\begin{split}%
\DT^{(\varepsilon c)}_{\mathrm{HL}}(\vec{r},\rs) &= D^{(\varepsilon c)}_{\mathrm{HL}}\Eins \delta(\vec{r}-\rs) ,
\end{split}\label{eq:Dec_HL}\\%
\begin{split}%
\DT^{(\varepsilon\varepsilon)}_{\mathrm{HL}}(\vec{r},\rs) &= D^{(\varepsilon\varepsilon)}_{\mathrm{HL}}\Eins \delta(\vec{r}-\rs) 
\end{split}\label{eq:Dee_HL}%
\end{align}}%
with the scalar constants $D^{(cc)}_{\mathrm{HL}}$, $D^{(c\varepsilon)}_{\mathrm{HL}}=D^{(\varepsilon c)}_{\mathrm{HL}}$,\footnote{The equality of the coefficients $D^{(c\varepsilon)}_{\mathrm{HL}}$ and $D^{(\varepsilon c)}_{\mathrm{HL}}$ follows from Onsager's principle \cite{LandauL1996}.} and $D^{(\varepsilon\varepsilon)}_{\mathrm{HL}}$ and with the constant viscous stress tensor $\mathrm{D}^{(g_{i}g_{j})}_{\mathrm{HL}}$ so that the EDDFT equations \eqref{eq:EDDFT_rhom}-\eqref{eq:EDDFT_e} become local.
The particular form of the viscous stress tensor $\mathrm{D}^{(g_{i}g_{j})}_{\mathrm{HL}}$ can be determined from symmetry considerations. 

Obviously, the EDDFT equations \eqref{eq:EDDFT_rhom}-\eqref{eq:EDDFT_e} should be invariant with respect to translations and rotations of the coordinate system. 
While Eqs.\ \eqref{eq:EDDFT_rhom}-\eqref{eq:EDDFT_e} are generally invariant with respect to translations of the coordinate system, their invariance with respect to rotations of the coordinate system requires the tensors \eqref{eq:Dcc_HL}, \eqref{eq:Dce_HL}, \eqref{eq:Dec_HL}, and \eqref{eq:Dee_HL} as well as the pressure tensor \eqref{eq:PI} to be proportional to the identity matrix $\Eins$. The pressure tensor can then be written as shown in Eq.\ \eqref{eq:pkI}. 
Furthermore, invariance with respect to rotations of the coordinate system requires the viscous stress tensor 
to have the form
\begin{equation}
\mathrm{D}^{(g_{i}g_{j})}_{\mathrm{HL},kl}=D^{(gg)}_{\mathrm{HL},1}\delta_{ij}\delta_{kl} 
+D^{(gg)}_{\mathrm{HL},2}\delta_{ik}\delta_{jl} +D^{(gg)}_{\mathrm{HL},3}\delta_{il}\delta_{jk}
\label{eq:ViscousStressTensor}%
\end{equation}
with the three constants $D^{(gg)}_{\mathrm{HL},1}$, $D^{(gg)}_{\mathrm{HL},2}$, and $D^{(gg)}_{\mathrm{HL},3}$. 
Since the viscous stress tensor $\mathrm{D}^{(g_{i}g_{j})}_{\mathrm{HL}}$ constitutes a dissipative contribution to the 
EDDFT equations \eqref{eq:EDDFT_rhom}-\eqref{eq:EDDFT_e}, it must vanish when the whole system only rotates with a constant angular velocity. With this argument, one can prove the equality $D^{(gg)}_{\mathrm{HL},1}=D^{(gg)}_{\mathrm{HL},3}$ between the constants $D^{(gg)}_{\mathrm{HL},1}$ and $D^{(gg)}_{\mathrm{HL},3}$ \cite{LandauL1987}. 
Inserting the hydrodynamic results for the diffusion tensors and for the pressure tensor into the EDDFT equations \eqref{eq:EDDFT_rhom}-\eqref{eq:EDDFT_e} and expressing them in terms of the thermodynamic conjugates related to the free energy (defined by \cref{eq:TC2}) leads to the \textit{hydrodynamic equations for suspensions}  
{\allowdisplaybreaks%
\begin{align}%
\begin{split}%
\dot{\rho}_{\mathrm{m}}\rt &= -\Nabla\!\cdot\!\vec{g}\rt ,
\end{split}\label{eq:EDDFT_rhom_HL1}\\%
\begin{split}%
\dot{c}\rt &= -\Nabla\!\cdot\!\big(c\rt\vec{g}^{\natural}\rt\big) \\
&\quad\:\! +D^{(cc)}_{\mathrm{HL}}\Laplace_{\vec{r}} (\beta \rt c^{\natural}\rt)
- D^{(c\varepsilon)}_{\mathrm{HL}}\Laplace_{\vec{r}} \beta\rt , 
\end{split}\label{eq:EDDFT_c_HL1}\raisetag{2.4em}\\ %
\begin{split}%
\dot{\vec{g}}\rt &= -\Nabla\!\cdot\!\big(\rhom \rt\vec{g}^{\natural}\rt
\!\:\!\otimes\!\:\!\vec{g}^{\natural}\rt\big) -\Nabla \cdot \Pi\rt \\
&\quad\:\! +D^{(gg)}_{\mathrm{HL},1}\Laplace_{\vec{r}}(\beta\rt \vec{g}^{\natural}\rt ) \\
&\quad\:\! +(D^{(gg)}_{\mathrm{HL},1}+D^{(gg)}_{\mathrm{HL},2})\Nabla\big(\Nabla\!\cdot\! \beta\rt\vec{g}^{\natural}\rt\big)\\
&\quad\:\! +\vec{F}\rt
,
\end{split}\label{eq:EDDFT_g_HL1}\raisetag{3em}\\%
\begin{split}%
\dot{\varepsilon}\rt &= -\Nabla\!\cdot\!\big((\varepsilon\rt\Eins + \Pi\rt\!\:\!)\vec{g}^{\natural}\rt\big) \\
&\quad\:\! +D^{(c\varepsilon)}_{\mathrm{HL}}\Laplace_{\vec{r}} \beta\rt c^{\natural}\rt 
-D^{(\varepsilon\varepsilon)}_{\mathrm{HL}}\Laplace_{\vec{r}} \beta\rt .
\end{split}\label{eq:EDDFT_e_HL1}\raisetag{2.4em}%
\end{align}}%
Using \cref{eq:EDDFT_rhom_HL1} and $\vec{g} = \rhom \vec{v}$ with the velocity field $\vec{v}$, we get
\begin{equation}
\begin{split}
\dot{\vec{g}} + \Nabla\cdot(\rho\vec{v}\otimes \vec{v})  
&= \dot{\rho}_m\vec{v} + \rhom \dot{\vec{v}} + \Nabla\cdot(\rho\vec{v}\otimes \vec{v}) \\
&= (-\Nabla \cdot \vec{g})\vec{v} + \rhom \dot{\vec{v}} + \Nabla\cdot(\rho\vec{v}\otimes \vec{v}) \\
&= -\Nabla \cdot (\rhom \vec{v})\vec{v} + \rhom  \dot{\vec{v}} \\
&\quad\:\! + (\Nabla\cdot(\rhom  \vec{v}))\vec{v} 
+ \rhom (\vec{v}\cdot\Nabla)\vec{v} \\
&= \rhom (\dot{\vec{v}}+ (\vec{v}\cdot\Nabla)\vec{v}).
\end{split}
\label{velocity}
\end{equation}
Moreover, we assume that the temperature $T = \frac{1}{\beta k_{\mathrm{B}}}$ is close to a reference temperature $T_0$, such that we can write
\begin{equation}
\beta = \frac{1}{k_{\mathrm{B}}T} \approx \frac{1}{k_{\mathrm{B}}T_0} - \frac{1}{k_{\mathrm{B}}T_0^2} (T-T_0) = \beta_0 - \frac{\beta_0}{T_0}(T - T_0). 
\label{temperature}
\end{equation}
Finally, we assume that we are so close to equilibrium that the thermodynamic conjugates $\rhom ^\natural$, $c^\natural$, and $\vec{g}^\natural$ are small enough to allow for the replacement
\begin{align}
\rhom ^\natural \Big(\beta_0 - \frac{\beta_0}{T_0}(T - T_0)\Big) &\approx  \rhom ^\natural\beta_0,\\
c^\natural \Big(\beta_0 - \frac{\beta_0}{T_0}(T - T_0)\Big) &\approx  c^\natural\beta_0,\\
\vec{g}^\natural \Big(\beta_0 - \frac{\beta_0}{T_0}(T - T_0)\Big) &\approx  \vec{g}^\natural\beta_0,
\end{align}
where we have dropped products of the small thermodynamic conjugates with the small factor $T-T_0$.
Using \cref{velocity,temperature,eq:TC3}, the hydrodynamic EDDFT equations \eqref{eq:EDDFT_rhom_HL1}-\eqref{eq:EDDFT_e_HL1} become the \textit{simpler hydrodynamic equations for suspensions}
{\allowdisplaybreaks%
\begin{align}%
\begin{split}%
\dot{\rho}_{\mathrm{m}}\rt &= -\Nabla\!\cdot\!(\rhom \rt \vec{v}\rt) ,
\end{split}\label{eq:EDDFT_rhom_HL}\\%
\begin{split}%
\dot{c}\rt &= -\Nabla\!\cdot\!\big(c\rt\vec{v}\rt\big) 
+\kappa_{\mathrm{D}}\Laplace \Fdif{\mathcal{F}_{\mathrm{dim}}}{c\rt} \\
&\quad\:\! +\kappa_{\mathrm{S}}\Laplace T\rt ,
\end{split}\label{eq:EDDFT_c_HL}\raisetag{1.8em}\\%
\begin{split}%
\rhom \rt\dot{\vec{v}}\rt &= -\rhom \rt(\vec{v}\rt\cdot\Nabla)\vec{v}\rt -\Nabla \cdot\Pi\rt \\
&\quad\:\! +\eta\Laplace\vec{v}\rt +(\eta+\lambda_{1})\Nabla\big(\Nabla\!\cdot\! \vec{v}\rt\big) \\
&\quad\:\! + \vec{F}\rt,
\end{split}\label{eq:EDDFT_g_HL}\raisetag{1.8em}\\%
\begin{split}%
\dot{\varepsilon}\rt &= -\Nabla\!\cdot\!\big((\varepsilon\rt\Eins+\Pi\rt\!\:\!)\vec{v}\rt\big) \\
&\quad\:\! +T_0\kappa_{\mathrm{S}}\Laplace \Fdif{\mathcal{F}_{\mathrm{dim}}}{c\rt} 
+\kappa_{\mathrm{H}}\Laplace T\rt 
\end{split}\label{eq:EDDFT_e_HL}%
\end{align}}%
with $\mathcal{F}_{\mathrm{dim}} = \mathcal{F}_{\mathrm{dim}}[\rhom ,c,(\beta_0)]$, where we defined the rescaled diffusion coefficient $\kappa_{\mathrm{D}}=\beta_0 D^{(cc)}_{\mathrm{HL}}$, the thermal diffusion coefficient $\kappa_{\mathrm{S}}=\beta_0 D^{(c\varepsilon)}_{\mathrm{HL}}/T_0$ (related to the Soret coefficient), the heat conductivity 
$\kappa_{\mathrm{H}}=\beta_0 D^{(\varepsilon\varepsilon)}_{\mathrm{HL}}/T_0$, the dynamic (shear) viscosity $\eta=\beta_0 D^{(gg)}_{\mathrm{HL},1}$, and the first Lam\'e constant $\lambda_{1}=\beta_0 D^{(gg)}_{\mathrm{HL},2}$ \cite{LandauL2005}. 
The first Lam\'e constant $\lambda_{1}$ can be expressed in terms of the dynamic viscosity $\eta$, which is also referred to as ``second Lam\'e constant'', and the volume viscosity $\zeta$ by the relation $\lambda_{1}=\zeta-2\eta/3$.

It should be noted at this point that, strictly speaking, it is not consistent to combine a hydrodynamic (Navier-Stokes-like) expression for the dissipative part with an organized drift that varies on microscopic scales (as it does if we model it via DDFT). A discussion of hydrodynamic equations on molecular scales can be found in Refs.\ \cite{TothGT2013,HeinonenAKYLA2016}. However, (a) the hydrodynamic model considered here is a first approximation to the correct microscopic model, and (b) the EDDFT approach gives the full microscopic expressions for the dissipative terms and one can simply not make the hydrodynamic approximation in cases where it is too inaccurate.

\subsection{One-component system}
For one-component atomic or molecular systems, the local concentration $c\rt$ is no longer a relevant variable and can be neglected. An example for such systems are metals.
In this special case, the EDDFT equations \eqref{eq:EDDFT_rhom}-\eqref{eq:EDDFT_e} reduce to the
\textit{EDDFT equations for one-component systems}
{\allowdisplaybreaks%
\begin{align}%
\begin{split}%
\dot{\rho}_{\mathrm{m}}\rt &= -\Nabla_{\vec{r}}\!\cdot\!\vec{g}\rt ,
\end{split}\label{eq:EDDFTm_rhom}\\%
\begin{split}%
\dot{\vec{g}}\rt &= -\big(\Nabla_{\vec{r}}\!\cdot\!\big(\rhom \rt\vec{g}^{\natural}\rt
\!\:\!\otimes\!\:\!\vec{g}^{\natural}\rt\big)\big)_{i} \\
&\quad\:\! -\big(\Nabla_{\vec{r}}\cdot\Pi\rt\big)_{i} + F_i \rt \\
&\quad\:\! -\sum^{3}_{j=1}\Nabla_{\vec{r}}\!\cdot\!\!\int_{\R^{3}}\!\!\!\!\!\:\!\dif^{3}r'\,\DT^{(g_{i}g_{j})}(\vec{r},\rs\!,t)
\Nabla_{\rs}g^{\flat}_{j}(\rs\!,t) ,
\end{split}\label{eq:EDDFTm_g}\raisetag{5em}\\%
\begin{split}%
\dot{\varepsilon}\rt &= -\Nabla_{\vec{r}}\!\cdot\!\big((\varepsilon\rt\Eins + \Pi\rt)\vec{g}^{\natural}\rt\big) \\
&\quad\:\! -\Nabla_{\vec{r}}\!\cdot\!\!\int_{\R^{3}}\!\!\!\!\!\:\!\dif^{3}r'\, \DT^{(\varepsilon\varepsilon)}(\vec{r},\rs\!,t) \Nabla_{\rs}\varepsilon^{\flat}(\rs\!,t) .
\end{split}\label{eq:EDDFTm_e}%
\end{align}}%
In the hydrodynamic limit, Eqs.\ \eqref{eq:EDDFTm_rhom}-\eqref{eq:EDDFTm_e} reduce to
{\allowdisplaybreaks%
\begin{align}%
\begin{split}%
\dot{\rho}_{\mathrm{m}}\rt &= -\Nabla\!\cdot\!(\rhom \rt\vec{v}\rt) ,
\end{split}\label{eq:EDDFTm_rhom_HL1T}\\%
\begin{split}%
\rhom\rt \dot{\vec{v}}\rt &=-\rhom\rt(\vec{v}\rt \cdot \Nabla)\vec{v}\rt -\Nabla \cdot\Pi\rt \\
&\quad\:\! +D^{(gg)}_{\mathrm{HL},1}\Laplace(\beta\rt \vec{v}\rt ) \\
&\quad\:\! +(D^{(gg)}_{\mathrm{HL},1}+D^{(gg)}_{\mathrm{HL},2})\Nabla\big(\Nabla\!\cdot\! \beta\rt\vec{v}\rt\big) \\
&\quad\:\! + \vec{F}\rt,
\end{split}\label{eq:EDDFTm_g_HL1T}\raisetag{4.5em}\\%
\begin{split}%
\dot{\varepsilon}\rt &= -\Nabla\!\cdot\!\big((\varepsilon\rt\Eins+\Pi\rt\!\:\!)\vec{v}\rt\big) \\
&\quad\:\!
-D^{(\varepsilon\varepsilon)}_{\mathrm{HL}}\Laplace \beta\rt .
\end{split}\label{eq:EDDFTm_e_HL1T}\raisetag{2.4em}%
\end{align}}%
Also assuming small deviations from a reference temperature (as done in \cref{hl}), Eqs.\ \eqref{eq:EDDFTm_rhom}-\eqref{eq:EDDFTm_e} become the \textit{hydrodynamic equations for one-component systems}
{\allowdisplaybreaks%
\begin{align}%
\begin{split}%
\dot{\rho}_{\mathrm{m}}\rt &= -\Nabla\!\cdot\!(\rhom \rt\vec{v}\rt) ,
\end{split}\label{eq:EDDFTm_rhom_HL}\\%
\begin{split}%
\rhom \rt\dot{\vec{v}}\rt &= -\rhom \rt(\vec{v}\rt\cdot\Nabla)\vec{v}\rt  \\
&\quad\:\! -\Nabla \cdot\Pi\rt +\eta\Laplace\vec{v}\rt \\
&\quad\:\!+(\eta+\lambda_{1})\Nabla\big(\Nabla\!\cdot\!\vec{v}\rt\big) + \vec{F}\rt ,
\end{split}\label{eq:EDDFTm_g_HL}\raisetag{2.5em}\\%
\begin{split}%
\dot{\varepsilon}\rt &= -\Nabla\!\cdot\!\big((\varepsilon\rt\Eins+\Pi\rt\!\:\!)\vec{v}\rt\big) \\
&\quad\:\! +\kappa_{\mathrm{H}}\Laplace T\rt 
\end{split}\label{eq:EDDFTm_e_HL}%
\end{align}}%
with the velocity field $\vec{v}\rt=\vec{g}\rt/\rhom \rt$. The energy density $\varepsilon$ can, using \cref{varepsilon2}, be related to the temperature $T\rt$ as
\begin{equation}
\varepsilon = \frac{3}{2} \frac{\rhom}{m} k_{\mathrm{B}}T + \frac{\rhom}{m} U_1 +\rhom \frac{\vec{v}^2}{2},
\label{varepsilon3}%
\end{equation}
where $m$ is the mass of the particles. 

\subsection{Relation to MCT}
Mode coupling theory (MCT) \cite{Goetze2009} is a successful and widely used microscopic theory of the glass transition. It provides an equation of motion for the density correlator, which, as discussed in our previous work \cite{WittkowskiLB2012,teVrugtW2019}, is easily obtained from the MZFT once the time evolution of the relevant variables is known. MCT has a complex relationship to DDFT, which was examined in detail by \citet{teVrugtLW2020}: Historically, improving MCT was among the main motivations for the development of DDFT \cite{Kawasaki1994,Kawasaki2009,FuchizakiK2002}. \citet{Archer2006}, who developed an early extension of DDFT to inertial dynamics, used it for a heuristic derivation of MCT. On the other hand, the applicability of DDFT to the glass transition has been questioned \cite{SchindlerWB2019,WittkowskiLB2012}.

Simple MCT predicts a transition to a nonergodic state, where the particles in a dense liquid are trapped and can therefore not access the whole phase space. Consequently, the glass does not freeze into a crystal, but instead reaches a state that is characterized by a strong dependence on the history of the system (memory) \cite{Kawasaki2009}. In practice, however, ergodicity is restored by thermal fluctuations. \citet{Kawasaki1994} has derived a stochastic DDFT in which thermal fluctuations are explicitly included to develop an improved description. He demonstrated that MCT arises as a special case.

The relation of MCT and (E)DDFT can be clarified within the framework of the MZFT, which is the basis of EDDFT, but also the standard way for the derivation of MCT \cite{Janssen2018}. As discussed in \cref{htheorem}, the MZFT generally leads to a formally exact transport equation for the relevant variables that contains a memory term. If the relevant variables are slow compared to all other variables in the system, this contribution can be replaced by a Markovian dissipative term that is local in time. This is what is done in the derivation of EDDFT \cite{WittkowskiLB2012}. In the case of MCT, however, memory effects are not ignored, instead one uses a particularly simple approximation for the form of the memory kernel (MCT approximation). Consequently, MCT allows to describe memory effects not present in EDDFT. 

On the other hand, the derivation of simple MCT also involves approximations (in particular using a linearized form of the MZFT), which is why EDDFT and MCT should be seen as complementary theories \cite{WittkowskiLB2012}. The situation is, as discussed by \citet{teVrugtLW2020}, somewhat different for the case of stochastic DDFT: Here, the set of relevant variables includes all nonlinear functions of the density. It is therefore larger, which is why it is more likely to cover all slow functions in the system and allows for a memoryless description. The equation of motion for the density correlator (MCT) then arises as a reduced description, which therefore involves memory effects.

\section{\label{applications}Applications}
In this section, we will discuss two important applications of EDDFT with momentum density, namely the derivation of a H-theorem and the calculation of the speed of sound.

\subsection{\label{htheorem}H-theorem}
An advantage of the EDDFT approach compared to existing derivations is that it provides a natural connection to thermodynamics, and in particular allows for the simple derivation of a H-theorem. 

We consider the entropy functional $\mathcal{S}$ given by \cref{entropyfunctional}. We summarize the relevant variables $\rhom $, $c$, $\vec{g}$, and $\varepsilon$ in a vector $\vec{R}$ whose time evolution is determined by EDDFT equations \eqref{eq:EDDFT_rhom}-\eqref{eq:EDDFT_e}. These consist of an organized and a disorganized drift. As shown by \citet{Grabert1978}, the organized drift does not contribute to the rate of change of the entropy. Consequently, we can restrict ourselves to the disorganized motion when computing its time evolution. It is given by
\begin{equation}
\begin{split}
\tdif{\mathcal{S}}{t} &= \int_{\R^{3}}\!\!\!\!\!\:\!\dif^{3}r \Fdif{\mathcal{S}}{\vec{R}\rt}\cdot\dot{\vec{R}}\rt \\
&= -\int_{\R^{3}}\!\!\!\!\!\:\!\dif^{3}r \Fdif{\mathcal{S}}{\vec{R}\rt} \int_{\R^{3}}\!\!\!\!\!\:\!\dif^{3}r' \\
&\quad\;\, \Nabla_{\vec{r}}\cdot\Big(\mathcal{D}^{(\vec{R}\vec{R})}(\vec{r},\vec{r}',t)\Nabla_{\vec{r}'}\vec{R}^\flat(\vec{r}',t)\Big)\\
&=-\int_{\R^{3}}\!\!\!\!\!\:\!\dif^{3}r \Fdif{\mathcal{S}}{\vec{R}\rt} \int_{\R^{3}}\!\!\!\!\!\:\!\dif^{3}r' \\
&\quad\;\, \Nabla_{\vec{r}}\cdot\Big(\mathcal{D}^{(\vec{R}\vec{R})}(\vec{r},\vec{r}',t)\Nabla_{\vec{r}'}\Fdif{\mathcal{S}}{\vec{R}(\vec{r}',t)}\Big) \\
&= \int_{\R^{3}}\!\!\!\!\!\:\!\dif^{3}r \Big(\Nabla_{\vec{r}}\Fdif{\mathcal{S}}{\vec{R}\rt}\Big)\cdot \int_{\R^{3}}\!\!\!\!\!\:\!\dif^{3}r' \\
&\quad\;\, \mathcal{D}^{(\vec{R}\vec{R})}(\vec{r},\vec{r}',t)\Big(\Nabla_{\vec{r}'}\Fdif{\mathcal{S}}{\vec{R}(\vec{r}',t)}\Big) \\ 
&\geq 0.
\end{split}
\end{equation}
In the last step, we have used integration by parts together with the fact that the diffusion tensor $\mathcal{D}^{(\vec{R}\vec{R})}$ is positive definite by the Wiener–Khinchin theorem since it is the integral of an autocorrelation function \cite{EspanolL2009}. This result generalizes the H-theorem \cite{MarconiT1999,Munakata1994,EspanolL2009} of standard overdamped one-component DDFT towards nonisothermal inertial mixtures. In principle, this line of argument can be employed for an arbitrary set of relevant variables \cite{AneroET2013}.

To appreciate this result, it is helpful to consider the microscopic origins of thermodynamic irreversibility (see \citet{teVrugt2020} for a detailed discussion). Since the microscopic Hamiltonian dynamics we have started from is reversible, the macroscopic irreversibility has to be a consequence of approximations. Here, the crucial step is the \textit{Markovian approximation}, which allows to write the disorganized motion in the EDDFT equations (which, if we had used the exact result, contains memory) in a form that is local in time. This approximation is justified if the relevant variables are slow compared to the microscopic dynamics \cite{teVrugtW2019d}. Standard DDFT -- and its H-theorem -- therefore assume that momentum density and energy density relax quickly compared to the number density \cite{teVrugtLW2020}, an assumption that is not justified for many systems of interest. In contrast, the H-theorem derived here is based on an EDDFT that assumes the full set of conserved variables (mass, momentum, and energy) to be slow, which, as known from hydrodynamics, is justified for a large variety of systems (as long as there are no orientational degrees of freedom, which would require additional order parameters \cite{teVrugtW2020b}). Consequently, the H-theorem of EDDFT has a much wider applicability than the H-theorem of standard DDFT.

\subsection{Speed of sound}
Extensions of DDFT involving inertia allow to calculate the speed of sound \cite{teVrugtLW2020}. This was first shown by \citet{Archer2006}, who derived the speed of sound from a DDFT for dense atomic liquids. However, his derivation did not provide the correct value of the speed of sound, since his derivation was based on a DDFT that assumes a constant temperature despite the fact that sound waves are an adiabatic process \cite{Archer2006}. In contrast, EDDFT describes nonisothermal systems and therefore allows to obtain the correct value of the speed of sound.

We will here demonstrate the calculation for a one-component system without external forces for simplicity. First, we assume that mass and energy density have only small deviations from a constant reference value $\rhomo$ and $\varepsilon_0$, respectively, i.e., that they can be written as
\begin{align}
\rhom\rt &= \rhomo + \delta \rhom\rt,\\
\varepsilon\rt &= \varepsilon_0 + \delta \varepsilon \rt.
\end{align}
We can then linearize the equation of state $p(\rhom,T)$ (assuming a scalar pressure) around the pressure $p_0$ in the homogeneous state, which gives \cite{HansenMD2009}
\begin{equation}
\begin{split}
p &= p_0 +\Big(\pdif{p}{\rhom}\Big)_T \delta \rhom + \Big(\pdif{p}{T}\Big)_{\rhom} \delta T \\
&= p_0 +\frac{1}{\rhomo \chi_T}\delta \rhom + \beta_V \delta T,   
\end{split}\label{equationofstate}
\end{equation}
where we have introduced the isothermal compressibility $\chi_T = (1/\rhomo) \big(\pdif{\rhom}{p}\big)_T$ and the thermal pressure coefficient $\beta_V = \big(\pdif{p}{T}\big)_{\rhom}$.
Second, we linearize the one-component EDDFT equations \eqref{eq:EDDFTm_rhom_HL}-\eqref{eq:EDDFTm_e_HL} (written in terms of $\vec{g} = \rhom \vec{v}$) around $(\rhom,\vec{g},\varepsilon)=(\rhomo,\vec{0},\varepsilon_0)$ and drop the dissipative contributions. The result can, using \cref{equationofstate}, be written as
\begin{align}
\delta\dot{\rho}_\mathrm{m} &= - \Nabla \cdot \vec{g},\label{eq:EDDFTm_linear_rhom}\\
\dot{\vec{g}} &= - \frac{1}{\rhomo \chi_T} \Nabla \delta \rhom - \beta_V \Nabla \delta T,\label{eq:EDDFTm_linear_g}\\
\delta\dot{\varepsilon} &= - e_0 \Nabla \cdot \vec{g},\label{eq:EDDFT_linear_e}
\end{align}
with the constant parameter
\begin{equation}
e_0 = \frac{\varepsilon_0 + p_0}{\rhomo}.    
\end{equation}
From \cref{eq:EDDFTm_linear_rhom,eq:EDDFT_linear_e}, we can infer 
\begin{equation}
e_0 \delta\dot{\rho}_\mathrm{m} = \delta\dot{\varepsilon}.
\label{derivatives}
\end{equation}
From thermodynamics, we know that at constant particle number the change of entropy $\dif S_{\mathrm{th}}$ is given by\footnote{We use the notation $S_{\mathrm{th}}$ for the entropy of macroscopic thermodynamics, depending on $\rho$ (or $V$) and $T$, to distinguish it from the entropy functional $\mathcal{S}$ derived earlier, which also depends on $\vec{g}$ (and $c$ in the case of mixtures).}
\begin{equation}
T \dif S_{\mathrm{th}} = \dif U + p \dif V,  
\end{equation}
where $U$ is the internal energy and $V$ is the volume. With $U = \varepsilon V$ (homogeneous system), we get
\begin{equation}
\begin{split}
T \dif S_{\mathrm{th}} &= \dif (\varepsilon V) + p \dif V = V\dif \varepsilon + (\varepsilon +p) \dif V  \\
&= V \dif \varepsilon -\frac{\varepsilon + p}{\rhom} V\dif \rhom,  
\end{split}\label{thermodynamicrelations}
\end{equation}
where in the last step we have used that a constant total mass $M$ implies
\begin{equation}
\begin{split}
\dif M &= \dif (\rhom V) = \rhom \dif V + V \dif \rhom = 0 \\
&\rightarrow \dif V =  -\frac{V}{\rhom} \dif \rhom.
\end{split}
\end{equation}
On the other hand, we can generally expand \cite{HansenMD2009}
\begin{equation}
\begin{split}
T \dif S_{\mathrm{th}} &= T \Big(\pdif{S_{\mathrm{th}}}{\rhom}\Big)_T \dif \rhom + T\Big(\pdif{S_{\mathrm{th}}}{T}\Big)_{\rhom}\dif T \\
&= - \frac{T V \beta_V}{\rhomo} \dif \rhom + \frac{\rhom V c_V}{m} \dif T,  
\end{split}\label{tds2}
\end{equation}
with the heat capacity per particle at constant volume $c_V = (mT/(V\rhom)) \big(\pdif{S_{\mathrm{th}}}{T}\big)_\rho$ and the particle mass $m$, having used
\begin{equation}
-\frac{\rhom}{V}\Big(\pdif{S_{\mathrm{th}}}{\rhom}\Big)_T = \Big(\pdif{S_{\mathrm{th}}}{V}\Big)_T = \Big(\pdif{p}{T}\Big)_V,   
\end{equation}
where the last equality is a Maxwell relation. Equations \eqref{thermodynamicrelations} and \eqref{tds2} give\footnote{We have assumed here that the laws of thermodynamics are locally valid in the sense that, e.g., the relations derived for a macroscopic temperature change $\dif T$ hold for the local temperature fluctuation $\delta T\rt$.}
\begin{equation}
\delta \varepsilon - e_0 \delta \rhom = - \frac{T \beta_V}{\rhomo} \delta \rhom + \frac{\rhomo c_V}{m} \delta T.
\label{constant}
\end{equation}
The left-hand side of \cref{constant} has to be constant because of \cref{derivatives}. Therefore, the right-hand side of \cref{constant} also has to be constant, giving
\begin{equation}
\delta T = \frac{m T \beta_V}{\rhomo^2 c_V} 
\delta \rho_\mathrm{m}.    
\label{deltarhodot}
\end{equation}
Finally, we take the time derivative of \cref{eq:EDDFTm_linear_rhom} and insert \cref{eq:EDDFTm_linear_g,deltarhodot} to obtain
\begin{equation}
\begin{split}
\delta \ddot{\rho}_{\mathrm{m}} &= \frac{1}{\rhomo \chi_T} \Laplace \delta \rhom + \beta_V \Laplace \delta T \\
&= \bigg(\frac{1}{\rhomo \chi_T} + \frac{m T \beta_V^2}{\rhomo^2 c_V} \bigg) \Laplace \delta \rhom.
\end{split}\label{waveequation}
\end{equation}
Comparing \cref{waveequation} with the general wave equation
\begin{equation}
\delta \ddot{\rho}_{\mathrm{m}} 
= c_\mathrm{s}^2 \Laplace \delta\rhom    
\end{equation}
shows that the speed of sound $c_\mathrm{s}$ is given by
\begin{equation}
\begin{split}
c_\mathrm{s}^2 = \frac{1}{\rhomo \chi_T} + \frac{m T \beta_V^2}{\rhomo^2 c_V} 
= \frac{\gamma}{\rhomo \chi_T},
\end{split}\label{speedofsound}
\end{equation}
where in the last step we have introduced the ratio of specific heats $\gamma = c_p/ c_V$ with the specific heat at constant pressure $c_p$ that is given by \cite{HansenMD2009}
\begin{equation}
c_p = c_V + \frac{m T \chi_T \beta_V^2}{\rhomo} .
\end{equation}
Equation \eqref{speedofsound} gives the correct value for the speed of sound reported in the literature \cite{HansenMD2009}. Therefore, unlike simpler forms of DDFT \cite{Archer2006}, EDDFT with momentum density allows for a physically accurate description of sound waves as adiabatic processes involving thermal fluctuations. Moreover, since in EDDFT the pressure can be connected to the excess free energy via the adiabatic approximation (see \cref{ddft}) it allows to study sound propagation also from a microscopic perspective.

\begin{figure*}
\begin{tabular}{|c|c|c|c|c|c|}\hline
     \textbf{Approximation} & \textbf{General variable} &\textbf{Mass density} & \textbf{Concentration} & \textbf{Momentum density} & \textbf{Energy density}  \\ \hline
     None & \eqref{eq:EDDFT_exact} & \eqref{eq:EDDFT_rhom} & - & - & - \\ \hline
     Markovian & \eqref{eq:EDDFT_allg} & - &  \eqref{eq:EDDFT_c}& \eqref{eq:EDDFT_g} & \eqref{eq:EDDFT_e}  \\ \hline
     Hydrodynamic limit (HL) & -& - & \eqref{eq:EDDFT_c_HL1} & \eqref{eq:EDDFT_g_HL1} &  \eqref{eq:EDDFT_e_HL1} \\ \hline
     One-component & - & - &-  &\eqref{eq:EDDFTm_g} &\eqref{eq:EDDFTm_e}\\ \hline
     One-component HL & - & - & -& \eqref{eq:EDDFTm_g_HL1T} & \eqref{eq:EDDFTm_e_HL1T}\\ \hline
\end{tabular}
\caption{Overview over the various levels of approximation in the dynamic equations. The transport equation  \eqref{eq:EDDFT_rhom} for the mass density is exact and left unchanged by any approximation, such that the identical \cref{eq:EDDFT_rhom_HL1,eq:EDDFTm_rhom,eq:EDDFTm_rhom_HL1T} are not listed here. Note that $c$ is not a relevant variable for one-component systems.}
\label{overview}
\end{figure*}
 
\section{\label{sec:conclusions}Conclusions and perspectives}
We have derived an extended dynamical density functional theory (EDDFT) for a two-component system that includes as relevant variables the total mass density, the concentration of one species, the momentum density and the energy density. Various approximations have been developed for the general transport equations, which include an adiabatic approximation, the hydrodynamic limit and the case of a one-component system (see \cref{overview} for an overview). We have derived the explicit form of the EDDFT entropy functional for hard spheres, which allows to make use of the well-developed DFT methods these systems. Moreover, we have demonstrated that, unlike previous DDFTs with inertia, our EDDFT gives the correct value for the speed of sound in liquids. 
The new EDDFT allows to describe inertial nonisothermal flow and is therefore a significant extension of standard DDFT. It allows for a large variety of industrial applications, such as multiphase flow \cite{TornbergE2000,BalachandarE2010}, metal alloys \cite{OforiFGEP2013}, and the flow of cement \cite{RousselFC2010,MonlouisVP2004}. Moreover, it is of fundamental interest for a variety of phenomena in soft matter physics \cite{BrandP2021,PleinerB2021,PleinerH2004}.

Possible extensions of this work include the incorporation of orientational degrees of freedom to describe liquid crystals \cite{WittkowskiLB2010,WittkowskiLB2011,WittkowskiLB2011b}. This would also allow for the development of an EDDFT with momentum density for active particles \cite{WensinkL2008,WittkowskiL2011}, which would be an important contribution to the growing field of underdamped active matter \cite{Loewen2020,ScholzJLL2018} and an important extension of existing field theories for such systems \cite{teVrugtJW2021,AroldS2020,teVrugt2026,teVrugtFHHTW2023,AroldS2020b}. Moreover, one could further improve the applicability in industry by adding chemical reactions \cite{LiuL2020,MonchoD2020,teVrugtBW2020,LutskoN2016}.

\acknowledgments{We thank Andrew J.\ Archer, Jens Bickmann, Pep Espa{\~n}ol, Rudolf Haussmann, J\"urgen Horbach, Julian Jeggle, Jim Lutsko, Adrian Paskert, Matthias Schmidt, and Uwe Thiele for helpful discussions. M.t.V.\, H.L.\, and R.W.\ are funded by the Deutsche Forschungsgemeinschaft (DFG, German Research Foundation) -- Project-IDs 464588647 -- SFB 1551 (M.t.V.), LO 418/25 (H.L.), and 535275785 (R.W.). R.W.\ is also supported by the SONOCRAFT project, funded by the European Innovation and Research Council (GA: 101187842).}

\section*{Author contributions}
H.L., H.R.B., and R.W.\ conceptualized the research.
M.t.V.\ and R.W.\ performed the derivations. 
H.L.\ and H.R.B.\ checked the derivations. 
M.t.V.\ and R.W.\ wrote the first version of the manuscript with input from H.L.\ and H.R.B.
All authors reviewed and edited the manuscript. 
H.L.\ and H.R.B.\ supervised the work.

\appendix
\section{\label{inversenabla}A continuity equation does not always describe conserved quantities}
Although, mathematically, it is straightforward to see using \cref{deltaidentity} that the operator $\Nabla^{-1}$ given by \cref{nablaminuseins} satisfies \cref{nablainversion}, it might raise the following physical question: Often, it is argued that, if the time evolution of a field $\phi$ satisfies a continuity equation
\begin{equation}
\dot{\phi}\rt = - \Nabla \cdot \vec{J}\rt  
\label{continuityequation}
\end{equation}
with a current $\vec{J}$, this implies that $\phi$ is a conserved quantity. However, given an arbitrary field theory
\begin{equation}
\dot{\phi}\rt = q\rt
\label{arbitraryfieldtheory}
\end{equation}
with an arbitrary right-hand side $q$, we can always define
\begin{equation}
\vec{J}\rt = - \Nabla^{-1}q\rt    
\label{current}
\end{equation}
to write \cref{arbitraryfieldtheory} in the form \cref{continuityequation}. Obviously, this does not mean that $\phi$ is always conserved. However, it requires a closer look to see why $\phi$ may not be conserved even though we can formally define a current $\vec{J}$ via \cref{current}.

To see this, consider the simple model
\begin{equation}
\dot{\phi}\rt = -\phi\rt    
\label{simplemodel}
\end{equation}
(with dimensionless time) where $\phi$ is some density. Obviously, the field $\phi$ is not conserved. Nevertheless, we may, using \cref{current}, define a current
\begin{equation}
\vec{J}\rt = \INT{}{}{^3r'}\frac{\vec{r}-\vec{r}'}{4\pi\norm{\vec{r}-\vec{r}'}}\phi(\vec{r}',t), 
\label{gausscurrent}
\end{equation}
such that $\phi$ satisfies the continuity equation \eqref{continuityequation}. We should now remember that the reason that \cref{continuityequation} usually describes conserved quantities is that it implies that the rate of change of the total amount $\Phi$ of $\phi$ in a certain volume $V$, given by 
\begin{equation}
\Phi(t) = \INT{V}{}{^3r}\phi\rt    
\end{equation}
is
\begin{equation}
\dot{\Phi}(t) = \oint_{\partial V}\!\!\!\! \vec{J}\cdot \dif \vec{S} 
\label{surface}
\end{equation}
with the infinitesimal surface element $\dif \vec{S}$ and the surface $\partial V$ of the volume $V$. Physically, \cref{surface} means that the amount of $\phi$ in $V$ only changes due to the flux of $\phi$ through the surface of $V$.

In our case, this flux is given by \cref{gausscurrent}. If $\phi$ were an electric charge density, $\vec{J}$ would be proportional to the electric field caused by this charge density. However, Gauss' law states that the flux of an electric field through a surface is proportional of the amount of charge inside the volume enclosed by this surface irrespective of how it is distributed. Consequently, no matter how the volume $V$ is chosen, there will always be a non-vanishing flux $\vec{J}$ through its surface as long as there is some amount of $\phi$ in it, meaning that the amount of $\phi$ in an arbitrary volume is \textit{not} conserved although the form \cref{continuityequation} might suggest it. In fact, since the flux though the surface is proportional $\Phi$, the rate of change of $\Phi$ is by Gauss' law proportional to $\Phi$ (which is also an obvious consequence of \cref{simplemodel}). This stands in contrast to, e.g., the continuity equation \eqref{eq:EDDFT_rhom} for the mass density $\rhom$. Here, the current is given by $\vec{g}= \rhom \vec{v}$, which vanishes at the surface of a volume if there are no particles moving though this surface, which will generally be the case if this volume is large enough. Consequently, in contrast to our field $\phi$, the mass density $\rhom$ \textit{is} conserved.

\section{\label{momentumconservation}Conservation of momentum}
Using an adiabatic closure for the momentum density is extremely common \cite{Archer2009,BurghardtB2006}. What is surprising, however, is that an apparent problem of the resulting transport equation \eqref{generalizedeuler} for the momentum density is almost never addressed\footnote{A notable exception is the article by \citet{EspanolD2015}, where the MZFT is applied to a fluid with an immersed nanoparticle. There, it is discussed how certain approximations can break translational invariance and thus momentum conservation. However, their article considers a different model. Moreover, momentum conservation was discussed by \citet{NakamuraY2009} for a stochastic DDFT with momentum density.}: It does (in the absence of external forces) not have the form of a continuity equation, such that it is not immediately clear whether the total momentum is conserved. In this subsection, we will show that this is indeed the case.

Since the proof is rather involved, we give it for a one-component system with density $c=\rhom/m$. What we wish to show is that
\begin{equation}
\dot{\vec{g}}\rt = - k_{\mathrm{B}}T_0 c\rt \Nabla \Fdif{\mathcal{F}_\mathrm{exc}}{c\rt}
\label{problematicterm}
\end{equation}
conserves the total momentum. If we perform a functional Taylor expansion of $\mathcal{F}_{\mathrm{exc}}$ around a homogeneous reference density $c_0$ as \cite{EmmerichEtAl2012,teVrugtLW2020}
\begin{equation}
\begin{split}
\mathcal{F}_{\mathrm{exc}} &= \mathcal{F}_{\mathrm{exc},0}- \sum_{k=1}^{\infty}\frac{1}{k!} \int_{\R^{3}}\!\!\!\!\:\!\dif^{3}r_1 \dotsb \int_{\R^{3}}\!\!\!\!\:\!\dif^{3}r_k\\
&\quad\, c^{(k)}(\vec{r}_1,\dotsc,\vec{r}_k) \Delta c(\vec{r}_1) \dotsb \Delta c(\vec{r}_k)
\end{split}
\label{functionaltaylorexpansion}
\end{equation}
with $\Delta c = c - c_0$, an irrelevant constant $\mathcal{F}_{\mathrm{exc},0}$ and the direct correlation function\footnote{The direct correlation function $c^{(k)}$ (with superscript) should not be confused with the density $c$ (without superscript).} 
\begin{equation}
c^{(k)}(\vec{r}_1,\dotsc,\vec{r}_k)  = \Fdif{^k\mathcal{F}_{\mathrm{exc}}}{c(\vec{r}_1)\dotsb \delta c(\vec{r}_k)}   
\end{equation}
we can, using \cref{problematicterm}, compute the change of the total momentum $\vec{P}$ as
\begin{equation}
\begin{split}
\dot{\vec{P}}(t)= & \sum_{k=1}^{\infty}\frac{k_{\mathrm{B}}T_0}{k!} \int_{\R^{3}}\!\!\!\!\:\!\dif^{3}r_1 \dotsb \int_{\R^{3}}\!\!\!\!\:\!\dif^{3}r_k\\
&\Delta c(\vec{r}_1) \Nabla_{\vec{r}_1} c^{(k)}(\vec{r}_1,\dotsc,\vec{r}_k) \Delta c(\vec{r}_2)\dotsb \Delta c(\vec{r}_k).
\end{split}
\label{dotp}
\end{equation}
Strictly speaking, the first factor in the second row of \cref{dotp} should be $c$ rather than $\Delta c$, but this replacement only requires adding a term $ c_0 \Nabla (\delta \mathcal{F}_{\mathrm{exc}}/\delta c)$ in \cref{problematicterm} that is the divergence of a current and thus does not harm momentum conservation. 
We can now use translational and rotational invariance, which implies that the direct correlation functions can be written as
\begin{equation}
\begin{split}
&c^{(k)}(\vec{r}_1,\dotsc,\vec{r}_k) = c^{(k)}(\norm{\vec{r}_2-\vec{r}_1},\dotsc,\norm{\vec{r}_k - \vec{r}_1})\\ 
&= c^{(k)}(r_1,\dotsc,r_{k-1})
\end{split}
\label{correlationinvariance}
\end{equation}
with $r_i = \norm{\vec{r}_{i+1}-\vec{r}_1}$. Equation
\eqref{correlationinvariance} implies that
\begin{equation}
\begin{split}
\Nabla_{\vec{r}_1}  c^{(k)}(\vec{r}_1,\dotsc,\vec{r}_k)  = \sum_{i=1}^{k-1} \frac{\vec{r}_{i+1} - \vec{r}_1}{r_i}\pdif{c^{(k)}}{r_i}.
\end{split}    
\label{relation}
\end{equation}
Inserting \cref{relation} into \cref{dotp}, we find
\begin{equation}
\begin{split}
\dot{\vec{P}}(t)= & \sum_{k=1}^{\infty}\sum_{i=1}^{k-1}\frac{k_{\mathrm{B}}T_0}{k!} \int_{\R^{3}}\!\!\!\!\:\!\dif^{3}r_1 \dotsb \int_{\R^{3}}\!\!\!\!\:\!\dif^{3}r_k\\
&\Delta c(\vec{r}_1)\dotsb \Delta c(\vec{r}_k)  \frac{\vec{r}_{i+1} - \vec{r}_1}{r_i}\pdif{c^{(k)}}{r_i}.
\end{split}
\label{dotp2}    
\end{equation}
Now since all position coordinates are integrated over (and therefore dummy variables), we could exchange $\vec{r}_1$ and $\vec{r}_i$ in every term in \cref{dotp2}. This gives the same result with a minus sign, showing $\dot{\vec{P}}(t)= - \dot{\vec{P}}(t)$ and thus $\dot{\vec{P}}(t) = \vec{0}$.

Note that this argument would not go through in the same way in the nonisothermal case, where (as can be seen from \cref{adiabaticapproximation}) we would also have to consider derivatives of $T$. This problem can be traced back to the fact that in the nonisothermal case $\mathcal{F}_{\mathrm{exc}}$ is a functional not just of $c$, but also of $T$, such that the expansion \eqref{functionaltaylorexpansion} would have to be generalized in a suitable way.

\section{\label{diffusiontensors}Diffusion tensors}
In the EDDFT equations \eqref{eq:EDDFT_rhom}-\eqref{eq:EDDFT_e}, $\DT^{(cc)}(\vec{r},\rs\!,t)$ is the diffusion tensor of the colloidal particles, $\DT^{(c\varepsilon)}(\vec{r},\rs\!,t)$ and $\DT^{(\varepsilon c)}(\vec{r},\rs\!,t)$ describe the coupling of particle diffusion and diffusive energy transport and are directly related to (inverse) thermodiffusion\footnote{The tensor $\DT^{(c\varepsilon)}(\vec{r},\rs\!,t)$ is associated with the Ludwig-Soret effect and $\DT^{(\varepsilon c)}(\vec{r},\rs\!,t)$ is associated with the Dufour effect \cite{WittkowskiLB2012}.}, $\DT^{(g_{i}g_{j})}(\vec{r},\rs\!,t)$ is a viscous stress tensor, and $\DT^{(\varepsilon\varepsilon)}(\vec{r},\rs\!,t)$ describes diffusive energy transport. 
Notice that due to Eq.\ \eqref{eq:QJ_rhom} all diffusion tensors $\DT^{(\mu\nu)}(\vec{r},\rs\!,t)$ with $\mu\equiv\rhom $ or $\nu\equiv\rhom $ vanish. 
Furthermore, the diffusion tensors with $(\mu,\nu)=(c,g_{i})$ or $(\mu,\nu)=(g_{i},c)$ vanish due to the different behavior of $c\rt$ and $g_{i}\rt$ under time reversal. For $\varepsilon\rt$ instead of $c\rt$ the situation is analogous. 
The remaining diffusion tensors appearing in the EDDFT equations 
\eqref{eq:EDDFT_rhom}-\eqref{eq:EDDFT_e} are explicitly given by\footnote{Equations \eqref{eq:D_cc}-\eqref{eq:D_ee} follow from Eq.\ \eqref{eq:Dijkl_allg} and Eqs.\ \eqref{eq:QJ_rhom}-\eqref{eq:QJ_e} by the Galilean transformation $\vec{p}^{\mu}_{i}(t)\to\vec{p}^{\mu}_{i}(t)+\vec{g}^{\natural}(\vec{r}^{\mu}_{i},t)m_{\mu}$. Notice that the trace $\Tr$ of an arbitrary expression is invariant under a Galilean transformation of this expression \cite{Grabert1982}.} 
{\allowdisplaybreaks%
\begin{align}%
\DT^{(cc)}(\vec{r},\rs\!,t) &= \!\int^{\infty}_{0}\!\!\!\!\!\!\dif t'\,\Tr\!\Big(\tilde{\rho}(t)
\Jw^{c}_{t}(\vec{r},t')\!\otimes\!\Jw^{c}_{t}(\rs\!,0)\!\Big) ,
\label{eq:D_cc}\raisetag{0.7em}\\%
\DT^{(c\varepsilon)}(\vec{r},\rs\!,t) &= \!\int^{\infty}_{0}\!\!\!\!\!\!\dif t'\,\Tr\!\Big(\tilde{\rho}(t)
\Jw^{c}_{t}(\vec{r},t')\!\otimes\!\Jw^{\varepsilon}_{t}(\rs\!,0)\!\Big) ,
\label{eq:D_ce}\raisetag{0.7em}\\%
\DT^{(g_{i}g_{j})}(\vec{r},\rs\!,t) &= \!\int^{\infty}_{0}\!\!\!\!\!\!\dif t'\,\Tr\!\Big(\tilde{\rho}(t)
\Jw^{g_{i}}_{t}(\vec{r},t')\!\otimes\!\Jw^{g_{j}}_{t}(\rs\!,0)\!\Big) ,
\label{eq:D_gg}\raisetag{0.7em}\\%
\DT^{(\varepsilon c)}(\vec{r},\rs\!,t) &= \!\int^{\infty}_{0}\!\!\!\!\!\!\dif t'\,\Tr\!\Big(\tilde{\rho}(t)
\Jw^{\varepsilon}_{t}(\vec{r},t')\!\otimes\!\Jw^{c}_{t}(\rs\!,0)\!\Big) ,
\label{eq:D_ec}\raisetag{0.7em}\\%
\DT^{(\varepsilon\varepsilon)}(\vec{r},\rs\!,t) &= \!\int^{\infty}_{0}\!\!\!\!\!\!\dif t'\,\Tr\!\Big(\tilde{\rho}(t)
\Jw^{\varepsilon}_{t}(\vec{r},t')\!\otimes\!\Jw^{\varepsilon}_{t}(\rs\!,0)\!\Big) 
\label{eq:D_ee}\raisetag{0.7em}%
\end{align}}%
with the Galilean-transformed relevant probability density 
\begin{equation}
\begin{split}
\tilde{\rho}(t) &= \frac{1}{Z(t)}\exp\!\bigg(\! -\int_{\R^{3}}\!\!\!\!\!\:\!\dif^{3}r \,\beta\rt\Big(\hat{\varepsilon}(\vec{r}) \\
&\quad\:\! -\Big(\rho^{\natural}_{\mathrm{m}}\rt
+\text{\footnotesize$\frac{1}{2}$}\big(\vec{g}^{\natural}\rt\big)^{2}\Big)\hat{\rho}_{\mathrm{m}}(\vec{r}) \\
&\quad\:\! -c^{\natural}\rt\hat{c}(\vec{r})\Big)\!\bigg)
\end{split}\label{eq:rho_GT}%
\end{equation}
\\
that is even in the momenta, the notation $(\Jw^{g_{i}}_{t}(\vec{r},\tau))_{k}=(\hat{\mathcal{J}}^{\vec{g}}_{t}(\vec{r},\tau))_{ik}$, and the Galilean-transformed projected currents\footnote{The current \eqref{eq:QJ_gW} contains the term $\widetilde{\Upsilon}_{t}(\vec{r},\tau)-\Pi_{1}\rt$ that is completely averaged and does not explicitly depend on the phase-space variables. To avoid that Eq.\ \eqref{eq:D_gg} diverges, this term has to be assumed to vanish for $\tau=0$. In fact, this term is zero for $\tau=0$ up to linear order in the averaged relevant variables, but there can be higher-order contributions that do not vanish. The circumstance that the term $\widetilde{\Upsilon}_{t}(\vec{r},\tau)-\Pi_{1}\rt$ vanishes only up to linear order in the averaged relevant variables is a consequence of the linearity of the projection operator \eqref{eq:Q_Operator} in the relevant variables.} 
\begin{align}%
\Jw^{c}_{t}(\vec{r},\tau) &= \hat{\vec{J}}^{c}(\vec{r},\tau)
-\hat{\vec{g}}(\vec{r},\tau)\frac{c\rt}{\rhom \rt} ,
\label{eq:QJ_cW}\\%
\notag \hat{\mathcal{J}}^{\vec{g}}_{t}(\vec{r},\tau) &= \hat{\mathrm{J}}^{\vec{g}}(\vec{r},\tau) \\
&\quad\:\! -\frac{\Eins}{\beta\:\! m_{\mathrm{s}}}\big(\hat{\rho}_{\mathrm{m}}(\vec{r},\tau)-(m_{\mathrm{c}}-m_{\mathrm{s}})\hat{c}(\vec{r},\tau)\big) \label{eq:QJ_gW}\\ 
&\quad\:\! -\Delta\hat{\widetilde{\Upsilon}}_{t}(\vec{r},\tau)-\Pi_{1}\rt ,
\notag\\%
\nonumber \Jw^{\varepsilon}_{t}(\vec{r},\tau) &= \hat{\mathcal{J}}^{\vec{g}}_{t}(\vec{r},\tau)\vec{g}^{\natural}\rt
+\hat{\vec{J}}^{\varepsilon}(\vec{r},\tau) 
-\frac{\hat{\vec{g}}(\vec{r},\tau)}{\rhom \rt}\\
&\quad\, \Big(\!\big(\varepsilon\rt-\text{\scriptsize$\frac{1}{2}$}\vec{g}\rt\!\cdot\!\vec{g}^{\natural}\rt\big)\Eins 
+\Pi\rt\!\Big) . \label{eq:QJ_eW} 
\end{align}%
Here, we introduced the Galilean-transformed generalized microscopic bulk modulus 
\begin{equation}
\begin{split}
\hat{\widetilde{\Upsilon}}_{t}(\vec{r},\tau)&=\int_{\R^{3}}\!\!\!\!\:\!\dif^{3}r'\Big(\hat{\rho}_{\mathrm{m}}(\rs\!,\tau)
\Fdif{\Pi_{1}\rt}{\rhom (\rs\!,t)}
+\hat{c}(\rs\!,\tau)\Fdif{\Pi_{1}\rt}{c(\rs\!,t)} 
\\ &\quad\:\! +\hat{e}(\rs\!,\tau)\Fdif{\Pi_{1}\rt}{\varepsilon(\rs\!,t)}\Big) ,
\end{split}\raisetag{2em}%
\end{equation}
its average $\widetilde{\Upsilon}_{t}(\vec{r},\tau)=\Tr(\rho(\tau)\hat{\widetilde{\Upsilon}}_{t}(\vec{r}))$, 
and the difference $\Delta\hat{\widetilde{\Upsilon}}_{t}(\vec{r},\tau)=\hat{\widetilde{\Upsilon}}_{t}(\vec{r},\tau)-\widetilde{\Upsilon}_{t}(\vec{r},\tau)$ with the Galilean-transformed energy density 
$\hat{e}\rt=\hat{\varepsilon}\rt+\hat{\vec{g}}\rt\!\cdot\!\vec{g}^{\natural}\rt+\frac{1}{2}\hat{\rho}_{\mathrm{m}}\rt(\vec{g}^{\natural}\rt)^{2}$, whose average is defined by 
$e\rt=\Tr(\tilde{\rho}(t)\hat{e}(\vec{r}))=\varepsilon\rt$.
The diffusion tensors \eqref{eq:D_cc}-\eqref{eq:D_ee} are given more explicitly by 
\begin{widetext}
{\allowdisplaybreaks%
\begin{align}%
\begin{split}%
\DT^{(cc)}(\vec{r},\rs\!,t) &= \mathrm{D}^{(\rhom \rhom )}(\vec{r},\rs\!,t)\frac{c\rt}{\rhom \rt}\frac{c(\rs\!,t)}{\rhom (\rs\!,t)} 
-\mathrm{D}^{(\rhom c)}(\vec{r},\rs\!,t)\frac{c\rt}{\rhom \rt}
-\mathrm{D}^{(c\rhom )}(\vec{r},\rs\!,t)\frac{c(\rs\!,t)}{\rhom (\rs\!,t)} \\
&\quad\:\!+\mathrm{D}^{(cc)}(\vec{r},\rs\!,t) ,
\end{split}\label{eq:D_cc_explicitly}\\%
\begin{split}%
\DT^{(c\varepsilon)}(\vec{r},\rs\!,t) &= \frac{c\rt}{\rhom \rt}
\Big(\mathrm{D}^{(\rhom \rhom )}(\vec{r},\rs\!,t)\mathrm{M}(\rs\!,t)
-\mathrm{D}^{(\rhom \varepsilon)}(\vec{r},\rs\!,t)\Big)
-\mathrm{D}^{(c\rhom )}(\vec{r},\rs\!,t)\mathrm{M}(\rs\!,t)
+\mathrm{D}^{(c\varepsilon)}(\vec{r},\rs\!,t) \\
&\quad\:\!+\!\int_{\R^{3}}\!\!\!\!\!\:\!\dif^{3}r''\:\!\Big(\frac{c\rt}{\rhom \rt}
\mathrm{D}^{(\rhom \rhom )}(\vec{r},\rss\!,t)
-\mathrm{D}^{(c\rhom )}(\vec{r},\rss\!,t)\!\Big)
\Lambda(\rss\!,\rs\!,t)\Fdif{\Pi_{1}(\rs\!,t)}{\varepsilon(\rss\!,t)} ,
\end{split}\label{eq:D_ce_explicitly}\\%
\begin{split}%
\DT^{(g_{i}g_{j})}(\vec{r},\rs\!,t) &= \frac{1}{(\beta\:\! m_{\mathrm{s}})^{2}}\Big(\CF^{(\rhom \rhom )}(\vec{r},\rs\!,t)
-m_{\mathrm{cs}}\big(\CF^{(\rhom c)}(\vec{r},\rs\!,t)+\CF^{(c\rhom )}(\vec{r},\rs\!,t)\big) 
+m^{2}_{\mathrm{cs}}\CF^{(cc)}(\vec{r},\rs\!,t)\Big) \hat{\mathfrak{e}}_{i}\!\otimes\!\hat{\mathfrak{e}}_{j} \\
&\quad\:\!-\frac{1}{\beta\:\! m_{\mathrm{s}}}\Big(\hat{\mathfrak{e}}_{i}\!\otimes\!\CV^{(\rhom g_{j})}(\vec{r},\rs\!,t)+\CV^{(g_{i}\rhom )}(\vec{r},\rs\!,t)\!\otimes\!\hat{\mathfrak{e}}_{j}\Big) \\ 
&\quad\:\!+\frac{m_{\mathrm{cs}}}{\beta\:\! m_{\mathrm{s}}}\Big(\hat{\mathfrak{e}}_{i}\!\otimes\!\CV^{(cg_{j})}(\vec{r},\rs\!,t)
+\CV^{(g_{i}c)}(\vec{r},\rs\!,t)\!\otimes\!\hat{\mathfrak{e}}_{j}\Big) 
+\mathrm{D}^{(g_{i}g_{j})}(\vec{r},\rs\!,t) \\
&\quad\:\!+\frac{1}{\beta\:\! m_{\mathrm{s}}}\!\!\!\!\!\!\sum_{\lambda\in\{\rhom ,c,\varepsilon\}}\!\!\!\!\!\!\int_{\R^{3}}\!\!\!\!\!\:\!\dif^{3}r''\:\!
\Big(\CF^{(\rhom \lambda)}(\vec{r},\rss\!,t)\hat{\mathfrak{e}}_{i}\!\otimes\!\vec{\chi}^{(j)}_{\lambda}(\rs\!,\rss\!,t) +\CF^{(\lambda\rhom )}(\rss\!,\rs\!,t)
\vec{\chi}^{(i)}_{\lambda}(\vec{r},\rss\!,t)\!\otimes\!\hat{\mathfrak{e}}_{j}\Big) \\
&\quad\:\!-\frac{m_{\mathrm{cs}}}{\beta\:\! m_{\mathrm{s}}}\!\!\!\!\!\!\sum_{\lambda\in\{\rhom ,c,\varepsilon\}}\!\!\!\!\!\!\int_{\R^{3}}\!\!\!\!\!\:\!\dif^{3}r''\:\!
\Big(\CF^{(c\lambda)}(\vec{r},\rss\!,t)\hat{\mathfrak{e}}_{i}\!\otimes\!\vec{\chi}^{(j)}_{\lambda}(\rs\!,\rss\!,t) +\CF^{(\lambda c)}(\rss\!,\rs\!,t)\vec{\chi}^{(i)}_{\lambda}(\vec{r},\rss\!,t)\!\otimes\!\hat{\mathfrak{e}}_{j}\Big) \\
&\quad\:\!-\!\!\!\sum_{\lambda\in\{\rhom ,c,\varepsilon\}}\!\!\!\!\!\!\int_{\R^{3}}\!\!\!\!\!\:\!\dif^{3}r''\:\!
\Big(\vec{\chi}^{(i)}_{\lambda}(\vec{r},\rss\!,t)\!\otimes\!\CV^{(\lambda g_{j})}(\rss\!,\rs\!,t)
+\CV^{(g_{i}\lambda)}(\vec{r},\rss\!,t)\!\otimes\!\vec{\chi}^{(j)}_{\lambda}(\rs\!,\rss\!,t)\Big) \\
&\quad\:\!+\!\!\!\!\!\sum_{\lambda,\mu\in\{\rhom ,c,\varepsilon\}}\!\!\!\!\!\!\!\!
\int_{\R^{3}}\!\!\!\!\!\:\!\dif^{3}r''\!\int_{\R^{3}}\!\!\!\!\!\:\!\dif^{3}r'''\, 
\CF^{(\lambda\mu)}(\rss\!,\rsss\!,t)\vec{\chi}^{(i)}_{\lambda}(\vec{r},\rss\!,t)
\!\otimes\!\vec{\chi}^{(j)}_{\mu}(\rs\!,\rsss\!,t) \\
&\quad\:\!+\!\int_{\R^{3}}\!\!\!\!\!\:\!\dif^{3}r''\!\int_{\R^{3}}\!\!\!\!\!\:\!\dif^{3}r'''\,
\vec{g}^{\natural}(\rss\!,t)\mathrm{D}^{(\rhom \rhom )}(\rss\!,\rsss\!,t)\vec{g}^{\natural}(\rsss\!,t)
\vec{\chi}^{(i)}_{\varepsilon}(\vec{r},\rss\!,t)\!\otimes\!\vec{\chi}^{(j)}_{\varepsilon}(\rs\!,\rsss\!,t) ,
\end{split}\label{eq:D_gg_explicitly}\\%
\begin{split}%
\DT^{(\varepsilon c)}(\vec{r},\rs\!,t) &= \frac{c(\rs\!,t)}{\rhom (\rs\!,t)}
\Big(\mathrm{M}\rt\mathrm{D}^{(\rhom \rhom )}(\vec{r},\rs\!,t)
-\mathrm{D}^{(\varepsilon\rhom )}(\vec{r},\rs\!,t)\Big)
-\mathrm{M}\rt\mathrm{D}^{(\rhom c)}(\vec{r},\rs\!,t)
+\mathrm{D}^{(\varepsilon c)}(\vec{r},\rs\!,t) \\
&\quad\:\!+\!\int_{\R^{3}}\!\!\!\!\!\:\!\dif^{3}r''\,\Fdif{\Pi_{1}\rt}{\varepsilon(\rss\!,t)}
\Lambda(\vec{r},\rss\!,t)\bigg(\frac{c(\rs\!,t)}{\rhom (\rs\!,t)}
\mathrm{D}^{(\rhom \rhom )}(\rss\!,\rs\!,t)
-\mathrm{D}^{(\rhom c)}(\rss\!,\rs\!,t)\!\bigg) \:\!,
\end{split}\label{eq:D_ec_explicitly}\\%
\begin{split}%
\DT^{(\varepsilon\varepsilon)}(\vec{r},\rs\!,t) &= \!\sum^{3}_{i,j=1}\!\vec{g}^{\natural}\rt
\DT^{(g_{i}g_{j})}(\vec{r},\rs\!,t)\vec{g}^{\natural}(\rs\!,t)\:\!\hat{\mathfrak{e}}_{i}\!\otimes\!\hat{\mathfrak{e}}_{j}
+\mathrm{M}\rt\mathrm{D}^{(\rhom \rhom )}(\vec{r},\rs\!,t)\mathrm{M}(\rs\!,t) \\
&\quad\:\!-\mathrm{M}\rt\mathrm{D}^{(\rhom \varepsilon)}(\vec{r},\rs\!,t)
-\mathrm{D}^{(\varepsilon\rhom )}(\vec{r},\rs\!,t)\mathrm{M}(\rs\!,t)
+\mathrm{D}^{(\varepsilon\varepsilon)}(\vec{r},\rs\!,t) \\
&\quad\:\!+\!\int_{\R^{3}}\!\!\!\!\!\:\!\dif^{3}r''\,\Fdif{\Pi_{1}\rt}{\varepsilon(\rss\!,t)}\Lambda(\vec{r},\rss\!,t) 
\Big(\mathrm{D}^{(\rhom \rhom )}(\rss\!,\rs\!,t)\mathrm{M}(\rs\!,t)-\mathrm{D}^{(\rhom \varepsilon)}(\rss\!,\rs\!,t)\Big) \\
&\quad\:\!+\!\int_{\R^{3}}\!\!\!\!\!\:\!\dif^{3}r''\:\!
\Big(\mathrm{M}\rt\mathrm{D}^{(\rhom \rhom )}(\vec{r},\rss\!,t)
-\mathrm{D}^{(\varepsilon\rhom )}(\vec{r},\rss\!,t)\Big)
\Lambda(\rss\!,\rs\!,t)\Fdif{\Pi_{1}(\rs\!,t)}{\varepsilon(\rss\!,t)} 
\end{split}\label{eq:D_ee_explicitly}%
\end{align}}%
\end{widetext}
with the Cartesian unit vectors $\hat{\mathfrak{e}}_{i}$ for $i\in\{1,2,3\}$, the reduced mass $m_{\mathrm{cs}}=m_{\mathrm{c}}-m_{\mathrm{s}}$, the functions
{\allowdisplaybreaks%
\begin{align}%
\begin{split}%
\vec{\chi}^{(i)}_{\rhom }(\vec{r},\rs\!,t)&=\Fdif{\vec{\Pi}^{(i)}_{1}\rt}{\rhom (\rs\!,t)}  +\frac{1}{2}\big(\vec{g}^{\natural}(\rs\!,t)\big)^{2}\Fdif{\vec{\Pi}^{(i)}_{1}\rt}{\varepsilon(\rs\!,t)} ,
\end{split}\label{eq:chi_rhom}\raisetag{1.8em}\\%
\begin{split}%
\vec{\chi}^{(i)}_{c}(\vec{r},\rs\!,t)&=\Fdif{\vec{\Pi}^{(i)}_{1}\rt}{c(\rs\!,t)} ,
\end{split}\label{eq:chi_c}\\%
\begin{split}%
\vec{\chi}^{(i)}_{\varepsilon}(\vec{r},\rs\!,t)&=\Fdif{\vec{\Pi}^{(i)}_{1}\rt}{\varepsilon(\rs\!,t)} ,
\end{split}\label{eq:chi_e}\\%
\begin{split}%
\Lambda(\vec{r},\rs\!,t)&=\vec{g}^{\natural}\rt\!\otimes\!\vec{g}^{\natural}(\rs\!,t) ,
\end{split}\label{eq:Lambda}\\%
\begin{split}%
\mathrm{M}\rt&=\frac{1}{\rhom \rt}\Big(\!\Big(\varepsilon\rt-\text{\footnotesize$\frac{1}{2}$}\vec{g}\rt\!\cdot\!\vec{g}^{\natural}\rt\Big)\Eins \\
&\quad\:\! +\Pi\rt\Big) ,
\end{split}\label{eq:M}\raisetag{2em}%
\end{align}}%
and the notation $(\vec{\Pi}^{(i)}_{1}\rt)_{k}=(\Pi_{1}\rt)_{ik}$.
Equations \eqref{eq:D_cc_explicitly}-\eqref{eq:D_ee_explicitly} express the diffusion tensors \eqref{eq:D_cc}-\eqref{eq:D_ee} in terms of the averaged relevant variables, the pressure tensor, 
the correlation functions 
{\allowdisplaybreaks%
\begin{align}%
\begin{split}%
\CF^{(\rhom \rhom )}(\vec{r},\rs\!,t) &=\!\int^{\infty}_{0}\!\!\!\!\!\!\dif t'\,\Tr\!\Big(\tilde{\rho}(t)
\hat{\rho}_{\mathrm{m}}(\vec{r},t')\hat{\rho}_{\mathrm{m}}(\rs\!,0)\!\Big) ,
\end{split}\label{eq:CF_rhomrhom}\raisetag{2em}\\%
\begin{split}%
\CF^{(\rhom c)}(\vec{r},\rs\!,t) &=\!\int^{\infty}_{0}\!\!\!\!\!\!\dif t'\,\Tr\!\Big(\tilde{\rho}(t)
\hat{\rho}_{\mathrm{m}}(\vec{r},t')\hat{c}(\rs\!,0)\!\Big) ,
\end{split}\label{eq:CF_rhomc}\\%
\begin{split}%
\CF^{(\rhom \varepsilon)}(\vec{r},\rs\!,t) &=\!\int^{\infty}_{0}\!\!\!\!\!\!\dif t'\,\Tr\!\Big(\tilde{\rho}(t)
\hat{\rho}_{\mathrm{m}}(\vec{r},t')\hat{\varepsilon}(\rs\!,0)\!\Big) ,
\end{split}\label{eq:CF_rhome}\\%
\begin{split}%
\CF^{(c\rhom )}(\vec{r},\rs\!,t) &=\!\int^{\infty}_{0}\!\!\!\!\!\!\dif t'\,\Tr\!\Big(\tilde{\rho}(t)
\hat{c}(\vec{r},t')\hat{\rho}_{\mathrm{m}}(\rs\!,0)\!\Big) ,
\end{split}\label{eq:CF_crhom}\\%
\begin{split}%
\CF^{(cc)}(\vec{r},\rs\!,t) &=\!\int^{\infty}_{0}\!\!\!\!\!\!\dif t'\,\Tr\!\Big(\tilde{\rho}(t)
\hat{c}(\vec{r},t')\hat{c}(\rs\!,0)\!\Big) ,
\end{split}\label{eq:CF_cc}\\%
\begin{split}%
\CF^{(c\varepsilon)}(\vec{r},\rs\!,t) &=\!\int^{\infty}_{0}\!\!\!\!\!\!\dif t'\,\Tr\!\Big(\tilde{\rho}(t)
\hat{c}(\vec{r},t')\hat{\varepsilon}(\rs\!,0)\!\Big) ,
\end{split}\label{eq:CF_ce}\\%
\begin{split}%
\CF^{(\varepsilon\rhom )}(\vec{r},\rs\!,t) &=\!\int^{\infty}_{0}\!\!\!\!\!\!\dif t'\,\Tr\!\Big(\tilde{\rho}(t)
\hat{\varepsilon}(\vec{r},t')\hat{\rho}_{\mathrm{m}}(\rs\!,0)\!\Big) ,
\end{split}\label{eq:CF_erhom}\\%
\begin{split}%
\CF^{(\varepsilon c)}(\vec{r},\rs\!,t) &=\!\int^{\infty}_{0}\!\!\!\!\!\!\dif t'\,\Tr\!\Big(\tilde{\rho}(t)
\hat{\varepsilon}(\vec{r},t')\hat{c}(\rs\!,0)\!\Big) ,
\end{split}\label{eq:CF_ec}\\%
\begin{split}%
\CF^{(\varepsilon\varepsilon)}(\vec{r},\rs\!,t) &=\!\int^{\infty}_{0}\!\!\!\!\!\!\dif t'\,\Tr\!\Big(\tilde{\rho}(t)
\hat{\varepsilon}(\vec{r},t')\hat{\varepsilon}(\rs\!,0)\!\Big) ,
\end{split}\label{eq:CF_ee}%
\end{align}}%
the correlation vectors 
{\allowdisplaybreaks%
\begin{align}%
\begin{split}%
\CV^{(\rhom g_{i})}(\vec{r},\rs\!,t) &=\!\int^{\infty}_{0}\!\!\!\!\!\!\dif t'\,\Tr\!\Big(\tilde{\rho}(t)
\hat{\rho}_{\mathrm{m}}(\vec{r},t')\hat{\vec{J}}^{g_{i}}(\rs\!,0)\!\Big) ,
\end{split}\label{eq:CV_rhomgi}\raisetag{2em}\\%
\begin{split}%
\CV^{(cg_{i})}(\vec{r},\rs\!,t) &=\!\int^{\infty}_{0}\!\!\!\!\!\!\dif t'\,\Tr\!\Big(\tilde{\rho}(t)
\hat{c}(\vec{r},t')\hat{\vec{J}}^{g_{i}}(\rs\!,0)\!\Big) ,
\end{split}\label{eq:CV_cgi}\\%
\begin{split}%
\CV^{(\varepsilon g_{i})}(\vec{r},\rs\!,t) &=\!\int^{\infty}_{0}\!\!\!\!\!\!\dif t'\,\Tr\!\Big(\tilde{\rho}(t)
\hat{\varepsilon}(\vec{r},t')\hat{\vec{J}}^{g_{i}}(\rs\!,0)\!\Big) ,
\end{split}\label{eq:CV_egi}\\%
\begin{split}%
\CV^{(g_{i}\rhom )}(\vec{r},\rs\!,t) &=\!\int^{\infty}_{0}\!\!\!\!\!\!\dif t'\,\Tr\!\Big(\tilde{\rho}(t)
\hat{\vec{J}}^{g_{i}}(\vec{r},t')\hat{\rho}_{\mathrm{m}}(\rs\!,0)\!\Big) ,
\end{split}\label{eq:CV_girhom}\raisetag{2em}\\%
\begin{split}%
\CV^{(g_{i}c)}(\vec{r},\rs\!,t) &=\!\int^{\infty}_{0}\!\!\!\!\!\!\dif t'\,\Tr\!\Big(\tilde{\rho}(t)
\hat{\vec{J}}^{g_{i}}(\vec{r},t')\hat{c}(\rs\!,0)\!\Big) ,
\end{split}\label{eq:CV_gic}\\%
\begin{split}%
\CV^{(g_{i}\varepsilon)}(\vec{r},\rs\!,t) &=\!\int^{\infty}_{0}\!\!\!\!\!\!\dif t'\,\Tr\!\Big(\tilde{\rho}(t)
\hat{\vec{J}}^{g_{i}}(\vec{r},t')\hat{\varepsilon}(\rs\!,0)\!\Big) ,
\end{split}\label{eq:CV_gie}%
\end{align}}%
and the basic diffusion tensors 
{\allowdisplaybreaks%
\begin{align}%
\begin{split}%
\mathrm{D}^{(\rhom \rhom )}(\vec{r},\rs\!,t) &=\!\int^{\infty}_{0}\!\!\!\!\!\!\dif t'\,\Tr\!\Big(\tilde{\rho}(t)
\hat{\vec{J}}^{\rhom }(\vec{r},t')\!\otimes\!\hat{\vec{J}}^{\rhom }(\rs\!,0)\!\Big) ,
\end{split}\label{eq:SD_rhomrhom}\\%
\begin{split}%
\mathrm{D}^{(\rhom c)}(\vec{r},\rs\!,t) &=\!\int^{\infty}_{0}\!\!\!\!\!\!\dif t'\,\Tr\!\Big(\tilde{\rho}(t)
\hat{\vec{J}}^{\rhom }(\vec{r},t')\!\otimes\!\hat{\vec{J}}^{c}(\rs\!,0)\!\Big) ,
\end{split}\label{eq:SD_rhomc}\\%
\begin{split}%
\mathrm{D}^{(\rhom \varepsilon)}(\vec{r},\rs\!,t) &=\!\int^{\infty}_{0}\!\!\!\!\!\!\dif t'\,\Tr\!\Big(\tilde{\rho}(t)
\hat{\vec{J}}^{\rhom }(\vec{r},t')\!\otimes\!\hat{\vec{J}}^{\varepsilon}(\rs\!,0)\!\Big) ,
\end{split}\label{eq:SD_rhome}\\%
\begin{split}%
\mathrm{D}^{(c\rhom )}(\vec{r},\rs\!,t) &=\!\int^{\infty}_{0}\!\!\!\!\!\!\dif t'\,\Tr\!\Big(\tilde{\rho}(t)
\hat{\vec{J}}^{c}(\vec{r},t')\!\otimes\!\hat{\vec{J}}^{\rhom }(\rs\!,0)\!\Big) ,
\end{split}\label{eq:SD_crhom}\\%
\begin{split}%
\mathrm{D}^{(cc)}(\vec{r},\rs\!,t) &=\!\int^{\infty}_{0}\!\!\!\!\!\!\dif t'\,\Tr\!\Big(\tilde{\rho}(t)
\hat{\vec{J}}^{c}(\vec{r},t')\!\otimes\!\hat{\vec{J}}^{c}(\rs\!,0)\!\Big) ,
\end{split}\label{eq:SD_cc}\\%
\begin{split}%
\mathrm{D}^{(c\varepsilon)}(\vec{r},\rs\!,t) &=\!\int^{\infty}_{0}\!\!\!\!\!\!\dif t'\,\Tr\!\Big(\tilde{\rho}(t)
\hat{\vec{J}}^{c}(\vec{r},t')\!\otimes\!\hat{\vec{J}}^{\varepsilon}(\rs\!,0)\!\Big) ,
\end{split}\label{eq:SD_ce}\\%
\begin{split}%
\mathrm{D}^{(g_{i}g_{j})}(\vec{r},\rs\!,t) &=\!\int^{\infty}_{0}\!\!\!\!\!\!\dif t'\,\Tr\!\Big(\tilde{\rho}(t)
\hat{\vec{J}}^{g_{i}}(\vec{r},t')\!\otimes\!\hat{\vec{J}}^{g_{j}}(\rs\!,0)\!\Big) ,
\end{split}\label{eq:SD_gigj}\\%
\begin{split}%
\mathrm{D}^{(\varepsilon\rhom )}(\vec{r},\rs\!,t) &=\!\int^{\infty}_{0}\!\!\!\!\!\!\dif t'\,\Tr\!\Big(\tilde{\rho}(t)
\hat{\vec{J}}^{\varepsilon}(\vec{r},t')\!\otimes\!\hat{\vec{J}}^{\rhom }(\rs\!,0)\!\Big) ,
\end{split}\label{eq:SD_erhom}\\%
\begin{split}%
\mathrm{D}^{(\varepsilon c)}(\vec{r},\rs\!,t) &=\!\int^{\infty}_{0}\!\!\!\!\!\!\dif t'\,\Tr\!\Big(\tilde{\rho}(t)
\hat{\vec{J}}^{\varepsilon}(\vec{r},t')\!\otimes\!\hat{\vec{J}}^{c}(\rs\!,0)\!\Big) ,
\end{split}\label{eq:SD_ec}\\%
\begin{split}%
\mathrm{D}^{(\varepsilon\varepsilon)}(\vec{r},\rs\!,t) &=\!\int^{\infty}_{0}\!\!\!\!\!\!\dif t'\,\Tr\!\Big(\tilde{\rho}(t)
\hat{\vec{J}}^{\varepsilon}(\vec{r},t')\!\otimes\!\hat{\vec{J}}^{\varepsilon}(\rs\!,0)\!\Big) 
\end{split}\label{eq:SD_ee}%
\end{align}}%
with $(\hat{\vec{J}}^{g_{i}}\rt)_{k}=(\hat{\mathrm{J}}^{\vec{g}}\rt)_{ik}$ and with the currents defined in Eqs.\ \eqref{eq:J_rhom}-\eqref{eq:J_e}.

\vspace*{5ex}
\section{\label{fourierlap}Fourier and Laplace transformation}
Here, we specify the conventions for Fourier- and Laplace transformations used in this work:

\subsection{Fourier transformation}
For a space-dependent function $X(\vec{r})$ the Fourier transformation is given by \cite{WittkowskiLB2012}
\begin{equation}
\begin{split}
\widetilde{X}(\vec{k}) &= \int_{\R^{3}}\!\!\!\!\!\:\!\dif^{3}r, X(\vec{r}) e^{-\ii\vec{k}\cdot\vec{r}} \;, \\
X(\vec{r}) &= \frac{1}{(2\pi)^{3}} \!\int_{\R^{3}}\!\!\!\!\!\:\!\dif^{3}k\!
\widetilde{X}(\vec{k}) e^{\ii\vec{k}\cdot\vec{r}} \\
\end{split}
\end{equation}
with $\vec{k}\in\R^{3}$. 

\subsection{Laplace transformation}
The Laplace transformation of a time-dependent function $X(t)$ is given by \cite{WittkowskiLB2012}
\begin{equation}
\begin{split}
\widetilde{X}(z) &= \int^{\infty}_{0}\!\!\!\!\!\!\!\:\!\dif t\, X(t) e^{-zt} \;, \\
X(t) &= \frac{1}{2\pi\:\!\ii} \!\int^{z_0+\ii\infty}_{z_0-\ii\infty}\!\!\!\!\!\!\!\!\!\!\!\!\!\!\dif z\,
\widetilde{X}(z) e^{zt} \\
\end{split}
\end{equation}
with $z\in\C$ and real part $\Re(z)>0$. Here, $z_0$ is a constant $z_0>z_{c}$, where $z_{c}$ is the convergence abscissa of $\widetilde{X}(z)$. The inverse Laplace transformation is also known as Bromwich integral
\cite{WittkowskiLB2012}.

\nocite{apsrev41Control}
\bibliographystyle{apsrev4-1}
\bibliography{refs,control}

\begin{thebibliography}{137}%
\makeatletter
\providecommand \@ifxundefined [1]{%
 \@ifx{#1\undefined}
}%
\providecommand \@ifnum [1]{%
 \ifnum #1\expandafter \@firstoftwo
 \else \expandafter \@secondoftwo
 \fi
}%
\providecommand \@ifx [1]{%
 \ifx #1\expandafter \@firstoftwo
 \else \expandafter \@secondoftwo
 \fi
}%
\providecommand \natexlab [1]{#1}%
\providecommand \enquote  [1]{``#1''}%
\providecommand \bibnamefont  [1]{#1}%
\providecommand \bibfnamefont [1]{#1}%
\providecommand \citenamefont [1]{#1}%
\providecommand \href@noop [0]{\@secondoftwo}%
\providecommand \href [0]{\begingroup \@sanitize@url \@href}%
\providecommand \@href[1]{\@@startlink{#1}\@@href}%
\providecommand \@@href[1]{\endgroup#1\@@endlink}%
\providecommand \@sanitize@url [0]{\catcode `\\12\catcode `\$12\catcode
  `\&12\catcode `\#12\catcode `\^12\catcode `\_12\catcode `\%12\relax}%
\providecommand \@@startlink[1]{}%
\providecommand \@@endlink[0]{}%
\providecommand \url  [0]{\begingroup\@sanitize@url \@url }%
\providecommand \@url [1]{\endgroup\@href {#1}{\urlprefix }}%
\providecommand \urlprefix  [0]{URL }%
\providecommand \Eprint [0]{\href }%
\providecommand \doibase [0]{http://dx.doi.org/}%
\providecommand \selectlanguage [0]{\@gobble}%
\providecommand \bibinfo  [0]{\@secondoftwo}%
\providecommand \bibfield  [0]{\@secondoftwo}%
\providecommand \translation [1]{[#1]}%
\providecommand \BibitemOpen [0]{}%
\providecommand \bibitemStop [0]{}%
\providecommand \bibitemNoStop [0]{.\EOS\space}%
\providecommand \EOS [0]{\spacefactor3000\relax}%
\providecommand \BibitemShut  [1]{\csname bibitem#1\endcsname}%
\let\auto@bib@innerbib\@empty
\bibitem [{\citenamefont {{Marini Bettolo Marconi}}\ and\ \citenamefont
  {Tarazona}(1999)}]{MarconiT1999}%
  \BibitemOpen
  \bibfield  {author} {\bibinfo {author} {\bibfnamefont {U.}~\bibnamefont
  {{Marini Bettolo Marconi}}}\ and\ \bibinfo {author} {\bibfnamefont
  {P.}~\bibnamefont {Tarazona}},\ }\bibfield  {title} {\enquote {\bibinfo
  {title} {Dynamic density functional theory of fluids},}\ }\href@noop {}
  {\bibfield  {journal} {\bibinfo  {journal} {Journal of Chemical Physics}\
  }\textbf {\bibinfo {volume} {110}},\ \bibinfo {pages} {8032--8044} (\bibinfo
  {year} {1999})}\BibitemShut {NoStop}%
\bibitem [{\citenamefont {{Marini Bettolo Marconi}}\ and\ \citenamefont
  {Tarazona}(2000)}]{MarconiT2000}%
  \BibitemOpen
  \bibfield  {author} {\bibinfo {author} {\bibfnamefont {U.}~\bibnamefont
  {{Marini Bettolo Marconi}}}\ and\ \bibinfo {author} {\bibfnamefont
  {P.}~\bibnamefont {Tarazona}},\ }\bibfield  {title} {\enquote {\bibinfo
  {title} {Dynamic density functional theory of fluids},}\ }\href@noop {}
  {\bibfield  {journal} {\bibinfo  {journal} {Journal of Physics: Condensed
  Matter}\ }\textbf {\bibinfo {volume} {12}},\ \bibinfo {pages} {413--418}
  (\bibinfo {year} {2000})}\BibitemShut {NoStop}%
\bibitem [{\citenamefont {Archer}\ and\ \citenamefont
  {Evans}(2004)}]{ArcherE2004}%
  \BibitemOpen
  \bibfield  {author} {\bibinfo {author} {\bibfnamefont {A.~J.}\ \bibnamefont
  {Archer}}\ and\ \bibinfo {author} {\bibfnamefont {R.}~\bibnamefont {Evans}},\
  }\bibfield  {title} {\enquote {\bibinfo {title} {Dynamical density functional
  theory and its application to spinodal decomposition},}\ }\href@noop {}
  {\bibfield  {journal} {\bibinfo  {journal} {Journal of Chemical Physics}\
  }\textbf {\bibinfo {volume} {121}},\ \bibinfo {pages} {4246--4254} (\bibinfo
  {year} {2004})}\BibitemShut {NoStop}%
\bibitem [{\citenamefont {Espa{\~n}ol}\ and\ \citenamefont
  {L{\"o}wen}(2009)}]{EspanolL2009}%
  \BibitemOpen
  \bibfield  {author} {\bibinfo {author} {\bibfnamefont {P.}~\bibnamefont
  {Espa{\~n}ol}}\ and\ \bibinfo {author} {\bibfnamefont {H.}~\bibnamefont
  {L{\"o}wen}},\ }\bibfield  {title} {\enquote {\bibinfo {title} {Derivation of
  dynamical density functional theory using the projection operator
  technique},}\ }\href@noop {} {\bibfield  {journal} {\bibinfo  {journal}
  {Journal of Chemical Physics}\ }\textbf {\bibinfo {volume} {131}},\ \bibinfo
  {pages} {244101} (\bibinfo {year} {2009})}\BibitemShut {NoStop}%
\bibitem [{\citenamefont {Evans}(1979)}]{Evans1979}%
  \BibitemOpen
  \bibfield  {author} {\bibinfo {author} {\bibfnamefont {R.}~\bibnamefont
  {Evans}},\ }\bibfield  {title} {\enquote {\bibinfo {title} {The nature of the
  liquid-vapour interface and other topics in the statistical mechanics of
  non-uniform, classical fluids},}\ }\href@noop {} {\bibfield  {journal}
  {\bibinfo  {journal} {Advances in Physics}\ }\textbf {\bibinfo {volume}
  {28}},\ \bibinfo {pages} {143--200} (\bibinfo {year} {1979})}\BibitemShut
  {NoStop}%
\bibitem [{\citenamefont {Munakata}(1989)}]{Munakata1989}%
  \BibitemOpen
  \bibfield  {author} {\bibinfo {author} {\bibfnamefont {T.}~\bibnamefont
  {Munakata}},\ }\bibfield  {title} {\enquote {\bibinfo {title} {A dynamical
  extension of the density functional theory},}\ }\href@noop {} {\bibfield
  {journal} {\bibinfo  {journal} {Journal of the Physical Society of Japan}\
  }\textbf {\bibinfo {volume} {58}},\ \bibinfo {pages} {2434--2438} (\bibinfo
  {year} {1989})}\BibitemShut {NoStop}%
\bibitem [{\citenamefont {Fraaije}(1993)}]{Fraaije1993}%
  \BibitemOpen
  \bibfield  {author} {\bibinfo {author} {\bibfnamefont {J.~G. E.~M.}\
  \bibnamefont {Fraaije}},\ }\bibfield  {title} {\enquote {\bibinfo {title}
  {Dynamic density functional theory for microphase separation kinetics of
  block copolymer melts},}\ }\href@noop {} {\bibfield  {journal} {\bibinfo
  {journal} {Journal of Chemical Physics}\ }\textbf {\bibinfo {volume} {99}},\
  \bibinfo {pages} {9202--9212} (\bibinfo {year} {1993})}\BibitemShut {NoStop}%
\bibitem [{\citenamefont {Kawasaki}(1994)}]{Kawasaki1994}%
  \BibitemOpen
  \bibfield  {author} {\bibinfo {author} {\bibfnamefont {K.}~\bibnamefont
  {Kawasaki}},\ }\bibfield  {title} {\enquote {\bibinfo {title} {Stochastic
  model of slow dynamics in supercooled liquids and dense colloidal
  suspensions},}\ }\href@noop {} {\bibfield  {journal} {\bibinfo  {journal}
  {Physica A: Statistical Mechanics and its Applications}\ }\textbf {\bibinfo
  {volume} {208}},\ \bibinfo {pages} {35--64} (\bibinfo {year}
  {1994})}\BibitemShut {NoStop}%
\bibitem [{\citenamefont {{te Vrugt}}\ \emph
  {et~al.}(2020{\natexlab{a}})\citenamefont {{te Vrugt}}, \citenamefont
  {L{\"o}wen},\ and\ \citenamefont {Wittkowski}}]{teVrugtLW2020}%
  \BibitemOpen
  \bibfield  {author} {\bibinfo {author} {\bibfnamefont {M.}~\bibnamefont {{te
  Vrugt}}}, \bibinfo {author} {\bibfnamefont {H.}~\bibnamefont {L{\"o}wen}}, \
  and\ \bibinfo {author} {\bibfnamefont {R.}~\bibnamefont {Wittkowski}},\
  }\bibfield  {title} {\enquote {\bibinfo {title} {Classical dynamical density
  functional theory: from fundamentals to applications},}\ }\href@noop {}
  {\bibfield  {journal} {\bibinfo  {journal} {Advances in Physics}\ }\textbf
  {\bibinfo {volume} {69}},\ \bibinfo {pages} {121--247} (\bibinfo {year}
  {2020}{\natexlab{a}})}\BibitemShut {NoStop}%
\bibitem [{\citenamefont {Angioletti-Uberti}\ \emph {et~al.}(2014)\citenamefont
  {Angioletti-Uberti}, \citenamefont {Ballauff},\ and\ \citenamefont
  {Dzubiella}}]{AngiolettiBD2014}%
  \BibitemOpen
  \bibfield  {author} {\bibinfo {author} {\bibfnamefont {S.}~\bibnamefont
  {Angioletti-Uberti}}, \bibinfo {author} {\bibfnamefont {M.}~\bibnamefont
  {Ballauff}}, \ and\ \bibinfo {author} {\bibfnamefont {J.}~\bibnamefont
  {Dzubiella}},\ }\bibfield  {title} {\enquote {\bibinfo {title} {Dynamic
  density functional theory of protein adsorption on polymer-coated
  nanoparticles},}\ }\href@noop {} {\bibfield  {journal} {\bibinfo  {journal}
  {Soft Matter}\ }\textbf {\bibinfo {volume} {10}},\ \bibinfo {pages}
  {7932--7945} (\bibinfo {year} {2014})}\BibitemShut {NoStop}%
\bibitem [{\citenamefont {Angioletti-Uberti}\ \emph {et~al.}(2018)\citenamefont
  {Angioletti-Uberti}, \citenamefont {Ballauff},\ and\ \citenamefont
  {Dzubiella}}]{AngiolettiBD2018}%
  \BibitemOpen
  \bibfield  {author} {\bibinfo {author} {\bibfnamefont {S.}~\bibnamefont
  {Angioletti-Uberti}}, \bibinfo {author} {\bibfnamefont {M.}~\bibnamefont
  {Ballauff}}, \ and\ \bibinfo {author} {\bibfnamefont {J.}~\bibnamefont
  {Dzubiella}},\ }\bibfield  {title} {\enquote {\bibinfo {title} {Competitive
  adsorption of multiple proteins to nanoparticles: the {V}roman effect
  revisited},}\ }\href@noop {} {\bibfield  {journal} {\bibinfo  {journal}
  {Molecular Physics}\ }\textbf {\bibinfo {volume} {116}},\ \bibinfo {pages}
  {3154--3163} (\bibinfo {year} {2018})}\BibitemShut {NoStop}%
\bibitem [{\citenamefont {Al-Saedi}\ \emph {et~al.}(2018)\citenamefont
  {Al-Saedi}, \citenamefont {Archer},\ and\ \citenamefont
  {Ward}}]{AlSaediHAW2018}%
  \BibitemOpen
  \bibfield  {author} {\bibinfo {author} {\bibfnamefont {H.~M.}\ \bibnamefont
  {Al-Saedi}}, \bibinfo {author} {\bibfnamefont {A.~J.}\ \bibnamefont
  {Archer}}, \ and\ \bibinfo {author} {\bibfnamefont {J.}~\bibnamefont
  {Ward}},\ }\bibfield  {title} {\enquote {\bibinfo {title} {Dynamical
  density-functional-theory-based modeling of tissue dynamics: application to
  tumor growth},}\ }\href@noop {} {\bibfield  {journal} {\bibinfo  {journal}
  {Physical Review E}\ }\textbf {\bibinfo {volume} {98}},\ \bibinfo {pages}
  {022407} (\bibinfo {year} {2018})}\BibitemShut {NoStop}%
\bibitem [{\citenamefont {Chauviere}\ \emph {et~al.}(2012)\citenamefont
  {Chauviere}, \citenamefont {Hatzikirou}, \citenamefont {Kevrekidis},
  \citenamefont {Lowengrub},\ and\ \citenamefont {Cristini}}]{ChauviereLC2012}%
  \BibitemOpen
  \bibfield  {author} {\bibinfo {author} {\bibfnamefont {A.}~\bibnamefont
  {Chauviere}}, \bibinfo {author} {\bibfnamefont {H.}~\bibnamefont
  {Hatzikirou}}, \bibinfo {author} {\bibfnamefont {I.~G.}\ \bibnamefont
  {Kevrekidis}}, \bibinfo {author} {\bibfnamefont {J.~S.}\ \bibnamefont
  {Lowengrub}}, \ and\ \bibinfo {author} {\bibfnamefont {V.}~\bibnamefont
  {Cristini}},\ }\bibfield  {title} {\enquote {\bibinfo {title} {Dynamic
  density functional theory of solid tumor growth: preliminary models},}\
  }\href@noop {} {\bibfield  {journal} {\bibinfo  {journal} {AIP Advances}\
  }\textbf {\bibinfo {volume} {2}},\ \bibinfo {pages} {011210} (\bibinfo {year}
  {2012})}\BibitemShut {NoStop}%
\bibitem [{\citenamefont {Werkhoven}\ \emph {et~al.}(2019)\citenamefont
  {Werkhoven}, \citenamefont {Samin},\ and\ \citenamefont {van
  Roij}}]{WerkhovenSvR2019}%
  \BibitemOpen
  \bibfield  {author} {\bibinfo {author} {\bibfnamefont {B.}~\bibnamefont
  {Werkhoven}}, \bibinfo {author} {\bibfnamefont {S.}~\bibnamefont {Samin}}, \
  and\ \bibinfo {author} {\bibfnamefont {R.}~\bibnamefont {van Roij}},\
  }\bibfield  {title} {\enquote {\bibinfo {title} {Dynamic {S}tern layers in
  charge-regulating electrokinetic systems: three regimes from an analytical
  approach},}\ }\href@noop {} {\bibfield  {journal} {\bibinfo  {journal}
  {European Physical Journal Special Topics}\ }\textbf {\bibinfo {volume}
  {227}},\ \bibinfo {pages} {2539--2557} (\bibinfo {year} {2019})}\BibitemShut
  {NoStop}%
\bibitem [{\citenamefont {Werkhoven}\ \emph {et~al.}(2018)\citenamefont
  {Werkhoven}, \citenamefont {Everts}, \citenamefont {Samin},\ and\
  \citenamefont {van Roij}}]{WerkhovenESvR2018}%
  \BibitemOpen
  \bibfield  {author} {\bibinfo {author} {\bibfnamefont {B.~L.}\ \bibnamefont
  {Werkhoven}}, \bibinfo {author} {\bibfnamefont {J.~C.}\ \bibnamefont
  {Everts}}, \bibinfo {author} {\bibfnamefont {S.}~\bibnamefont {Samin}}, \
  and\ \bibinfo {author} {\bibfnamefont {R.}~\bibnamefont {van Roij}},\
  }\bibfield  {title} {\enquote {\bibinfo {title} {Flow-induced surface charge
  heterogeneity in electrokinetics due to {S}tern-layer conductance coupled to
  reaction kinetics},}\ }\href@noop {} {\bibfield  {journal} {\bibinfo
  {journal} {Physical Review Letters}\ }\textbf {\bibinfo {volume} {120}},\
  \bibinfo {pages} {264502} (\bibinfo {year} {2018})}\BibitemShut {NoStop}%
\bibitem [{\citenamefont {Gao}\ and\ \citenamefont {Xiao}(2018)}]{GaoX2018}%
  \BibitemOpen
  \bibfield  {author} {\bibinfo {author} {\bibfnamefont {H.}~\bibnamefont
  {Gao}}\ and\ \bibinfo {author} {\bibfnamefont {C.}~\bibnamefont {Xiao}},\
  }\bibfield  {title} {\enquote {\bibinfo {title} {Viscosity effects for ion
  transport in electrochemical systems},}\ }\href@noop {} {\bibfield  {journal}
  {\bibinfo  {journal} {Europhysics Letters}\ }\textbf {\bibinfo {volume}
  {124}},\ \bibinfo {pages} {58002} (\bibinfo {year} {2018})}\BibitemShut
  {NoStop}%
\bibitem [{\citenamefont {Qing}\ \emph {et~al.}(2020)\citenamefont {Qing},
  \citenamefont {Lei}, \citenamefont {Zhao}, \citenamefont {Qiu}, \citenamefont
  {Ma}, \citenamefont {Xu},\ and\ \citenamefont {Zhao}}]{QingLZTQMXZ2020}%
  \BibitemOpen
  \bibfield  {author} {\bibinfo {author} {\bibfnamefont {L.}~\bibnamefont
  {Qing}}, \bibinfo {author} {\bibfnamefont {J.}~\bibnamefont {Lei}}, \bibinfo
  {author} {\bibfnamefont {T.}~\bibnamefont {Zhao}}, \bibinfo {author}
  {\bibfnamefont {G.}~\bibnamefont {Qiu}}, \bibinfo {author} {\bibfnamefont
  {M.}~\bibnamefont {Ma}}, \bibinfo {author} {\bibfnamefont {Z.}~\bibnamefont
  {Xu}}, \ and\ \bibinfo {author} {\bibfnamefont {S.}~\bibnamefont {Zhao}},\
  }\bibfield  {title} {\enquote {\bibinfo {title} {Effects of kinetic
  dielectric decrement on ion diffusion and capacitance in electrochemical
  systems},}\ }\href@noop {} {\bibfield  {journal} {\bibinfo  {journal}
  {Langmuir}\ }\textbf {\bibinfo {volume} {36}},\ \bibinfo {pages} {4055--4064}
  (\bibinfo {year} {2020})}\BibitemShut {NoStop}%
\bibitem [{\citenamefont {{te Vrugt}}\ \emph
  {et~al.}(2020{\natexlab{b}})\citenamefont {{te Vrugt}}, \citenamefont
  {Bickmann},\ and\ \citenamefont {Wittkowski}}]{teVrugtBW2020}%
  \BibitemOpen
  \bibfield  {author} {\bibinfo {author} {\bibfnamefont {M.}~\bibnamefont {{te
  Vrugt}}}, \bibinfo {author} {\bibfnamefont {J.}~\bibnamefont {Bickmann}}, \
  and\ \bibinfo {author} {\bibfnamefont {R.}~\bibnamefont {Wittkowski}},\
  }\bibfield  {title} {\enquote {\bibinfo {title} {Effects of social distancing
  and isolation on epidemic spreading modeled via dynamical density functional
  theory},}\ }\href@noop {} {\bibfield  {journal} {\bibinfo  {journal} {Nature
  Communications}\ }\textbf {\bibinfo {volume} {11}},\ \bibinfo {pages} {5576}
  (\bibinfo {year} {2020}{\natexlab{b}})}\BibitemShut {NoStop}%
\bibitem [{\citenamefont {{te Vrugt}}\ \emph
  {et~al.}(2021{\natexlab{a}})\citenamefont {{te Vrugt}}, \citenamefont
  {Bickmann},\ and\ \citenamefont {Wittkowski}}]{teVrugtBW2020b}%
  \BibitemOpen
  \bibfield  {author} {\bibinfo {author} {\bibfnamefont {M.}~\bibnamefont {{te
  Vrugt}}}, \bibinfo {author} {\bibfnamefont {J.}~\bibnamefont {Bickmann}}, \
  and\ \bibinfo {author} {\bibfnamefont {R.}~\bibnamefont {Wittkowski}},\
  }\bibfield  {title} {\enquote {\bibinfo {title} {Containing a pandemic:
  {N}onpharmaceutical interventions and the ``second wave''},}\ }\href@noop {}
  {\bibfield  {journal} {\bibinfo  {journal} {Journal of Physics
  Communications}\ }\textbf {\bibinfo {volume} {5}},\ \bibinfo {pages} {055008}
  (\bibinfo {year} {2021}{\natexlab{a}})}\BibitemShut {NoStop}%
\bibitem [{\citenamefont {Diaw}\ and\ \citenamefont
  {Murillo}(2015)}]{DiawM2015}%
  \BibitemOpen
  \bibfield  {author} {\bibinfo {author} {\bibfnamefont {A.}~\bibnamefont
  {Diaw}}\ and\ \bibinfo {author} {\bibfnamefont {M.~S.}\ \bibnamefont
  {Murillo}},\ }\bibfield  {title} {\enquote {\bibinfo {title} {Generalized
  hydrodynamics model for strongly coupled plasmas},}\ }\href@noop {}
  {\bibfield  {journal} {\bibinfo  {journal} {Physical Review E}\ }\textbf
  {\bibinfo {volume} {92}},\ \bibinfo {pages} {013107} (\bibinfo {year}
  {2015})}\BibitemShut {NoStop}%
\bibitem [{\citenamefont {Diaw}\ and\ \citenamefont
  {Murillo}(2016)}]{DiawM2016}%
  \BibitemOpen
  \bibfield  {author} {\bibinfo {author} {\bibfnamefont {A.}~\bibnamefont
  {Diaw}}\ and\ \bibinfo {author} {\bibfnamefont {M.~S.}\ \bibnamefont
  {Murillo}},\ }\bibfield  {title} {\enquote {\bibinfo {title} {A dynamic
  density functional theory approach to diffusion in white dwarfs and neutron
  star envelopes},}\ }\href@noop {} {\bibfield  {journal} {\bibinfo  {journal}
  {Astrophysical Journal}\ }\textbf {\bibinfo {volume} {829}},\ \bibinfo
  {pages} {16} (\bibinfo {year} {2016})}\BibitemShut {NoStop}%
\bibitem [{\citenamefont {Diaw}\ and\ \citenamefont
  {Murillo}(2017)}]{DiawM2017}%
  \BibitemOpen
  \bibfield  {author} {\bibinfo {author} {\bibfnamefont {A.}~\bibnamefont
  {Diaw}}\ and\ \bibinfo {author} {\bibfnamefont {M.~S.}\ \bibnamefont
  {Murillo}},\ }\bibfield  {title} {\enquote {\bibinfo {title} {A viscous
  quantum hydrodynamics model based on dynamic density functional theory},}\
  }\href@noop {} {\bibfield  {journal} {\bibinfo  {journal} {Scientific
  Reports}\ }\textbf {\bibinfo {volume} {7}},\ \bibinfo {pages} {15352}
  (\bibinfo {year} {2017})}\BibitemShut {NoStop}%
\bibitem [{\citenamefont {Fraaije}\ \emph {et~al.}(1997)\citenamefont
  {Fraaije}, \citenamefont {van Vlimmeren}, \citenamefont {Maurits},
  \citenamefont {Postma}, \citenamefont {Evers}, \citenamefont {Hoffmann},
  \citenamefont {Altevogt},\ and\ \citenamefont
  {Goldbeck-Wood}}]{FraaijevVMPEHAGW1997}%
  \BibitemOpen
  \bibfield  {author} {\bibinfo {author} {\bibfnamefont {J.~G. E.~M.}\
  \bibnamefont {Fraaije}}, \bibinfo {author} {\bibfnamefont {B.~A.~C.}\
  \bibnamefont {van Vlimmeren}}, \bibinfo {author} {\bibfnamefont {N.~M.}\
  \bibnamefont {Maurits}}, \bibinfo {author} {\bibfnamefont {M.}~\bibnamefont
  {Postma}}, \bibinfo {author} {\bibfnamefont {O.~A.}\ \bibnamefont {Evers}},
  \bibinfo {author} {\bibfnamefont {C.}~\bibnamefont {Hoffmann}}, \bibinfo
  {author} {\bibfnamefont {P.}~\bibnamefont {Altevogt}}, \ and\ \bibinfo
  {author} {\bibfnamefont {G.}~\bibnamefont {Goldbeck-Wood}},\ }\bibfield
  {title} {\enquote {\bibinfo {title} {The dynamic mean-field density
  functional method and its application to the mesoscopic dynamics of quenched
  block copolymer melts},}\ }\href@noop {} {\bibfield  {journal} {\bibinfo
  {journal} {Journal of Chemical Physics}\ }\textbf {\bibinfo {volume} {106}},\
  \bibinfo {pages} {4260--4269} (\bibinfo {year} {1997})}\BibitemShut {NoStop}%
\bibitem [{\citenamefont {Mantha}\ \emph {et~al.}(2020)\citenamefont {Mantha},
  \citenamefont {Qi},\ and\ \citenamefont {Schmid}}]{ManthaQS2020}%
  \BibitemOpen
  \bibfield  {author} {\bibinfo {author} {\bibfnamefont {S.}~\bibnamefont
  {Mantha}}, \bibinfo {author} {\bibfnamefont {S.}~\bibnamefont {Qi}}, \ and\
  \bibinfo {author} {\bibfnamefont {F.}~\bibnamefont {Schmid}},\ }\bibfield
  {title} {\enquote {\bibinfo {title} {Bottom-up construction of dynamic
  density functional theories for inhomogeneous polymer systems from
  microscopic simulations},}\ }\href@noop {} {\bibfield  {journal} {\bibinfo
  {journal} {Macromolecules}\ }\textbf {\bibinfo {volume} {53}},\ \bibinfo
  {pages} {3409--3423} (\bibinfo {year} {2020})}\BibitemShut {NoStop}%
\bibitem [{\citenamefont {Knoll}\ \emph {et~al.}(2002)\citenamefont {Knoll},
  \citenamefont {Horvat}, \citenamefont {Lyakhova}, \citenamefont {Krausch},
  \citenamefont {Sevink}, \citenamefont {Zvelindovsky},\ and\ \citenamefont
  {Magerle}}]{KnollHLKSZM2002}%
  \BibitemOpen
  \bibfield  {author} {\bibinfo {author} {\bibfnamefont {A.}~\bibnamefont
  {Knoll}}, \bibinfo {author} {\bibfnamefont {A.}~\bibnamefont {Horvat}},
  \bibinfo {author} {\bibfnamefont {K.~S.}\ \bibnamefont {Lyakhova}}, \bibinfo
  {author} {\bibfnamefont {G.}~\bibnamefont {Krausch}}, \bibinfo {author}
  {\bibfnamefont {G.~J.~A.}\ \bibnamefont {Sevink}}, \bibinfo {author}
  {\bibfnamefont {A.~V.}\ \bibnamefont {Zvelindovsky}}, \ and\ \bibinfo
  {author} {\bibfnamefont {R.}~\bibnamefont {Magerle}},\ }\bibfield  {title}
  {\enquote {\bibinfo {title} {Phase behavior in thin films of cylinder-forming
  block copolymers},}\ }\href@noop {} {\bibfield  {journal} {\bibinfo
  {journal} {Physical Review Letters}\ }\textbf {\bibinfo {volume} {89}},\
  \bibinfo {pages} {035501} (\bibinfo {year} {2002})}\BibitemShut {NoStop}%
\bibitem [{\citenamefont {Knoll}\ \emph {et~al.}(2004)\citenamefont {Knoll},
  \citenamefont {Lyakhova}, \citenamefont {Horvat}, \citenamefont {Krausch},
  \citenamefont {Sevink}, \citenamefont {Zvelindovsky},\ and\ \citenamefont
  {Magerle}}]{KnollLHKSZM2004}%
  \BibitemOpen
  \bibfield  {author} {\bibinfo {author} {\bibfnamefont {A.}~\bibnamefont
  {Knoll}}, \bibinfo {author} {\bibfnamefont {K.~S.}\ \bibnamefont {Lyakhova}},
  \bibinfo {author} {\bibfnamefont {A.}~\bibnamefont {Horvat}}, \bibinfo
  {author} {\bibfnamefont {G.}~\bibnamefont {Krausch}}, \bibinfo {author}
  {\bibfnamefont {G.~J.~A.}\ \bibnamefont {Sevink}}, \bibinfo {author}
  {\bibfnamefont {A.~V.}\ \bibnamefont {Zvelindovsky}}, \ and\ \bibinfo
  {author} {\bibfnamefont {R.}~\bibnamefont {Magerle}},\ }\bibfield  {title}
  {\enquote {\bibinfo {title} {Direct imaging and mesoscale modelling of phase
  transitions in a nanostructured fluid},}\ }\href@noop {} {\bibfield
  {journal} {\bibinfo  {journal} {Nature Materials}\ }\textbf {\bibinfo
  {volume} {3}},\ \bibinfo {pages} {886--891} (\bibinfo {year}
  {2004})}\BibitemShut {NoStop}%
\bibitem [{\citenamefont {Ludwigs}\ \emph {et~al.}(2003)\citenamefont
  {Ludwigs}, \citenamefont {B{\"o}ker}, \citenamefont {Voronov}, \citenamefont
  {Rehse}, \citenamefont {Magerle},\ and\ \citenamefont
  {Krausch}}]{LudwigsBVRMK2003}%
  \BibitemOpen
  \bibfield  {author} {\bibinfo {author} {\bibfnamefont {S.}~\bibnamefont
  {Ludwigs}}, \bibinfo {author} {\bibfnamefont {A.}~\bibnamefont {B{\"o}ker}},
  \bibinfo {author} {\bibfnamefont {A.}~\bibnamefont {Voronov}}, \bibinfo
  {author} {\bibfnamefont {N.}~\bibnamefont {Rehse}}, \bibinfo {author}
  {\bibfnamefont {R.}~\bibnamefont {Magerle}}, \ and\ \bibinfo {author}
  {\bibfnamefont {G.}~\bibnamefont {Krausch}},\ }\bibfield  {title} {\enquote
  {\bibinfo {title} {Self-assembly of functional nanostructures from {ABC}
  triblock copolymers},}\ }\href@noop {} {\bibfield  {journal} {\bibinfo
  {journal} {Nature Materials}\ }\textbf {\bibinfo {volume} {2}},\ \bibinfo
  {pages} {744--747} (\bibinfo {year} {2003})}\BibitemShut {NoStop}%
\bibitem [{\citenamefont {Ofori-Opoku}\ \emph {et~al.}(2013)\citenamefont
  {Ofori-Opoku}, \citenamefont {Fallah}, \citenamefont {Greenwood},
  \citenamefont {Esmaeili},\ and\ \citenamefont {Provatas}}]{OforiFGEP2013}%
  \BibitemOpen
  \bibfield  {author} {\bibinfo {author} {\bibfnamefont {N.}~\bibnamefont
  {Ofori-Opoku}}, \bibinfo {author} {\bibfnamefont {V.}~\bibnamefont {Fallah}},
  \bibinfo {author} {\bibfnamefont {M.}~\bibnamefont {Greenwood}}, \bibinfo
  {author} {\bibfnamefont {S.}~\bibnamefont {Esmaeili}}, \ and\ \bibinfo
  {author} {\bibfnamefont {N.}~\bibnamefont {Provatas}},\ }\bibfield  {title}
  {\enquote {\bibinfo {title} {Multicomponent phase-field crystal model for
  structural transformations in metal alloys},}\ }\href@noop {} {\bibfield
  {journal} {\bibinfo  {journal} {Physical Review B}\ }\textbf {\bibinfo
  {volume} {87}},\ \bibinfo {pages} {134105} (\bibinfo {year}
  {2013})}\BibitemShut {NoStop}%
\bibitem [{\citenamefont {Roussel}\ \emph {et~al.}(2010)\citenamefont
  {Roussel}, \citenamefont {Lema{\^\i}tre}, \citenamefont {Flatt},\ and\
  \citenamefont {Coussot}}]{RousselFC2010}%
  \BibitemOpen
  \bibfield  {author} {\bibinfo {author} {\bibfnamefont {N.}~\bibnamefont
  {Roussel}}, \bibinfo {author} {\bibfnamefont {A.}~\bibnamefont
  {Lema{\^\i}tre}}, \bibinfo {author} {\bibfnamefont {R.~J.}\ \bibnamefont
  {Flatt}}, \ and\ \bibinfo {author} {\bibfnamefont {P.}~\bibnamefont
  {Coussot}},\ }\bibfield  {title} {\enquote {\bibinfo {title} {Steady state
  flow of cement suspensions: {A} micromechanical state of the art},}\
  }\href@noop {} {\bibfield  {journal} {\bibinfo  {journal} {Cement and
  Concrete Research}\ }\textbf {\bibinfo {volume} {40}},\ \bibinfo {pages}
  {77--84} (\bibinfo {year} {2010})}\BibitemShut {NoStop}%
\bibitem [{\citenamefont {Monlouis-Bonnaire}\ \emph {et~al.}(2004)\citenamefont
  {Monlouis-Bonnaire}, \citenamefont {Verdier},\ and\ \citenamefont
  {Perrin}}]{MonlouisVP2004}%
  \BibitemOpen
  \bibfield  {author} {\bibinfo {author} {\bibfnamefont {J.~P.}\ \bibnamefont
  {Monlouis-Bonnaire}}, \bibinfo {author} {\bibfnamefont {J.}~\bibnamefont
  {Verdier}}, \ and\ \bibinfo {author} {\bibfnamefont {B.}~\bibnamefont
  {Perrin}},\ }\bibfield  {title} {\enquote {\bibinfo {title} {Prediction of
  the relative permeability to gas flow of cement-based materials},}\
  }\href@noop {} {\bibfield  {journal} {\bibinfo  {journal} {Cement and
  Concrete Research}\ }\textbf {\bibinfo {volume} {34}},\ \bibinfo {pages}
  {737--744} (\bibinfo {year} {2004})}\BibitemShut {NoStop}%
\bibitem [{\citenamefont {Bhagat}\ \emph {et~al.}(2020)\citenamefont {Bhagat},
  \citenamefont {Wykes}, \citenamefont {Dalziel},\ and\ \citenamefont
  {Linden}}]{BhagatWDL2020}%
  \BibitemOpen
  \bibfield  {author} {\bibinfo {author} {\bibfnamefont {R.~K.}\ \bibnamefont
  {Bhagat}}, \bibinfo {author} {\bibfnamefont {M.~S.~D.}\ \bibnamefont
  {Wykes}}, \bibinfo {author} {\bibfnamefont {S.~B.}\ \bibnamefont {Dalziel}},
  \ and\ \bibinfo {author} {\bibfnamefont {P.~F.}\ \bibnamefont {Linden}},\
  }\bibfield  {title} {\enquote {\bibinfo {title} {Effects of ventilation on
  the indoor spread of {COVID}-19},}\ }\href@noop {} {\bibfield  {journal}
  {\bibinfo  {journal} {Journal of Fluid Mechanics}\ }\textbf {\bibinfo
  {volume} {903}} (\bibinfo {year} {2020})}\BibitemShut {NoStop}%
\bibitem [{\citenamefont {Abkarian}\ \emph {et~al.}(2020)\citenamefont
  {Abkarian}, \citenamefont {Mendez}, \citenamefont {Xue}, \citenamefont
  {Yang},\ and\ \citenamefont {Stone}}]{AbkarianMXYS2020}%
  \BibitemOpen
  \bibfield  {author} {\bibinfo {author} {\bibfnamefont {M.}~\bibnamefont
  {Abkarian}}, \bibinfo {author} {\bibfnamefont {S.}~\bibnamefont {Mendez}},
  \bibinfo {author} {\bibfnamefont {N.}~\bibnamefont {Xue}}, \bibinfo {author}
  {\bibfnamefont {F.}~\bibnamefont {Yang}}, \ and\ \bibinfo {author}
  {\bibfnamefont {H.~A.}\ \bibnamefont {Stone}},\ }\bibfield  {title} {\enquote
  {\bibinfo {title} {Speech can produce jet-like transport relevant to
  asymptomatic spreading of virus},}\ }\href@noop {} {\bibfield  {journal}
  {\bibinfo  {journal} {Proceedings of the National Academy of Sciences
  U.S.A.}\ }\textbf {\bibinfo {volume} {117}},\ \bibinfo {pages} {25237--25245}
  (\bibinfo {year} {2020})}\BibitemShut {NoStop}%
\bibitem [{\citenamefont {{Marini Bettolo Marconi}}\ and\ \citenamefont
  {Melchionna}(2014)}]{MarconiM2014}%
  \BibitemOpen
  \bibfield  {author} {\bibinfo {author} {\bibfnamefont {U.}~\bibnamefont
  {{Marini Bettolo Marconi}}}\ and\ \bibinfo {author} {\bibfnamefont
  {S.}~\bibnamefont {Melchionna}},\ }\bibfield  {title} {\enquote {\bibinfo
  {title} {Kinetic density functional theory: a microscopic approach to fluid
  mechanics},}\ }\href@noop {} {\bibfield  {journal} {\bibinfo  {journal}
  {Communications in Theoretical Physics}\ }\textbf {\bibinfo {volume} {62}},\
  \bibinfo {pages} {596--606} (\bibinfo {year} {2014})}\BibitemShut {NoStop}%
\bibitem [{\citenamefont {Archer}(2009)}]{Archer2009}%
  \BibitemOpen
  \bibfield  {author} {\bibinfo {author} {\bibfnamefont {A.~J.}\ \bibnamefont
  {Archer}},\ }\bibfield  {title} {\enquote {\bibinfo {title} {Dynamical
  density functional theory for molecular and colloidal fluids: a microscopic
  approach to fluid mechanics},}\ }\href@noop {} {\bibfield  {journal}
  {\bibinfo  {journal} {Journal of Chemical Physics}\ }\textbf {\bibinfo
  {volume} {130}},\ \bibinfo {pages} {014509} (\bibinfo {year}
  {2009})}\BibitemShut {NoStop}%
\bibitem [{\citenamefont {Stierle}\ and\ \citenamefont
  {Gross}(2021)}]{StierleG2021}%
  \BibitemOpen
  \bibfield  {author} {\bibinfo {author} {\bibfnamefont {R.}~\bibnamefont
  {Stierle}}\ and\ \bibinfo {author} {\bibfnamefont {J.}~\bibnamefont
  {Gross}},\ }\bibfield  {title} {\enquote {\bibinfo {title} {Hydrodynamic
  density functional theory for mixtures from a variational principle and its
  application to droplet coalescence},}\ }\href@noop {} {\bibfield  {journal}
  {\bibinfo  {journal} {Journal of Chemical Physics}\ }\textbf {\bibinfo
  {volume} {155}} (\bibinfo {year} {2021})}\BibitemShut {NoStop}%
\bibitem [{\citenamefont {Archer}(2006)}]{Archer2006}%
  \BibitemOpen
  \bibfield  {author} {\bibinfo {author} {\bibfnamefont {A.~J.}\ \bibnamefont
  {Archer}},\ }\bibfield  {title} {\enquote {\bibinfo {title} {Dynamical
  density functional theory for dense atomic liquids},}\ }\href@noop {}
  {\bibfield  {journal} {\bibinfo  {journal} {Journal of Physics: Condensed
  Matter}\ }\textbf {\bibinfo {volume} {18}},\ \bibinfo {pages} {5617--5628}
  (\bibinfo {year} {2006})}\BibitemShut {NoStop}%
\bibitem [{\citenamefont {{Marini Bettolo Marconi}}\ and\ \citenamefont
  {Tarazona}(2006)}]{MarconiT2006}%
  \BibitemOpen
  \bibfield  {author} {\bibinfo {author} {\bibfnamefont {U.}~\bibnamefont
  {{Marini Bettolo Marconi}}}\ and\ \bibinfo {author} {\bibfnamefont
  {P.}~\bibnamefont {Tarazona}},\ }\bibfield  {title} {\enquote {\bibinfo
  {title} {Nonequilibrium inertial dynamics of colloidal systems},}\
  }\href@noop {} {\bibfield  {journal} {\bibinfo  {journal} {Journal of
  Chemical Physics}\ }\textbf {\bibinfo {volume} {124}},\ \bibinfo {pages}
  {164901} (\bibinfo {year} {2006})}\BibitemShut {NoStop}%
\bibitem [{\citenamefont {{Marini Bettolo Marconi}}\ and\ \citenamefont
  {Melchionna}(2007)}]{MarconiM2007}%
  \BibitemOpen
  \bibfield  {author} {\bibinfo {author} {\bibfnamefont {U.}~\bibnamefont
  {{Marini Bettolo Marconi}}}\ and\ \bibinfo {author} {\bibfnamefont
  {S.}~\bibnamefont {Melchionna}},\ }\bibfield  {title} {\enquote {\bibinfo
  {title} {Phase-space approach to dynamical density functional theory},}\
  }\href@noop {} {\bibfield  {journal} {\bibinfo  {journal} {Journal of
  Chemical Physics}\ }\textbf {\bibinfo {volume} {126}},\ \bibinfo {pages}
  {184109} (\bibinfo {year} {2007})}\BibitemShut {NoStop}%
\bibitem [{\citenamefont {Chan}\ and\ \citenamefont
  {Finken}(2005)}]{ChanF2005}%
  \BibitemOpen
  \bibfield  {author} {\bibinfo {author} {\bibfnamefont {G.~K.-L.}\
  \bibnamefont {Chan}}\ and\ \bibinfo {author} {\bibfnamefont {R.}~\bibnamefont
  {Finken}},\ }\bibfield  {title} {\enquote {\bibinfo {title} {Time-dependent
  density functional theory of classical fluids},}\ }\href@noop {} {\bibfield
  {journal} {\bibinfo  {journal} {Physical Review Letters}\ }\textbf {\bibinfo
  {volume} {94}},\ \bibinfo {pages} {183001} (\bibinfo {year}
  {2005})}\BibitemShut {NoStop}%
\bibitem [{\citenamefont {Qiao}\ \emph {et~al.}(2021)\citenamefont {Qiao},
  \citenamefont {Zhao}, \citenamefont {Yu}, \citenamefont {Qing}, \citenamefont
  {Bao}, \citenamefont {Zhao},\ and\ \citenamefont {Liu}}]{QiaoZYQBZL2021}%
  \BibitemOpen
  \bibfield  {author} {\bibinfo {author} {\bibfnamefont {C.}~\bibnamefont
  {Qiao}}, \bibinfo {author} {\bibfnamefont {T.}~\bibnamefont {Zhao}}, \bibinfo
  {author} {\bibfnamefont {X.}~\bibnamefont {Yu}}, \bibinfo {author}
  {\bibfnamefont {L.}~\bibnamefont {Qing}}, \bibinfo {author} {\bibfnamefont
  {B.}~\bibnamefont {Bao}}, \bibinfo {author} {\bibfnamefont {S.}~\bibnamefont
  {Zhao}}, \ and\ \bibinfo {author} {\bibfnamefont {H.}~\bibnamefont {Liu}},\
  }\bibfield  {title} {\enquote {\bibinfo {title} {On the relation between
  dynamical density functional theory and {N}avier-{S}tokes equation},}\
  }\href@noop {} {\bibfield  {journal} {\bibinfo  {journal} {Chemical
  Engineering Science}\ }\textbf {\bibinfo {volume} {230}},\ \bibinfo {pages}
  {116203} (\bibinfo {year} {2021})}\BibitemShut {NoStop}%
\bibitem [{\citenamefont {Schmidt}(2018)}]{Schmidt2018}%
  \BibitemOpen
  \bibfield  {author} {\bibinfo {author} {\bibfnamefont {M.}~\bibnamefont
  {Schmidt}},\ }\bibfield  {title} {\enquote {\bibinfo {title} {Power
  functional theory for {N}ewtonian many-body dynamics},}\ }\href@noop {}
  {\bibfield  {journal} {\bibinfo  {journal} {Journal of Chemical Physics}\
  }\textbf {\bibinfo {volume} {148}},\ \bibinfo {pages} {044502} (\bibinfo
  {year} {2018})}\BibitemShut {NoStop}%
\bibitem [{\citenamefont {Boltzmann}(1872)}]{Boltzmann1872}%
  \BibitemOpen
  \bibfield  {author} {\bibinfo {author} {\bibfnamefont {L.}~\bibnamefont
  {Boltzmann}},\ }\bibfield  {title} {\enquote {\bibinfo {title} {Weitere
  {S}tudien \"uber das {W}\"armegleichgewicht unter {G}asmolek\"ulen},}\
  }\href@noop {} {\bibfield  {journal} {\bibinfo  {journal} {Sitzungberichte
  der Akademie der Wissenschaften zu Wien, mathematisch-naturwissenschaftliche
  Klasse}\ }\textbf {\bibinfo {volume} {66}},\ \bibinfo {pages} {275--370}
  (\bibinfo {year} {1872})}\BibitemShut {NoStop}%
\bibitem [{\citenamefont {Brown}\ \emph {et~al.}(2009)\citenamefont {Brown},
  \citenamefont {Myrvold},\ and\ \citenamefont {Uffink}}]{BrownUM2009}%
  \BibitemOpen
  \bibfield  {author} {\bibinfo {author} {\bibfnamefont {H.~R.}\ \bibnamefont
  {Brown}}, \bibinfo {author} {\bibfnamefont {W.}~\bibnamefont {Myrvold}}, \
  and\ \bibinfo {author} {\bibfnamefont {J.}~\bibnamefont {Uffink}},\
  }\bibfield  {title} {\enquote {\bibinfo {title} {Boltzmann's {H-}theorem, its
  discontents, and the birth of statistical mechanics},}\ }\href@noop {}
  {\bibfield  {journal} {\bibinfo  {journal} {Studies in History and Philosophy
  of Modern Physics}\ }\textbf {\bibinfo {volume} {40}},\ \bibinfo {pages}
  {174--191} (\bibinfo {year} {2009})}\BibitemShut {NoStop}%
\bibitem [{\citenamefont {{Marini Bettolo Marconi}}(2011)}]{Marconi2011}%
  \BibitemOpen
  \bibfield  {author} {\bibinfo {author} {\bibfnamefont {U.}~\bibnamefont
  {{Marini Bettolo Marconi}}},\ }\bibfield  {title} {\enquote {\bibinfo {title}
  {Non-local kinetic theory of inhomogeneous liquid mixtures},}\ }\href@noop {}
  {\bibfield  {journal} {\bibinfo  {journal} {Molecular Physics}\ }\textbf
  {\bibinfo {volume} {109}},\ \bibinfo {pages} {1265--1274} (\bibinfo {year}
  {2011})}\BibitemShut {NoStop}%
\bibitem [{\citenamefont {{Marini Bettolo Marconi}}\ and\ \citenamefont
  {Melchionna}(2009)}]{MarconiM2009}%
  \BibitemOpen
  \bibfield  {author} {\bibinfo {author} {\bibfnamefont {U.}~\bibnamefont
  {{Marini Bettolo Marconi}}}\ and\ \bibinfo {author} {\bibfnamefont
  {S.}~\bibnamefont {Melchionna}},\ }\bibfield  {title} {\enquote {\bibinfo
  {title} {Kinetic theory of correlated fluids: from dynamic density functional
  to {L}attice {B}oltzmann methods},}\ }\href@noop {} {\bibfield  {journal}
  {\bibinfo  {journal} {Journal of Chemical Physics}\ }\textbf {\bibinfo
  {volume} {131}},\ \bibinfo {pages} {014105} (\bibinfo {year}
  {2009})}\BibitemShut {NoStop}%
\bibitem [{\citenamefont {Goddard}\ \emph {et~al.}(2021)\citenamefont
  {Goddard}, \citenamefont {Hurst},\ and\ \citenamefont
  {Ocone}}]{GoddardHO2020}%
  \BibitemOpen
  \bibfield  {author} {\bibinfo {author} {\bibfnamefont {B.~D.}\ \bibnamefont
  {Goddard}}, \bibinfo {author} {\bibfnamefont {T.~D.}\ \bibnamefont {Hurst}},
  \ and\ \bibinfo {author} {\bibfnamefont {R.}~\bibnamefont {Ocone}},\
  }\bibfield  {title} {\enquote {\bibinfo {title} {Modelling inelastic granular
  media using dynamical density functional theory},}\ }\href@noop {} {\bibfield
   {journal} {\bibinfo  {journal} {Journal of Statistical Physics}\ }\textbf
  {\bibinfo {volume} {183}},\ \bibinfo {pages} {6} (\bibinfo {year}
  {2021})}\BibitemShut {NoStop}%
\bibitem [{\citenamefont {Zhao}\ and\ \citenamefont {Wu}(2011)}]{ZhaoW2011}%
  \BibitemOpen
  \bibfield  {author} {\bibinfo {author} {\bibfnamefont {S.-L.}\ \bibnamefont
  {Zhao}}\ and\ \bibinfo {author} {\bibfnamefont {J.}~\bibnamefont {Wu}},\
  }\bibfield  {title} {\enquote {\bibinfo {title} {Self-consistent equations
  governing the dynamics of nonequilibrium colloidal systems},}\ }\href@noop {}
  {\bibfield  {journal} {\bibinfo  {journal} {Journal of Chemical Physics}\
  }\textbf {\bibinfo {volume} {134}},\ \bibinfo {pages} {054514} (\bibinfo
  {year} {2011})}\BibitemShut {NoStop}%
\bibitem [{\citenamefont {Goddard}\ \emph {et~al.}(2013)\citenamefont
  {Goddard}, \citenamefont {Nold}, \citenamefont {Savva}, \citenamefont
  {Yatsyshin},\ and\ \citenamefont {Kalliadasis}}]{GoddardNSYK2013}%
  \BibitemOpen
  \bibfield  {author} {\bibinfo {author} {\bibfnamefont {B.~D.}\ \bibnamefont
  {Goddard}}, \bibinfo {author} {\bibfnamefont {A.}~\bibnamefont {Nold}},
  \bibinfo {author} {\bibfnamefont {N.}~\bibnamefont {Savva}}, \bibinfo
  {author} {\bibfnamefont {P.}~\bibnamefont {Yatsyshin}}, \ and\ \bibinfo
  {author} {\bibfnamefont {S.}~\bibnamefont {Kalliadasis}},\ }\bibfield
  {title} {\enquote {\bibinfo {title} {Unification of dynamic density
  functional theory for colloidal fluids to include inertia and hydrodynamic
  interactions: derivation and numerical experiments},}\ }\href@noop {}
  {\bibfield  {journal} {\bibinfo  {journal} {Journal of Physics: Condensed
  Matter}\ }\textbf {\bibinfo {volume} {25}},\ \bibinfo {pages} {035101}
  (\bibinfo {year} {2013})}\BibitemShut {NoStop}%
\bibitem [{\citenamefont {{Marini Bettolo Marconi}}\ and\ \citenamefont
  {Melchionna}(2011{\natexlab{a}})}]{MarconiM2011}%
  \BibitemOpen
  \bibfield  {author} {\bibinfo {author} {\bibfnamefont {U.}~\bibnamefont
  {{Marini Bettolo Marconi}}}\ and\ \bibinfo {author} {\bibfnamefont
  {S.}~\bibnamefont {Melchionna}},\ }\bibfield  {title} {\enquote {\bibinfo
  {title} {Dynamics of fluid mixtures in nanospaces},}\ }\href@noop {}
  {\bibfield  {journal} {\bibinfo  {journal} {Journal of Chemical Physics}\
  }\textbf {\bibinfo {volume} {134}},\ \bibinfo {pages} {064118} (\bibinfo
  {year} {2011}{\natexlab{a}})}\BibitemShut {NoStop}%
\bibitem [{\citenamefont {{Marini Bettolo Marconi}}\ and\ \citenamefont
  {Melchionna}(2011{\natexlab{b}})}]{MarconiM2011b}%
  \BibitemOpen
  \bibfield  {author} {\bibinfo {author} {\bibfnamefont {U.}~\bibnamefont
  {{Marini Bettolo Marconi}}}\ and\ \bibinfo {author} {\bibfnamefont
  {S.}~\bibnamefont {Melchionna}},\ }\bibfield  {title} {\enquote {\bibinfo
  {title} {Multicomponent diffusion in nanosystems},}\ }\href@noop {}
  {\bibfield  {journal} {\bibinfo  {journal} {Journal of Chemical Physics}\
  }\textbf {\bibinfo {volume} {135}},\ \bibinfo {pages} {044104} (\bibinfo
  {year} {2011}{\natexlab{b}})}\BibitemShut {NoStop}%
\bibitem [{\citenamefont {Monteferrante}\ \emph {et~al.}(2014)\citenamefont
  {Monteferrante}, \citenamefont {Melchionna},\ and\ \citenamefont {{Marini
  Bettolo Marconi}}}]{MonteferranteMM2014}%
  \BibitemOpen
  \bibfield  {author} {\bibinfo {author} {\bibfnamefont {M.}~\bibnamefont
  {Monteferrante}}, \bibinfo {author} {\bibfnamefont {S.}~\bibnamefont
  {Melchionna}}, \ and\ \bibinfo {author} {\bibfnamefont {U.}~\bibnamefont
  {{Marini Bettolo Marconi}}},\ }\bibfield  {title} {\enquote {\bibinfo {title}
  {Lattice {B}oltzmann method for mixtures at variable {S}chmidt number},}\
  }\href@noop {} {\bibfield  {journal} {\bibinfo  {journal} {Journal of
  Chemical Physics}\ }\textbf {\bibinfo {volume} {141}},\ \bibinfo {pages}
  {014102} (\bibinfo {year} {2014})}\BibitemShut {NoStop}%
\bibitem [{\citenamefont {{Marini Bettolo Marconi}}\ and\ \citenamefont
  {Melchionna}(2012)}]{MarconiM2012}%
  \BibitemOpen
  \bibfield  {author} {\bibinfo {author} {\bibfnamefont {U.}~\bibnamefont
  {{Marini Bettolo Marconi}}}\ and\ \bibinfo {author} {\bibfnamefont
  {S.}~\bibnamefont {Melchionna}},\ }\bibfield  {title} {\enquote {\bibinfo
  {title} {Charge transport in nanochannels: a molecular theory},}\ }\href@noop
  {} {\bibfield  {journal} {\bibinfo  {journal} {Langmuir}\ }\textbf {\bibinfo
  {volume} {28}},\ \bibinfo {pages} {13727--13740} (\bibinfo {year}
  {2012})}\BibitemShut {NoStop}%
\bibitem [{\citenamefont {Mills-Williams}\ \emph {et~al.}(2024)\citenamefont
  {Mills-Williams}, \citenamefont {Goddard},\ and\ \citenamefont
  {Archer}}]{MillsGA2024}%
  \BibitemOpen
  \bibfield  {author} {\bibinfo {author} {\bibfnamefont {R.~D.}\ \bibnamefont
  {Mills-Williams}}, \bibinfo {author} {\bibfnamefont {B.~D.}\ \bibnamefont
  {Goddard}}, \ and\ \bibinfo {author} {\bibfnamefont {A.~J.}\ \bibnamefont
  {Archer}},\ }\bibfield  {title} {\enquote {\bibinfo {title} {Dynamic density
  functional theory with inertia and background flow},}\ }\href@noop {}
  {\bibfield  {journal} {\bibinfo  {journal} {Journal of Chemical Physics}\
  }\textbf {\bibinfo {volume} {160}} (\bibinfo {year} {2024})}\BibitemShut
  {NoStop}%
\bibitem [{\citenamefont {Nakajima}(1958)}]{Nakajima1958}%
  \BibitemOpen
  \bibfield  {author} {\bibinfo {author} {\bibfnamefont {S.}~\bibnamefont
  {Nakajima}},\ }\bibfield  {title} {\enquote {\bibinfo {title} {On quantum
  theory of transport phenomena: steady diffusion},}\ }\href@noop {} {\bibfield
   {journal} {\bibinfo  {journal} {Progress of Theoretical Physics}\ }\textbf
  {\bibinfo {volume} {20}},\ \bibinfo {pages} {948--959} (\bibinfo {year}
  {1958})}\BibitemShut {NoStop}%
\bibitem [{\citenamefont {Zwanzig}(1960)}]{Zwanzig1960}%
  \BibitemOpen
  \bibfield  {author} {\bibinfo {author} {\bibfnamefont {R.}~\bibnamefont
  {Zwanzig}},\ }\bibfield  {title} {\enquote {\bibinfo {title} {Ensemble method
  in the theory of irreversibility},}\ }\href@noop {} {\bibfield  {journal}
  {\bibinfo  {journal} {Journal of Chemical Physics}\ }\textbf {\bibinfo
  {volume} {33}},\ \bibinfo {pages} {1338--1341} (\bibinfo {year}
  {1960})}\BibitemShut {NoStop}%
\bibitem [{\citenamefont {Mori}(1965)}]{Mori1965}%
  \BibitemOpen
  \bibfield  {author} {\bibinfo {author} {\bibfnamefont {H.}~\bibnamefont
  {Mori}},\ }\bibfield  {title} {\enquote {\bibinfo {title} {Transport,
  collective motion, and {B}rownian motion},}\ }\href@noop {} {\bibfield
  {journal} {\bibinfo  {journal} {Progress of Theoretical Physics}\ }\textbf
  {\bibinfo {volume} {33}},\ \bibinfo {pages} {423--455} (\bibinfo {year}
  {1965})}\BibitemShut {NoStop}%
\bibitem [{\citenamefont {Grabert}(1982)}]{Grabert1982}%
  \BibitemOpen
  \bibfield  {author} {\bibinfo {author} {\bibfnamefont {H.}~\bibnamefont
  {Grabert}},\ }\href@noop {} {\emph {\bibinfo {title} {Projection Operator
  Techniques in Nonequilibrium Statistical Mechanics}}},\ \bibinfo {edition}
  {1st}\ ed.,\ \bibinfo {series} {Springer Tracts in Modern Physics},
  Vol.~\bibinfo {volume} {95}\ (\bibinfo  {publisher} {Springer-Verlag},\
  \bibinfo {address} {Berlin},\ \bibinfo {year} {1982})\BibitemShut {NoStop}%
\bibitem [{\citenamefont {{te Vrugt}}\ and\ \citenamefont
  {Wittkowski}(2019)}]{teVrugtW2019}%
  \BibitemOpen
  \bibfield  {author} {\bibinfo {author} {\bibfnamefont {M.}~\bibnamefont {{te
  Vrugt}}}\ and\ \bibinfo {author} {\bibfnamefont {R.}~\bibnamefont
  {Wittkowski}},\ }\bibfield  {title} {\enquote {\bibinfo {title}
  {Mori-{Z}wanzig projection operator formalism for far-from-equilibrium
  systems with time-dependent {H}amiltonians},}\ }\href@noop {} {\bibfield
  {journal} {\bibinfo  {journal} {Physical Review E}\ }\textbf {\bibinfo
  {volume} {99}},\ \bibinfo {pages} {062118} (\bibinfo {year}
  {2019})}\BibitemShut {NoStop}%
\bibitem [{\citenamefont {{te Vrugt}}\ and\ \citenamefont
  {Wittkowski}(2020{\natexlab{a}})}]{teVrugtW2019d}%
  \BibitemOpen
  \bibfield  {author} {\bibinfo {author} {\bibfnamefont {M.}~\bibnamefont {{te
  Vrugt}}}\ and\ \bibinfo {author} {\bibfnamefont {R.}~\bibnamefont
  {Wittkowski}},\ }\bibfield  {title} {\enquote {\bibinfo {title} {Projection
  operators in statistical mechanics: a pedagogical approach},}\ }\href@noop {}
  {\bibfield  {journal} {\bibinfo  {journal} {European Journal of Physics}\
  }\textbf {\bibinfo {volume} {41}},\ \bibinfo {pages} {045101} (\bibinfo
  {year} {2020}{\natexlab{a}})}\BibitemShut {NoStop}%
\bibitem [{\citenamefont {Netz}(2024)}]{Netz2024}%
  \BibitemOpen
  \bibfield  {author} {\bibinfo {author} {\bibfnamefont {R.~R.}\ \bibnamefont
  {Netz}},\ }\bibfield  {title} {\enquote {\bibinfo {title} {Derivation of the
  nonequilibrium generalized {L}angevin equation from a time-dependent
  many-body hamiltonian},}\ }\href@noop {} {\bibfield  {journal} {\bibinfo
  {journal} {Physical Review E}\ }\textbf {\bibinfo {volume} {110}},\ \bibinfo
  {pages} {014123} (\bibinfo {year} {2024})}\BibitemShut {NoStop}%
\bibitem [{\citenamefont {Meyer}\ \emph {et~al.}(2017)\citenamefont {Meyer},
  \citenamefont {Voigtmann},\ and\ \citenamefont {Schilling}}]{MeyerVS2017}%
  \BibitemOpen
  \bibfield  {author} {\bibinfo {author} {\bibfnamefont {H.}~\bibnamefont
  {Meyer}}, \bibinfo {author} {\bibfnamefont {T.}~\bibnamefont {Voigtmann}}, \
  and\ \bibinfo {author} {\bibfnamefont {T.}~\bibnamefont {Schilling}},\
  }\bibfield  {title} {\enquote {\bibinfo {title} {On the non-stationary
  generalized {L}angevin equation},}\ }\href@noop {} {\bibfield  {journal}
  {\bibinfo  {journal} {Journal of Chemical Physics}\ }\textbf {\bibinfo
  {volume} {147}} (\bibinfo {year} {2017})}\BibitemShut {NoStop}%
\bibitem [{\citenamefont {Jung}\ and\ \citenamefont {Jung}(2023)}]{JungJ2023}%
  \BibitemOpen
  \bibfield  {author} {\bibinfo {author} {\bibfnamefont {B.}~\bibnamefont
  {Jung}}\ and\ \bibinfo {author} {\bibfnamefont {G.}~\bibnamefont {Jung}},\
  }\bibfield  {title} {\enquote {\bibinfo {title} {Dynamic coarse-graining of
  linear and non-linear systems: {M}ori--{Z}wanzig formalism and beyond},}\
  }\href@noop {} {\bibfield  {journal} {\bibinfo  {journal} {Journal of
  Chemical Physics}\ }\textbf {\bibinfo {volume} {159}} (\bibinfo {year}
  {2023})}\BibitemShut {NoStop}%
\bibitem [{\citenamefont {Yoshimori}(1999)}]{Yoshimori1999}%
  \BibitemOpen
  \bibfield  {author} {\bibinfo {author} {\bibfnamefont {A.}~\bibnamefont
  {Yoshimori}},\ }\bibfield  {title} {\enquote {\bibinfo {title} {Nonlinear
  {L}angevin equations and the time dependent density functional method},}\
  }\href@noop {} {\bibfield  {journal} {\bibinfo  {journal} {Physical Review
  E}\ }\textbf {\bibinfo {volume} {59}},\ \bibinfo {pages} {6535--6540}
  (\bibinfo {year} {1999})}\BibitemShut {NoStop}%
\bibitem [{\citenamefont {Munakata}(2003)}]{Munakata2003}%
  \BibitemOpen
  \bibfield  {author} {\bibinfo {author} {\bibfnamefont {T.}~\bibnamefont
  {Munakata}},\ }\bibfield  {title} {\enquote {\bibinfo {title} {Markovian
  approximation and dynamic density functional theory for classical dense
  liquids},}\ }\href@noop {} {\bibfield  {journal} {\bibinfo  {journal}
  {Physical Review E}\ }\textbf {\bibinfo {volume} {67}},\ \bibinfo {pages}
  {022101} (\bibinfo {year} {2003})}\BibitemShut {NoStop}%
\bibitem [{\citenamefont {Yoshimori}(2005)}]{Yoshimori2005}%
  \BibitemOpen
  \bibfield  {author} {\bibinfo {author} {\bibfnamefont {A.}~\bibnamefont
  {Yoshimori}},\ }\bibfield  {title} {\enquote {\bibinfo {title} {Microscopic
  derivation of time-dependent density functional methods},}\ }\href@noop {}
  {\bibfield  {journal} {\bibinfo  {journal} {Physical Review E}\ }\textbf
  {\bibinfo {volume} {71}},\ \bibinfo {pages} {031203} (\bibinfo {year}
  {2005})}\BibitemShut {NoStop}%
\bibitem [{\citenamefont {Espa\~{n}ol}\ and\ \citenamefont
  {V{\'a}zquez}(2002)}]{EspanolV2002}%
  \BibitemOpen
  \bibfield  {author} {\bibinfo {author} {\bibfnamefont {P.}~\bibnamefont
  {Espa\~{n}ol}}\ and\ \bibinfo {author} {\bibfnamefont {F.}~\bibnamefont
  {V{\'a}zquez}},\ }\bibfield  {title} {\enquote {\bibinfo {title} {Coarse
  graining from coarse-grained descriptions},}\ }\href@noop {} {\bibfield
  {journal} {\bibinfo  {journal} {Philosophical Transactions of the Royal
  Society of London. Series A: Mathematical, Physical and Engineering
  Sciences}\ }\textbf {\bibinfo {volume} {360}},\ \bibinfo {pages} {383--394}
  (\bibinfo {year} {2002})}\BibitemShut {NoStop}%
\bibitem [{\citenamefont {Wittkowski}\ \emph {et~al.}(2012)\citenamefont
  {Wittkowski}, \citenamefont {L{\"o}wen},\ and\ \citenamefont
  {Brand}}]{WittkowskiLB2012}%
  \BibitemOpen
  \bibfield  {author} {\bibinfo {author} {\bibfnamefont {R.}~\bibnamefont
  {Wittkowski}}, \bibinfo {author} {\bibfnamefont {H.}~\bibnamefont
  {L{\"o}wen}}, \ and\ \bibinfo {author} {\bibfnamefont {H.~R.}\ \bibnamefont
  {Brand}},\ }\bibfield  {title} {\enquote {\bibinfo {title} {Extended
  dynamical density functional theory for colloidal mixtures with temperature
  gradients},}\ }\href@noop {} {\bibfield  {journal} {\bibinfo  {journal}
  {Journal of Chemical Physics}\ }\textbf {\bibinfo {volume} {137}},\ \bibinfo
  {pages} {224904} (\bibinfo {year} {2012})}\BibitemShut {NoStop}%
\bibitem [{\citenamefont {Anero}\ \emph {et~al.}(2013)\citenamefont {Anero},
  \citenamefont {Espa\~{n}ol},\ and\ \citenamefont {Tarazona}}]{AneroET2013}%
  \BibitemOpen
  \bibfield  {author} {\bibinfo {author} {\bibfnamefont {J.~G.}\ \bibnamefont
  {Anero}}, \bibinfo {author} {\bibfnamefont {P.}~\bibnamefont {Espa\~{n}ol}},
  \ and\ \bibinfo {author} {\bibfnamefont {P.}~\bibnamefont {Tarazona}},\
  }\bibfield  {title} {\enquote {\bibinfo {title} {Functional thermo-dynamics:
  a generalization of dynamic density functional theory to non-isothermal
  situations},}\ }\href@noop {} {\bibfield  {journal} {\bibinfo  {journal}
  {Journal of Chemical Physics}\ }\textbf {\bibinfo {volume} {139}},\ \bibinfo
  {pages} {034106} (\bibinfo {year} {2013})}\BibitemShut {NoStop}%
\bibitem [{\citenamefont {Jia}\ and\ \citenamefont {Kusaka}(2021)}]{JiaK2021}%
  \BibitemOpen
  \bibfield  {author} {\bibinfo {author} {\bibfnamefont {W.}~\bibnamefont
  {Jia}}\ and\ \bibinfo {author} {\bibfnamefont {I.}~\bibnamefont {Kusaka}},\
  }\bibfield  {title} {\enquote {\bibinfo {title} {Density functional study of
  non-isothermal hard sphere fluids},}\ }\href@noop {} {\bibfield  {journal}
  {\bibinfo  {journal} {Molecular Physics}\ }\textbf {\bibinfo {volume}
  {119}},\ \bibinfo {pages} {e1875077} (\bibinfo {year} {2021})}\BibitemShut
  {NoStop}%
\bibitem [{\citenamefont {Kaufman}\ and\ \citenamefont
  {te~Vrugt}(2026)}]{KaufmantV2026}%
  \BibitemOpen
  \bibfield  {author} {\bibinfo {author} {\bibfnamefont {R.}~\bibnamefont
  {Kaufman}}\ and\ \bibinfo {author} {\bibfnamefont {M.}~\bibnamefont
  {te~Vrugt}},\ }\bibfield  {title} {\enquote {\bibinfo {title} {Time reversal
  and the heat equation},}\ }\href@noop {} {\bibfield  {journal} {\bibinfo
  {journal} {Continuum Mechanics and Thermodynamics}\ }\textbf {\bibinfo
  {volume} {38}},\ \bibinfo {pages} {13} (\bibinfo {year} {2026})}\BibitemShut
  {NoStop}%
\bibitem [{\citenamefont {Camargo}\ \emph {et~al.}(2019)\citenamefont
  {Camargo}, \citenamefont {{de la Torre}}, \citenamefont {Delgado-Buscalioni},
  \citenamefont {Chejne},\ and\ \citenamefont
  {Espa{\~n}ol}}]{CamargodlTDBCE2019}%
  \BibitemOpen
  \bibfield  {author} {\bibinfo {author} {\bibfnamefont {D.}~\bibnamefont
  {Camargo}}, \bibinfo {author} {\bibfnamefont {J.~A.}\ \bibnamefont {{de la
  Torre}}}, \bibinfo {author} {\bibfnamefont {R.}~\bibnamefont
  {Delgado-Buscalioni}}, \bibinfo {author} {\bibfnamefont {F.}~\bibnamefont
  {Chejne}}, \ and\ \bibinfo {author} {\bibfnamefont {P.}~\bibnamefont
  {Espa{\~n}ol}},\ }\bibfield  {title} {\enquote {\bibinfo {title} {Boundary
  conditions derived from a microscopic theory of hydrodynamics near solids},}\
  }\href@noop {} {\bibfield  {journal} {\bibinfo  {journal} {Journal of
  Chemical Physics}\ }\textbf {\bibinfo {volume} {150}},\ \bibinfo {pages}
  {144104} (\bibinfo {year} {2019})}\BibitemShut {NoStop}%
\bibitem [{\citenamefont {Duque-Zumajo}\ \emph
  {et~al.}(2019{\natexlab{a}})\citenamefont {Duque-Zumajo}, \citenamefont
  {Camargo}, \citenamefont {{de la Torre}}, \citenamefont {Chejne},\ and\
  \citenamefont {Espa{\~n}ol}}]{DuqueZumajoCdlTCE2019}%
  \BibitemOpen
  \bibfield  {author} {\bibinfo {author} {\bibfnamefont {D.}~\bibnamefont
  {Duque-Zumajo}}, \bibinfo {author} {\bibfnamefont {D.}~\bibnamefont
  {Camargo}}, \bibinfo {author} {\bibfnamefont {J.~A.}\ \bibnamefont {{de la
  Torre}}}, \bibinfo {author} {\bibfnamefont {F.}~\bibnamefont {Chejne}}, \
  and\ \bibinfo {author} {\bibfnamefont {P.}~\bibnamefont {Espa{\~n}ol}},\
  }\bibfield  {title} {\enquote {\bibinfo {title} {Discrete hydrodynamics near
  solid planar walls},}\ }\href@noop {} {\bibfield  {journal} {\bibinfo
  {journal} {Physical Review E}\ }\textbf {\bibinfo {volume} {99}},\ \bibinfo
  {pages} {052130} (\bibinfo {year} {2019}{\natexlab{a}})}\BibitemShut
  {NoStop}%
\bibitem [{\citenamefont {Duque-Zumajo}\ \emph
  {et~al.}(2019{\natexlab{b}})\citenamefont {Duque-Zumajo}, \citenamefont {{de
  la Torre}}, \citenamefont {Camargo},\ and\ \citenamefont
  {Espa{\~n}ol}}]{DuqueZumajodlTCE2019}%
  \BibitemOpen
  \bibfield  {author} {\bibinfo {author} {\bibfnamefont {D.}~\bibnamefont
  {Duque-Zumajo}}, \bibinfo {author} {\bibfnamefont {J.~A.}\ \bibnamefont {{de
  la Torre}}}, \bibinfo {author} {\bibfnamefont {D.}~\bibnamefont {Camargo}}, \
  and\ \bibinfo {author} {\bibfnamefont {P.}~\bibnamefont {Espa{\~n}ol}},\
  }\bibfield  {title} {\enquote {\bibinfo {title} {Discrete hydrodynamics near
  solid walls: non-{M}arkovian effects and the slip boundary condition},}\
  }\href@noop {} {\bibfield  {journal} {\bibinfo  {journal} {Physical Review
  E}\ }\textbf {\bibinfo {volume} {100}},\ \bibinfo {pages} {062133} (\bibinfo
  {year} {2019}{\natexlab{b}})}\BibitemShut {NoStop}%
\bibitem [{\citenamefont {Camargo}\ \emph {et~al.}(2018)\citenamefont
  {Camargo}, \citenamefont {{de la Torre}}, \citenamefont {Duque-Zumajo},
  \citenamefont {Espa{\~n}ol}, \citenamefont {Delgado-Buscalioni},\ and\
  \citenamefont {Chejne}}]{CamargodlTDZEDBC2018}%
  \BibitemOpen
  \bibfield  {author} {\bibinfo {author} {\bibfnamefont {D.}~\bibnamefont
  {Camargo}}, \bibinfo {author} {\bibfnamefont {J.~A.}\ \bibnamefont {{de la
  Torre}}}, \bibinfo {author} {\bibfnamefont {D.}~\bibnamefont {Duque-Zumajo}},
  \bibinfo {author} {\bibfnamefont {P.}~\bibnamefont {Espa{\~n}ol}}, \bibinfo
  {author} {\bibfnamefont {R.}~\bibnamefont {Delgado-Buscalioni}}, \ and\
  \bibinfo {author} {\bibfnamefont {F.}~\bibnamefont {Chejne}},\ }\bibfield
  {title} {\enquote {\bibinfo {title} {Nanoscale hydrodynamics near solids},}\
  }\href@noop {} {\bibfield  {journal} {\bibinfo  {journal} {Journal of
  Chemical Physics}\ }\textbf {\bibinfo {volume} {148}},\ \bibinfo {pages}
  {064107} (\bibinfo {year} {2018})}\BibitemShut {NoStop}%
\bibitem [{\citenamefont {Dur{\'a}n-Olivencia}\ \emph
  {et~al.}(2017)\citenamefont {Dur{\'a}n-Olivencia}, \citenamefont {Yatsyshin},
  \citenamefont {Goddard},\ and\ \citenamefont {Kalliadasis}}]{DuranYGK2017}%
  \BibitemOpen
  \bibfield  {author} {\bibinfo {author} {\bibfnamefont {M.~A.}\ \bibnamefont
  {Dur{\'a}n-Olivencia}}, \bibinfo {author} {\bibfnamefont {P.}~\bibnamefont
  {Yatsyshin}}, \bibinfo {author} {\bibfnamefont {B.~D.}\ \bibnamefont
  {Goddard}}, \ and\ \bibinfo {author} {\bibfnamefont {S.}~\bibnamefont
  {Kalliadasis}},\ }\bibfield  {title} {\enquote {\bibinfo {title} {General
  framework for fluctuating dynamic density functional theory},}\ }\href@noop
  {} {\bibfield  {journal} {\bibinfo  {journal} {New Journal of Physics}\
  }\textbf {\bibinfo {volume} {19}},\ \bibinfo {pages} {123022} (\bibinfo
  {year} {2017})}\BibitemShut {NoStop}%
\bibitem [{\citenamefont {Haussmann}(2016)}]{Haussmann2016}%
  \BibitemOpen
  \bibfield  {author} {\bibinfo {author} {\bibfnamefont {R.}~\bibnamefont
  {Haussmann}},\ }\bibfield  {title} {\enquote {\bibinfo {title} {The way from
  microscopic many-particle theory to macroscopic hydrodynamics},}\ }\href@noop
  {} {\bibfield  {journal} {\bibinfo  {journal} {Journal of Physics: Condensed
  Matter}\ }\textbf {\bibinfo {volume} {28}},\ \bibinfo {pages} {113001}
  (\bibinfo {year} {2016})}\BibitemShut {NoStop}%
\bibitem [{\citenamefont {Haussmann}(2022)}]{Haussmann2022}%
  \BibitemOpen
  \bibfield  {author} {\bibinfo {author} {\bibfnamefont {R.}~\bibnamefont
  {Haussmann}},\ }\bibfield  {title} {\enquote {\bibinfo {title} {Microscopic
  density-functional approach to nonlinear elasticity theory},}\ }\href@noop {}
  {\bibfield  {journal} {\bibinfo  {journal} {Journal of Statistical Mechanics:
  Theory and Experiment}\ }\textbf {\bibinfo {volume} {2022}},\ \bibinfo
  {pages} {053210} (\bibinfo {year} {2022})}\BibitemShut {NoStop}%
\bibitem [{\citenamefont {Wittkowski}\ \emph {et~al.}(2013)\citenamefont
  {Wittkowski}, \citenamefont {L{\"o}wen},\ and\ \citenamefont
  {Brand}}]{WittkowskiLB2013}%
  \BibitemOpen
  \bibfield  {author} {\bibinfo {author} {\bibfnamefont {R.}~\bibnamefont
  {Wittkowski}}, \bibinfo {author} {\bibfnamefont {H.}~\bibnamefont
  {L{\"o}wen}}, \ and\ \bibinfo {author} {\bibfnamefont {H.~R.}\ \bibnamefont
  {Brand}},\ }\bibfield  {title} {\enquote {\bibinfo {title} {Microscopic
  approach to entropy production},}\ }\href@noop {} {\bibfield  {journal}
  {\bibinfo  {journal} {Journal of Physics A: Mathematical and Theoretical}\
  }\textbf {\bibinfo {volume} {46}},\ \bibinfo {pages} {355003} (\bibinfo
  {year} {2013})}\BibitemShut {NoStop}%
\bibitem [{\citenamefont {Piccirelli}(1968)}]{Piccirelli1968}%
  \BibitemOpen
  \bibfield  {author} {\bibinfo {author} {\bibfnamefont {R.~A.}\ \bibnamefont
  {Piccirelli}},\ }\bibfield  {title} {\enquote {\bibinfo {title} {Theory of
  the dynamics of simple fluids for large spatial gradients and long memory},}\
  }\href@noop {} {\bibfield  {journal} {\bibinfo  {journal} {Physical Review}\
  }\textbf {\bibinfo {volume} {175}},\ \bibinfo {pages} {77} (\bibinfo {year}
  {1968})}\BibitemShut {NoStop}%
\bibitem [{\citenamefont {G{\"o}tze}(2009)}]{Goetze2009}%
  \BibitemOpen
  \bibfield  {author} {\bibinfo {author} {\bibfnamefont {W.}~\bibnamefont
  {G{\"o}tze}},\ }\href@noop {} {\emph {\bibinfo {title} {Complex Dynamics of
  Glass-Forming Liquids: A Mode-Coupling Theory}}},\ \bibinfo {edition} {1st}\
  ed.,\ \bibinfo {series} {International Series of Monographs on Physics},
  Vol.\ \bibinfo {volume} {143}\ (\bibinfo  {publisher} {Oxford University
  Press},\ \bibinfo {address} {Oxford},\ \bibinfo {year} {2009})\BibitemShut
  {NoStop}%
\bibitem [{\citenamefont {Janssen}(2018)}]{Janssen2018}%
  \BibitemOpen
  \bibfield  {author} {\bibinfo {author} {\bibfnamefont {L.~M.~C.}\
  \bibnamefont {Janssen}},\ }\bibfield  {title} {\enquote {\bibinfo {title}
  {Mode-coupling theory of the glass transition: a primer},}\ }\href@noop {}
  {\bibfield  {journal} {\bibinfo  {journal} {Frontiers in Physics}\ }\textbf
  {\bibinfo {volume} {6}},\ \bibinfo {pages} {97} (\bibinfo {year}
  {2018})}\BibitemShut {NoStop}%
\bibitem [{\citenamefont {{te Vrugt}}(2021)}]{teVrugt2020}%
  \BibitemOpen
  \bibfield  {author} {\bibinfo {author} {\bibfnamefont {M.}~\bibnamefont {{te
  Vrugt}}},\ }\bibfield  {title} {\enquote {\bibinfo {title} {The five problems
  of irreversibility},}\ }\href@noop {} {\bibfield  {journal} {\bibinfo
  {journal} {Studies in History and Philosophy of Science}\ }\textbf {\bibinfo
  {volume} {87}},\ \bibinfo {pages} {136--146} (\bibinfo {year}
  {2021})}\BibitemShut {NoStop}%
\bibitem [{\citenamefont {Landau}\ and\ \citenamefont
  {Lifshitz}(1987)}]{LandauL1987}%
  \BibitemOpen
  \bibfield  {author} {\bibinfo {author} {\bibfnamefont {L.~D.}\ \bibnamefont
  {Landau}}\ and\ \bibinfo {author} {\bibfnamefont {E.~M.}\ \bibnamefont
  {Lifshitz}},\ }\href@noop {} {\emph {\bibinfo {title} {Fluid Mechanics}}},\
  \bibinfo {edition} {2nd}\ ed.,\ \bibinfo {series} {Landau and Lifshitz:
  Course of Theoretical Physics}, Vol.~\bibinfo {volume} {6}\ (\bibinfo
  {publisher} {Butterworth-Heinemann},\ \bibinfo {address} {Oxford},\ \bibinfo
  {year} {1987})\BibitemShut {NoStop}%
\bibitem [{\citenamefont {Zwanzig}\ and\ \citenamefont
  {Mountain}(1965)}]{ZwanzigM1965}%
  \BibitemOpen
  \bibfield  {author} {\bibinfo {author} {\bibfnamefont {R.}~\bibnamefont
  {Zwanzig}}\ and\ \bibinfo {author} {\bibfnamefont {R.~D.}\ \bibnamefont
  {Mountain}},\ }\bibfield  {title} {\enquote {\bibinfo {title} {High-frequency
  elastic moduli of simple fluids},}\ }\href@noop {} {\bibfield  {journal}
  {\bibinfo  {journal} {Journal of Chemical Physics}\ }\textbf {\bibinfo
  {volume} {43}},\ \bibinfo {pages} {4464--4471} (\bibinfo {year}
  {1965})}\BibitemShut {NoStop}%
\bibitem [{\citenamefont {Forster}(1974)}]{Forster1974}%
  \BibitemOpen
  \bibfield  {author} {\bibinfo {author} {\bibfnamefont {D.}~\bibnamefont
  {Forster}},\ }\bibfield  {title} {\enquote {\bibinfo {title} {Hydrodynamics
  and correlation functions in ordered systems: nematic liquid crystals},}\
  }\href@noop {} {\bibfield  {journal} {\bibinfo  {journal} {Annals of
  Physics}\ }\textbf {\bibinfo {volume} {84}},\ \bibinfo {pages} {505--534}
  (\bibinfo {year} {1974})}\BibitemShut {NoStop}%
\bibitem [{\citenamefont {Forster}(1990)}]{Forster1990}%
  \BibitemOpen
  \bibfield  {author} {\bibinfo {author} {\bibfnamefont {D.}~\bibnamefont
  {Forster}},\ }\href@noop {} {\emph {\bibinfo {title} {Hydrodynamic
  Fluctuations, Broken Symmetry, and Correlation Functions}}},\ \bibinfo
  {edition} {1st}\ ed.,\ Frontiers in Physics\ (\bibinfo  {publisher} {Addison
  Wesley},\ \bibinfo {address} {Redwood City},\ \bibinfo {year}
  {1990})\BibitemShut {NoStop}%
\bibitem [{\citenamefont {Zwanzig}(2001)}]{Zwanzig2001}%
  \BibitemOpen
  \bibfield  {author} {\bibinfo {author} {\bibfnamefont {R.}~\bibnamefont
  {Zwanzig}},\ }\href@noop {} {\emph {\bibinfo {title} {Nonequilibrium
  Statistical Mechanics}}},\ \bibinfo {edition} {3rd}\ ed.\ (\bibinfo
  {publisher} {Oxford University Press},\ \bibinfo {address} {New York},\
  \bibinfo {year} {2001})\BibitemShut {NoStop}%
\bibitem [{\citenamefont {Wajnryb}\ \emph {et~al.}(1995)\citenamefont
  {Wajnryb}, \citenamefont {Altenberger},\ and\ \citenamefont
  {Dahler}}]{WajnrybAD1995}%
  \BibitemOpen
  \bibfield  {author} {\bibinfo {author} {\bibfnamefont {E.}~\bibnamefont
  {Wajnryb}}, \bibinfo {author} {\bibfnamefont {A.~R.}\ \bibnamefont
  {Altenberger}}, \ and\ \bibinfo {author} {\bibfnamefont {J.~S.}\ \bibnamefont
  {Dahler}},\ }\bibfield  {title} {\enquote {\bibinfo {title} {Uniqueness of
  the microscopic stress tensor},}\ }\href@noop {} {\bibfield  {journal}
  {\bibinfo  {journal} {Journal of Chemical Physics}\ }\textbf {\bibinfo
  {volume} {103}},\ \bibinfo {pages} {9782--9787} (\bibinfo {year}
  {1995})}\BibitemShut {NoStop}%
\bibitem [{\citenamefont {{de las Heras}}\ and\ \citenamefont
  {Schmidt}(2018)}]{delasHerasS2018}%
  \BibitemOpen
  \bibfield  {author} {\bibinfo {author} {\bibfnamefont {D.}~\bibnamefont {{de
  las Heras}}}\ and\ \bibinfo {author} {\bibfnamefont {M.}~\bibnamefont
  {Schmidt}},\ }\bibfield  {title} {\enquote {\bibinfo {title} {Velocity
  gradient power functional for {B}rownian dynamics},}\ }\href@noop {}
  {\bibfield  {journal} {\bibinfo  {journal} {Physical Review Letters}\
  }\textbf {\bibinfo {volume} {120}},\ \bibinfo {pages} {028001} (\bibinfo
  {year} {2018})}\BibitemShut {NoStop}%
\bibitem [{\citenamefont {Grabert}(1978)}]{Grabert1978}%
  \BibitemOpen
  \bibfield  {author} {\bibinfo {author} {\bibfnamefont {H.}~\bibnamefont
  {Grabert}},\ }\bibfield  {title} {\enquote {\bibinfo {title} {Nonlinear
  transport and dynamics of fluctuations},}\ }\href@noop {} {\bibfield
  {journal} {\bibinfo  {journal} {Journal of Statistical Physics}\ }\textbf
  {\bibinfo {volume} {19}},\ \bibinfo {pages} {479--497} (\bibinfo {year}
  {1978})}\BibitemShut {NoStop}%
\bibitem [{\citenamefont {Meyer}\ \emph {et~al.}(2019)\citenamefont {Meyer},
  \citenamefont {Voigtmann},\ and\ \citenamefont {Schilling}}]{MeyerVS2019}%
  \BibitemOpen
  \bibfield  {author} {\bibinfo {author} {\bibfnamefont {H.}~\bibnamefont
  {Meyer}}, \bibinfo {author} {\bibfnamefont {T.}~\bibnamefont {Voigtmann}}, \
  and\ \bibinfo {author} {\bibfnamefont {T.}~\bibnamefont {Schilling}},\
  }\bibfield  {title} {\enquote {\bibinfo {title} {On the dynamics of reaction
  coordinates in classical, time-dependent, many-body processes},}\ }\href@noop
  {} {\bibfield  {journal} {\bibinfo  {journal} {Journal of Chemical Physics}\
  }\textbf {\bibinfo {volume} {150}},\ \bibinfo {pages} {174118} (\bibinfo
  {year} {2019})}\BibitemShut {NoStop}%
\bibitem [{\citenamefont {Schilling}(2022)}]{Schilling2022}%
  \BibitemOpen
  \bibfield  {author} {\bibinfo {author} {\bibfnamefont {T.}~\bibnamefont
  {Schilling}},\ }\bibfield  {title} {\enquote {\bibinfo {title}
  {Coarse-grained modelling out of equilibrium},}\ }\href@noop {} {\bibfield
  {journal} {\bibinfo  {journal} {Physics Reports}\ }\textbf {\bibinfo {volume}
  {972}},\ \bibinfo {pages} {1--45} (\bibinfo {year} {2022})}\BibitemShut
  {NoStop}%
\bibitem [{\citenamefont {Kawasaki}(2000)}]{Kawasaki2000}%
  \BibitemOpen
  \bibfield  {author} {\bibinfo {author} {\bibfnamefont {K.}~\bibnamefont
  {Kawasaki}},\ }\bibfield  {title} {\enquote {\bibinfo {title} {Theoretical
  methods dealing with slow dynamics},}\ }\href@noop {} {\bibfield  {journal}
  {\bibinfo  {journal} {Journal of Physics: Condensed Matter}\ }\textbf
  {\bibinfo {volume} {12}},\ \bibinfo {pages} {6343--6351} (\bibinfo {year}
  {2000})}\BibitemShut {NoStop}%
\bibitem [{\citenamefont {Kawasaki}(2006)}]{Kawasaki2006b}%
  \BibitemOpen
  \bibfield  {author} {\bibinfo {author} {\bibfnamefont {K.}~\bibnamefont
  {Kawasaki}},\ }\bibfield  {title} {\enquote {\bibinfo {title} {Interpolation
  of stochastic and deterministic reduced dynamics},}\ }\href@noop {}
  {\bibfield  {journal} {\bibinfo  {journal} {Physica A: Statistical Mechanics
  and its Applications}\ }\textbf {\bibinfo {volume} {362}},\ \bibinfo {pages}
  {249--260} (\bibinfo {year} {2006})}\BibitemShut {NoStop}%
\bibitem [{\citenamefont {te~Vrugt}(2022)}]{teVrugt2022}%
  \BibitemOpen
  \bibfield  {author} {\bibinfo {author} {\bibfnamefont {M.}~\bibnamefont
  {te~Vrugt}},\ }\bibfield  {title} {\enquote {\bibinfo {title} {Understanding
  probability and irreversibility in the {M}ori-{Z}wanzig projection operator
  formalism},}\ }\href@noop {} {\bibfield  {journal} {\bibinfo  {journal}
  {European Journal for Philosophy of Science}\ }\textbf {\bibinfo {volume}
  {12}},\ \bibinfo {pages} {41} (\bibinfo {year} {2022})}\BibitemShut {NoStop}%
\bibitem [{\citenamefont {Kadanoff}\ and\ \citenamefont
  {Baym}(1989)}]{KadanoffB1989}%
  \BibitemOpen
  \bibfield  {author} {\bibinfo {author} {\bibfnamefont {L.~P.}\ \bibnamefont
  {Kadanoff}}\ and\ \bibinfo {author} {\bibfnamefont {G.}~\bibnamefont
  {Baym}},\ }\href@noop {} {\emph {\bibinfo {title} {Quantum Statistical
  Mechanics: Green's Function Methods in Equilibrium and Nonequilibrium
  Problems}}},\ \bibinfo {edition} {reprint}\ ed.,\ Advanced Book Classics\
  (\bibinfo  {publisher} {Addison Wesley},\ \bibinfo {address} {Redwood City},\
  \bibinfo {year} {1989})\BibitemShut {NoStop}%
\bibitem [{\citenamefont {Hansen}\ and\ \citenamefont
  {McDonald}(2009)}]{HansenMD2009}%
  \BibitemOpen
  \bibfield  {author} {\bibinfo {author} {\bibfnamefont {J.-P.}\ \bibnamefont
  {Hansen}}\ and\ \bibinfo {author} {\bibfnamefont {I.~R.}\ \bibnamefont
  {McDonald}},\ }\href@noop {} {\emph {\bibinfo {title} {Theory of Simple
  Liquids: with Applications to Soft Matter}}},\ \bibinfo {edition} {4th}\ ed.\
  (\bibinfo  {publisher} {Elsevier Academic Press},\ \bibinfo {address}
  {Oxford},\ \bibinfo {year} {2009})\BibitemShut {NoStop}%
\bibitem [{\citenamefont {{Hohenberg}}\ and\ \citenamefont
  {{Martin}}(1965)}]{HohenbergM1965}%
  \BibitemOpen
  \bibfield  {author} {\bibinfo {author} {\bibfnamefont {P.~C.}\ \bibnamefont
  {{Hohenberg}}}\ and\ \bibinfo {author} {\bibfnamefont {P.~C.}\ \bibnamefont
  {{Martin}}},\ }\bibfield  {title} {\enquote {\bibinfo {title} {Microscopic
  theory of superfluid helium},}\ }\href@noop {} {\bibfield  {journal}
  {\bibinfo  {journal} {Annals of Physics}\ }\textbf {\bibinfo {volume} {34}},\
  \bibinfo {pages} {291--359} (\bibinfo {year} {1965})}\BibitemShut {NoStop}%
\bibitem [{\citenamefont {Khalatnikov}(1989)}]{Khalatnikov1989}%
  \BibitemOpen
  \bibfield  {author} {\bibinfo {author} {\bibfnamefont {I.~M.}\ \bibnamefont
  {Khalatnikov}},\ }\href@noop {} {\emph {\bibinfo {title} {An Introduction to
  the Theory of Superfluidity}}},\ \bibinfo {edition} {2nd}\ ed.,\ Frontiers in
  Physics\ (\bibinfo  {publisher} {Addison Wesley},\ \bibinfo {address}
  {Redwood City},\ \bibinfo {year} {1989})\BibitemShut {NoStop}%
\bibitem [{\citenamefont {Burghardt}\ and\ \citenamefont
  {Bagchi}(2006)}]{BurghardtB2006}%
  \BibitemOpen
  \bibfield  {author} {\bibinfo {author} {\bibfnamefont {I.}~\bibnamefont
  {Burghardt}}\ and\ \bibinfo {author} {\bibfnamefont {B.}~\bibnamefont
  {Bagchi}},\ }\bibfield  {title} {\enquote {\bibinfo {title} {On the
  non-adiabatic dynamics of solvation: a molecular hydrodynamic formulation},}\
  }\href@noop {} {\bibfield  {journal} {\bibinfo  {journal} {Chemical Physics}\
  }\textbf {\bibinfo {volume} {329}},\ \bibinfo {pages} {343--356} (\bibinfo
  {year} {2006})}\BibitemShut {NoStop}%
\bibitem [{\citenamefont {Roth}(2010)}]{Roth2010}%
  \BibitemOpen
  \bibfield  {author} {\bibinfo {author} {\bibfnamefont {R.}~\bibnamefont
  {Roth}},\ }\bibfield  {title} {\enquote {\bibinfo {title} {Fundamental
  measure theory for hard-sphere mixtures: a review},}\ }\href@noop {}
  {\bibfield  {journal} {\bibinfo  {journal} {Journal of Physics: Condensed
  Matter}\ }\textbf {\bibinfo {volume} {22}},\ \bibinfo {pages} {063102}
  (\bibinfo {year} {2010})}\BibitemShut {NoStop}%
\bibitem [{\citenamefont {{Rosenfeld}}(1989)}]{Rosenfeld1989}%
  \BibitemOpen
  \bibfield  {author} {\bibinfo {author} {\bibfnamefont {Y.}~\bibnamefont
  {{Rosenfeld}}},\ }\bibfield  {title} {\enquote {\bibinfo {title} {Free-energy
  model for the inhomogeneous hard-sphere fluid mixture and density-functional
  theory of freezing},}\ }\href@noop {} {\bibfield  {journal} {\bibinfo
  {journal} {Physical Review Letters}\ }\textbf {\bibinfo {volume} {63}},\
  \bibinfo {pages} {980--983} (\bibinfo {year} {1989})}\BibitemShut {NoStop}%
\bibitem [{\citenamefont {Galenko}\ \emph {et~al.}(2015)\citenamefont
  {Galenko}, \citenamefont {Sanches},\ and\ \citenamefont
  {Elder}}]{GalenkoSE2015}%
  \BibitemOpen
  \bibfield  {author} {\bibinfo {author} {\bibfnamefont {P.~K.}\ \bibnamefont
  {Galenko}}, \bibinfo {author} {\bibfnamefont {F.~I.}\ \bibnamefont
  {Sanches}}, \ and\ \bibinfo {author} {\bibfnamefont {K.~R.}\ \bibnamefont
  {Elder}},\ }\bibfield  {title} {\enquote {\bibinfo {title} {Traveling wave
  profiles for a crystalline front invading liquid states: analytical and
  numerical solutions},}\ }\href@noop {} {\bibfield  {journal} {\bibinfo
  {journal} {Physica D: Nonlinear Phenomena}\ }\textbf {\bibinfo {volume}
  {308}},\ \bibinfo {pages} {1--10} (\bibinfo {year} {2015})}\BibitemShut
  {NoStop}%
\bibitem [{\citenamefont {te~Vrugt}(2026)}]{teVrugt2026}%
  \BibitemOpen
  \bibfield  {author} {\bibinfo {author} {\bibfnamefont {M.}~\bibnamefont
  {te~Vrugt}},\ }\bibfield  {title} {\enquote {\bibinfo {title} {Microscopic
  field theory for active {B}rownian particles with translational and
  rotational inertia},}\ }\href@noop {} {\bibfield  {journal} {\bibinfo
  {journal} {arXiv:2602.11916}\ } (\bibinfo {year} {2026})}\BibitemShut
  {NoStop}%
\bibitem [{\citenamefont {Tschopp}\ \emph {et~al.}(2022)\citenamefont
  {Tschopp}, \citenamefont {Samm{\"u}ller}, \citenamefont {Hermann},
  \citenamefont {Schmidt},\ and\ \citenamefont {Brader}}]{TschoppSHSB2022}%
  \BibitemOpen
  \bibfield  {author} {\bibinfo {author} {\bibfnamefont {S.~M.}\ \bibnamefont
  {Tschopp}}, \bibinfo {author} {\bibfnamefont {F.}~\bibnamefont
  {Samm{\"u}ller}}, \bibinfo {author} {\bibfnamefont {S.}~\bibnamefont
  {Hermann}}, \bibinfo {author} {\bibfnamefont {M.}~\bibnamefont {Schmidt}}, \
  and\ \bibinfo {author} {\bibfnamefont {J.~M.}\ \bibnamefont {Brader}},\
  }\bibfield  {title} {\enquote {\bibinfo {title} {Force density functional
  theory in-and out-of-equilibrium},}\ }\href@noop {} {\bibfield  {journal}
  {\bibinfo  {journal} {Physical Review E}\ }\textbf {\bibinfo {volume}
  {106}},\ \bibinfo {pages} {014115} (\bibinfo {year} {2022})}\BibitemShut
  {NoStop}%
\bibitem [{\citenamefont {{Rauscher}}\ \emph {et~al.}(2007)\citenamefont
  {{Rauscher}}, \citenamefont {{Dom{\'{\i}}nguez}}, \citenamefont
  {{Kr{\"u}ger}},\ and\ \citenamefont {{Penna}}}]{RauscherDKP2007}%
  \BibitemOpen
  \bibfield  {author} {\bibinfo {author} {\bibfnamefont {M.}~\bibnamefont
  {{Rauscher}}}, \bibinfo {author} {\bibfnamefont {A.}~\bibnamefont
  {{Dom{\'{\i}}nguez}}}, \bibinfo {author} {\bibfnamefont {M.}~\bibnamefont
  {{Kr{\"u}ger}}}, \ and\ \bibinfo {author} {\bibfnamefont {F.}~\bibnamefont
  {{Penna}}},\ }\bibfield  {title} {\enquote {\bibinfo {title} {A dynamic
  density functional theory for particles in a flowing solvent},}\ }\href@noop
  {} {\bibfield  {journal} {\bibinfo  {journal} {Journal of Chemical Physics}\
  }\textbf {\bibinfo {volume} {127}},\ \bibinfo {pages} {244906} (\bibinfo
  {year} {2007})}\BibitemShut {NoStop}%
\bibitem [{\citenamefont {Landau}\ and\ \citenamefont
  {Lifshitz}(1996)}]{LandauL1996}%
  \BibitemOpen
  \bibfield  {author} {\bibinfo {author} {\bibfnamefont {L.~D.}\ \bibnamefont
  {Landau}}\ and\ \bibinfo {author} {\bibfnamefont {E.~M.}\ \bibnamefont
  {Lifshitz}},\ }\href@noop {} {\emph {\bibinfo {title} {Statistical Physics
  {I}}}},\ \bibinfo {edition} {3rd}\ ed.,\ \bibinfo {series} {Landau and
  Lifshitz: Course of Theoretical Physics}, Vol.~\bibinfo {volume} {5}\
  (\bibinfo  {publisher} {Butterworth-Heinemann},\ \bibinfo {address}
  {Oxford},\ \bibinfo {year} {1996})\BibitemShut {NoStop}%
\bibitem [{\citenamefont {Landau}\ and\ \citenamefont
  {Lifshitz}(2005)}]{LandauL2005}%
  \BibitemOpen
  \bibfield  {author} {\bibinfo {author} {\bibfnamefont {L.~D.}\ \bibnamefont
  {Landau}}\ and\ \bibinfo {author} {\bibfnamefont {E.~M.}\ \bibnamefont
  {Lifshitz}},\ }\href@noop {} {\emph {\bibinfo {title} {Theory of
  Elasticity}}},\ \bibinfo {edition} {3rd}\ ed.,\ \bibinfo {series} {Landau and
  Lifshitz: Course of Theoretical Physics}, Vol.~\bibinfo {volume} {7}\
  (\bibinfo  {publisher} {Butterworth-Heinemann},\ \bibinfo {address}
  {Oxford},\ \bibinfo {year} {2005})\BibitemShut {NoStop}%
\bibitem [{\citenamefont {T{\'o}th}\ \emph {et~al.}(2013)\citenamefont
  {T{\'o}th}, \citenamefont {Gr{\'a}n{\'a}sy},\ and\ \citenamefont
  {Tegze}}]{TothGT2013}%
  \BibitemOpen
  \bibfield  {author} {\bibinfo {author} {\bibfnamefont {G.~I.}\ \bibnamefont
  {T{\'o}th}}, \bibinfo {author} {\bibfnamefont {L.}~\bibnamefont
  {Gr{\'a}n{\'a}sy}}, \ and\ \bibinfo {author} {\bibfnamefont {G.}~\bibnamefont
  {Tegze}},\ }\bibfield  {title} {\enquote {\bibinfo {title} {Nonlinear
  hydrodynamic theory of crystallization},}\ }\href@noop {} {\bibfield
  {journal} {\bibinfo  {journal} {Journal of Physics: Condensed Matter}\
  }\textbf {\bibinfo {volume} {26}},\ \bibinfo {pages} {055001} (\bibinfo
  {year} {2013})}\BibitemShut {NoStop}%
\bibitem [{\citenamefont {Heinonen}\ \emph {et~al.}(2016)\citenamefont
  {Heinonen}, \citenamefont {Achim}, \citenamefont {Kosterlitz}, \citenamefont
  {Ying}, \citenamefont {Lowengrub},\ and\ \citenamefont
  {Ala-Nissila}}]{HeinonenAKYLA2016}%
  \BibitemOpen
  \bibfield  {author} {\bibinfo {author} {\bibfnamefont {V.}~\bibnamefont
  {Heinonen}}, \bibinfo {author} {\bibfnamefont {C.~V.}\ \bibnamefont {Achim}},
  \bibinfo {author} {\bibfnamefont {J.~M.}\ \bibnamefont {Kosterlitz}},
  \bibinfo {author} {\bibfnamefont {S.-C.}\ \bibnamefont {Ying}}, \bibinfo
  {author} {\bibfnamefont {J.}~\bibnamefont {Lowengrub}}, \ and\ \bibinfo
  {author} {\bibfnamefont {T.}~\bibnamefont {Ala-Nissila}},\ }\bibfield
  {title} {\enquote {\bibinfo {title} {Consistent hydrodynamics for phase field
  crystals},}\ }\href@noop {} {\bibfield  {journal} {\bibinfo  {journal}
  {Physical Review Letters}\ }\textbf {\bibinfo {volume} {116}},\ \bibinfo
  {pages} {024303} (\bibinfo {year} {2016})}\BibitemShut {NoStop}%
\bibitem [{\citenamefont {Kawasaki}(2009)}]{Kawasaki2009}%
  \BibitemOpen
  \bibfield  {author} {\bibinfo {author} {\bibfnamefont {K.}~\bibnamefont
  {Kawasaki}},\ }\bibfield  {title} {\enquote {\bibinfo {title} {A mini-review
  of structural glasses---a personal view---},}\ }\href@noop {} {\bibfield
  {journal} {\bibinfo  {journal} {Forma}\ }\textbf {\bibinfo {volume} {24}},\
  \bibinfo {pages} {3--9} (\bibinfo {year} {2009})}\BibitemShut {NoStop}%
\bibitem [{\citenamefont {Fuchizaki}\ and\ \citenamefont
  {Kawasaki}(2002)}]{FuchizakiK2002}%
  \BibitemOpen
  \bibfield  {author} {\bibinfo {author} {\bibfnamefont {K.}~\bibnamefont
  {Fuchizaki}}\ and\ \bibinfo {author} {\bibfnamefont {K.}~\bibnamefont
  {Kawasaki}},\ }\bibfield  {title} {\enquote {\bibinfo {title} {Dynamical
  density functional theory for glassy behaviour},}\ }\href@noop {} {\bibfield
  {journal} {\bibinfo  {journal} {Journal of Physics: Condensed Matter}\
  }\textbf {\bibinfo {volume} {14}},\ \bibinfo {pages} {12203--12222} (\bibinfo
  {year} {2002})}\BibitemShut {NoStop}%
\bibitem [{\citenamefont {Schindler}\ \emph {et~al.}(2019)\citenamefont
  {Schindler}, \citenamefont {Wittmann},\ and\ \citenamefont
  {Brader}}]{SchindlerWB2019}%
  \BibitemOpen
  \bibfield  {author} {\bibinfo {author} {\bibfnamefont {T.}~\bibnamefont
  {Schindler}}, \bibinfo {author} {\bibfnamefont {R.}~\bibnamefont {Wittmann}},
  \ and\ \bibinfo {author} {\bibfnamefont {J.~M.}\ \bibnamefont {Brader}},\
  }\bibfield  {title} {\enquote {\bibinfo {title} {Particle-conserving dynamics
  on the single-particle level},}\ }\href@noop {} {\bibfield  {journal}
  {\bibinfo  {journal} {Physical Review E}\ }\textbf {\bibinfo {volume} {99}},\
  \bibinfo {pages} {012605} (\bibinfo {year} {2019})}\BibitemShut {NoStop}%
\bibitem [{\citenamefont {Munakata}(1994)}]{Munakata1994}%
  \BibitemOpen
  \bibfield  {author} {\bibinfo {author} {\bibfnamefont {T.}~\bibnamefont
  {Munakata}},\ }\bibfield  {title} {\enquote {\bibinfo {title} {Time-dependent
  density-functional theory with {H} theorems},}\ }\href@noop {} {\bibfield
  {journal} {\bibinfo  {journal} {Physical Review E}\ }\textbf {\bibinfo
  {volume} {50}},\ \bibinfo {pages} {2347--2350} (\bibinfo {year}
  {1994})}\BibitemShut {NoStop}%
\bibitem [{\citenamefont {{te Vrugt}}\ and\ \citenamefont
  {Wittkowski}(2020{\natexlab{b}})}]{teVrugtW2020b}%
  \BibitemOpen
  \bibfield  {author} {\bibinfo {author} {\bibfnamefont {M.}~\bibnamefont {{te
  Vrugt}}}\ and\ \bibinfo {author} {\bibfnamefont {R.}~\bibnamefont
  {Wittkowski}},\ }\bibfield  {title} {\enquote {\bibinfo {title} {Relations
  between angular and {C}artesian orientational expansions},}\ }\href@noop {}
  {\bibfield  {journal} {\bibinfo  {journal} {AIP Advances}\ }\textbf {\bibinfo
  {volume} {10}},\ \bibinfo {pages} {035106} (\bibinfo {year}
  {2020}{\natexlab{b}})}\BibitemShut {NoStop}%
\bibitem [{\citenamefont {Tornberg}\ and\ \citenamefont
  {Engquist}(2000)}]{TornbergE2000}%
  \BibitemOpen
  \bibfield  {author} {\bibinfo {author} {\bibfnamefont {A.-K.}\ \bibnamefont
  {Tornberg}}\ and\ \bibinfo {author} {\bibfnamefont {B.}~\bibnamefont
  {Engquist}},\ }\bibfield  {title} {\enquote {\bibinfo {title} {A finite
  element based level-set method for multiphase flow applications},}\
  }\href@noop {} {\bibfield  {journal} {\bibinfo  {journal} {Computing and
  Visualization in Science}\ }\textbf {\bibinfo {volume} {3}},\ \bibinfo
  {pages} {93--101} (\bibinfo {year} {2000})}\BibitemShut {NoStop}%
\bibitem [{\citenamefont {Balachandar}\ and\ \citenamefont
  {Eaton}(2010)}]{BalachandarE2010}%
  \BibitemOpen
  \bibfield  {author} {\bibinfo {author} {\bibfnamefont {S.}~\bibnamefont
  {Balachandar}}\ and\ \bibinfo {author} {\bibfnamefont {J.~K.}\ \bibnamefont
  {Eaton}},\ }\bibfield  {title} {\enquote {\bibinfo {title} {Turbulent
  dispersed multiphase flow},}\ }\href@noop {} {\bibfield  {journal} {\bibinfo
  {journal} {Annual Review of Fluid Mechanics}\ }\textbf {\bibinfo {volume}
  {42}},\ \bibinfo {pages} {111--133} (\bibinfo {year} {2010})}\BibitemShut
  {NoStop}%
\bibitem [{\citenamefont {Brand}\ and\ \citenamefont
  {Pleiner}(2021)}]{BrandP2021}%
  \BibitemOpen
  \bibfield  {author} {\bibinfo {author} {\bibfnamefont {H.~R.}\ \bibnamefont
  {Brand}}\ and\ \bibinfo {author} {\bibfnamefont {H.}~\bibnamefont
  {Pleiner}},\ }\bibfield  {title} {\enquote {\bibinfo {title} {Two-fluid model
  for the breakdown of flow alignment in nematic liquid crystals},}\
  }\href@noop {} {\bibfield  {journal} {\bibinfo  {journal} {Physical Review
  E}\ }\textbf {\bibinfo {volume} {103}},\ \bibinfo {pages} {012705} (\bibinfo
  {year} {2021})}\BibitemShut {NoStop}%
\bibitem [{\citenamefont {Pleiner}\ and\ \citenamefont
  {Brand}(2021)}]{PleinerB2021}%
  \BibitemOpen
  \bibfield  {author} {\bibinfo {author} {\bibfnamefont {H.}~\bibnamefont
  {Pleiner}}\ and\ \bibinfo {author} {\bibfnamefont {H.~R.}\ \bibnamefont
  {Brand}},\ }\bibfield  {title} {\enquote {\bibinfo {title} {A two-fluid model
  for the formation of clusters close to a continuous or almost continuous
  transition},}\ }\href@noop {} {\bibfield  {journal} {\bibinfo  {journal}
  {Rheologica Acta}\ }\textbf {\bibinfo {volume} {60}},\ \bibinfo {pages}
  {675--690} (\bibinfo {year} {2021})}\BibitemShut {NoStop}%
\bibitem [{\citenamefont {Pleiner}\ and\ \citenamefont
  {Harden}(2004)}]{PleinerH2004}%
  \BibitemOpen
  \bibfield  {author} {\bibinfo {author} {\bibfnamefont {H.}~\bibnamefont
  {Pleiner}}\ and\ \bibinfo {author} {\bibfnamefont {J.~L.}\ \bibnamefont
  {Harden}},\ }\bibfield  {title} {\enquote {\bibinfo {title} {General
  nonlinear 2‐fluid hydrodynamics of complex fluids and soft matter},}\
  }\href@noop {} {\bibfield  {journal} {\bibinfo  {journal} {AIP Conference
  Proceedings}\ }\textbf {\bibinfo {volume} {708}},\ \bibinfo {pages} {46--51}
  (\bibinfo {year} {2004})}\BibitemShut {NoStop}%
\bibitem [{\citenamefont {Wittkowski}\ \emph {et~al.}(2010)\citenamefont
  {Wittkowski}, \citenamefont {L{\"o}wen},\ and\ \citenamefont
  {Brand}}]{WittkowskiLB2010}%
  \BibitemOpen
  \bibfield  {author} {\bibinfo {author} {\bibfnamefont {R.}~\bibnamefont
  {Wittkowski}}, \bibinfo {author} {\bibfnamefont {H.}~\bibnamefont
  {L{\"o}wen}}, \ and\ \bibinfo {author} {\bibfnamefont {H.~R.}\ \bibnamefont
  {Brand}},\ }\bibfield  {title} {\enquote {\bibinfo {title} {Derivation of a
  three-dimensional phase-field-crystal model for liquid crystals from density
  functional theory},}\ }\href@noop {} {\bibfield  {journal} {\bibinfo
  {journal} {Physical Review E}\ }\textbf {\bibinfo {volume} {82}},\ \bibinfo
  {pages} {031708} (\bibinfo {year} {2010})}\BibitemShut {NoStop}%
\bibitem [{\citenamefont {Wittkowski}\ \emph
  {et~al.}(2011{\natexlab{a}})\citenamefont {Wittkowski}, \citenamefont
  {L{\"o}wen},\ and\ \citenamefont {Brand}}]{WittkowskiLB2011}%
  \BibitemOpen
  \bibfield  {author} {\bibinfo {author} {\bibfnamefont {R.}~\bibnamefont
  {Wittkowski}}, \bibinfo {author} {\bibfnamefont {H.}~\bibnamefont
  {L{\"o}wen}}, \ and\ \bibinfo {author} {\bibfnamefont {H.~R.}\ \bibnamefont
  {Brand}},\ }\bibfield  {title} {\enquote {\bibinfo {title} {Polar liquid
  crystals in two spatial dimensions: the bridge from microscopic to
  macroscopic modeling},}\ }\href@noop {} {\bibfield  {journal} {\bibinfo
  {journal} {Physical Review E}\ }\textbf {\bibinfo {volume} {83}},\ \bibinfo
  {pages} {061706} (\bibinfo {year} {2011}{\natexlab{a}})}\BibitemShut
  {NoStop}%
\bibitem [{\citenamefont {Wittkowski}\ \emph
  {et~al.}(2011{\natexlab{b}})\citenamefont {Wittkowski}, \citenamefont
  {L{\"o}wen},\ and\ \citenamefont {Brand}}]{WittkowskiLB2011b}%
  \BibitemOpen
  \bibfield  {author} {\bibinfo {author} {\bibfnamefont {R.}~\bibnamefont
  {Wittkowski}}, \bibinfo {author} {\bibfnamefont {H.}~\bibnamefont
  {L{\"o}wen}}, \ and\ \bibinfo {author} {\bibfnamefont {H.~R.}\ \bibnamefont
  {Brand}},\ }\bibfield  {title} {\enquote {\bibinfo {title} {Microscopic and
  macroscopic theories for the dynamics of polar liquid crystals},}\
  }\href@noop {} {\bibfield  {journal} {\bibinfo  {journal} {Physical Review
  E}\ }\textbf {\bibinfo {volume} {84}},\ \bibinfo {pages} {041708} (\bibinfo
  {year} {2011}{\natexlab{b}})}\BibitemShut {NoStop}%
\bibitem [{\citenamefont {Wensink}\ and\ \citenamefont
  {L{\"o}wen}(2008)}]{WensinkL2008}%
  \BibitemOpen
  \bibfield  {author} {\bibinfo {author} {\bibfnamefont {H.~H.}\ \bibnamefont
  {Wensink}}\ and\ \bibinfo {author} {\bibfnamefont {H.}~\bibnamefont
  {L{\"o}wen}},\ }\bibfield  {title} {\enquote {\bibinfo {title} {Aggregation
  of self-propelled colloidal rods near confining walls},}\ }\href@noop {}
  {\bibfield  {journal} {\bibinfo  {journal} {Physical Review E}\ }\textbf
  {\bibinfo {volume} {78}},\ \bibinfo {pages} {031409} (\bibinfo {year}
  {2008})}\BibitemShut {NoStop}%
\bibitem [{\citenamefont {Wittkowski}\ and\ \citenamefont
  {L{\"o}wen}(2011)}]{WittkowskiL2011}%
  \BibitemOpen
  \bibfield  {author} {\bibinfo {author} {\bibfnamefont {R.}~\bibnamefont
  {Wittkowski}}\ and\ \bibinfo {author} {\bibfnamefont {H.}~\bibnamefont
  {L{\"o}wen}},\ }\bibfield  {title} {\enquote {\bibinfo {title} {Dynamical
  density functional theory for colloidal particles with arbitrary shape},}\
  }\href@noop {} {\bibfield  {journal} {\bibinfo  {journal} {Molecular
  Physics}\ }\textbf {\bibinfo {volume} {109}},\ \bibinfo {pages} {2935--2943}
  (\bibinfo {year} {2011})}\BibitemShut {NoStop}%
\bibitem [{\citenamefont {L{\"o}wen}(2020)}]{Loewen2020}%
  \BibitemOpen
  \bibfield  {author} {\bibinfo {author} {\bibfnamefont {H.}~\bibnamefont
  {L{\"o}wen}},\ }\bibfield  {title} {\enquote {\bibinfo {title} {Inertial
  effects of self-propelled particles: From active {B}rownian to active
  {L}angevin motion},}\ }\href@noop {} {\bibfield  {journal} {\bibinfo
  {journal} {Journal of Chemical Physics}\ }\textbf {\bibinfo {volume} {152}},\
  \bibinfo {pages} {040901} (\bibinfo {year} {2020})}\BibitemShut {NoStop}%
\bibitem [{\citenamefont {Scholz}\ \emph {et~al.}(2018)\citenamefont {Scholz},
  \citenamefont {Jahanshahi}, \citenamefont {Ldov},\ and\ \citenamefont
  {L{\"o}wen}}]{ScholzJLL2018}%
  \BibitemOpen
  \bibfield  {author} {\bibinfo {author} {\bibfnamefont {C.}~\bibnamefont
  {Scholz}}, \bibinfo {author} {\bibfnamefont {S.}~\bibnamefont {Jahanshahi}},
  \bibinfo {author} {\bibfnamefont {A.}~\bibnamefont {Ldov}}, \ and\ \bibinfo
  {author} {\bibfnamefont {H.}~\bibnamefont {L{\"o}wen}},\ }\bibfield  {title}
  {\enquote {\bibinfo {title} {Inertial delay of self-propelled particles},}\
  }\href@noop {} {\bibfield  {journal} {\bibinfo  {journal} {Nature
  Communications}\ }\textbf {\bibinfo {volume} {9}},\ \bibinfo {pages} {5156}
  (\bibinfo {year} {2018})}\BibitemShut {NoStop}%
\bibitem [{\citenamefont {{te Vrugt}}\ \emph
  {et~al.}(2021{\natexlab{b}})\citenamefont {{te Vrugt}}, \citenamefont
  {Jeggle},\ and\ \citenamefont {Wittkowski}}]{teVrugtJW2021}%
  \BibitemOpen
  \bibfield  {author} {\bibinfo {author} {\bibfnamefont {M.}~\bibnamefont {{te
  Vrugt}}}, \bibinfo {author} {\bibfnamefont {J.}~\bibnamefont {Jeggle}}, \
  and\ \bibinfo {author} {\bibfnamefont {R.}~\bibnamefont {Wittkowski}},\
  }\bibfield  {title} {\enquote {\bibinfo {title} {Jerky active matter: a phase
  field crystal model with translational and orientational memory},}\
  }\href@noop {} {\bibfield  {journal} {\bibinfo  {journal} {New Journal of
  Physics}\ }\textbf {\bibinfo {volume} {23}},\ \bibinfo {pages} {063023}
  (\bibinfo {year} {2021}{\natexlab{b}})}\BibitemShut {NoStop}%
\bibitem [{\citenamefont {Arold}\ and\ \citenamefont
  {Schmiedeberg}(2020{\natexlab{a}})}]{AroldS2020}%
  \BibitemOpen
  \bibfield  {author} {\bibinfo {author} {\bibfnamefont {D.}~\bibnamefont
  {Arold}}\ and\ \bibinfo {author} {\bibfnamefont {M.}~\bibnamefont
  {Schmiedeberg}},\ }\bibfield  {title} {\enquote {\bibinfo {title} {Active
  phase field crystal systems with inertial delay and underdamped dynamics},}\
  }\href@noop {} {\bibfield  {journal} {\bibinfo  {journal} {European Physical
  Journal E}\ }\textbf {\bibinfo {volume} {43}},\ \bibinfo {pages} {47}
  (\bibinfo {year} {2020}{\natexlab{a}})}\BibitemShut {NoStop}%
\bibitem [{\citenamefont {{te Vrugt}}\ \emph {et~al.}(2023)\citenamefont {{te
  Vrugt}}, \citenamefont {Frohoff-H{\"u}lsmann}, \citenamefont {Heifetz},
  \citenamefont {Thiele},\ and\ \citenamefont {Wittkowski}}]{teVrugtFHHTW2023}%
  \BibitemOpen
  \bibfield  {author} {\bibinfo {author} {\bibfnamefont {M.}~\bibnamefont {{te
  Vrugt}}}, \bibinfo {author} {\bibfnamefont {T.}~\bibnamefont
  {Frohoff-H{\"u}lsmann}}, \bibinfo {author} {\bibfnamefont {E.}~\bibnamefont
  {Heifetz}}, \bibinfo {author} {\bibfnamefont {U.}~\bibnamefont {Thiele}}, \
  and\ \bibinfo {author} {\bibfnamefont {R.}~\bibnamefont {Wittkowski}},\
  }\bibfield  {title} {\enquote {\bibinfo {title} {From a microscopic inertial
  active matter model to the {S}chr\"odinger equation},}\ }\href@noop {}
  {\bibfield  {journal} {\bibinfo  {journal} {Nature Communications}\ }\textbf
  {\bibinfo {volume} {14}},\ \bibinfo {pages} {1302} (\bibinfo {year}
  {2023})}\BibitemShut {NoStop}%
\bibitem [{\citenamefont {Arold}\ and\ \citenamefont
  {Schmiedeberg}(2020{\natexlab{b}})}]{AroldS2020b}%
  \BibitemOpen
  \bibfield  {author} {\bibinfo {author} {\bibfnamefont {D.}~\bibnamefont
  {Arold}}\ and\ \bibinfo {author} {\bibfnamefont {M.}~\bibnamefont
  {Schmiedeberg}},\ }\bibfield  {title} {\enquote {\bibinfo {title} {Mean field
  approach of dynamical pattern formation in underdamped active matter with
  short-ranged alignment and distant anti-alignment interactions},}\
  }\href@noop {} {\bibfield  {journal} {\bibinfo  {journal} {Journal of
  Physics: Condensed Matter}\ }\textbf {\bibinfo {volume} {32}},\ \bibinfo
  {pages} {315403} (\bibinfo {year} {2020}{\natexlab{b}})}\BibitemShut
  {NoStop}%
\bibitem [{\citenamefont {Liu}\ and\ \citenamefont {Liu}(2020)}]{LiuL2020}%
  \BibitemOpen
  \bibfield  {author} {\bibinfo {author} {\bibfnamefont {Y.}~\bibnamefont
  {Liu}}\ and\ \bibinfo {author} {\bibfnamefont {H.}~\bibnamefont {Liu}},\
  }\bibfield  {title} {\enquote {\bibinfo {title} {Development of
  reaction-diffusion {DFT} and its application to catalytic oxidation of {NO}
  in porous materials},}\ }\href@noop {} {\bibfield  {journal} {\bibinfo
  {journal} {AIChE Journal}\ }\textbf {\bibinfo {volume} {66}},\ \bibinfo
  {pages} {e16824} (\bibinfo {year} {2020})}\BibitemShut {NoStop}%
\bibitem [{\citenamefont {Moncho-Jord\'a}\ and\ \citenamefont
  {Dzubiella}(2020)}]{MonchoD2020}%
  \BibitemOpen
  \bibfield  {author} {\bibinfo {author} {\bibfnamefont {A.}~\bibnamefont
  {Moncho-Jord\'a}}\ and\ \bibinfo {author} {\bibfnamefont {J.}~\bibnamefont
  {Dzubiella}},\ }\bibfield  {title} {\enquote {\bibinfo {title} {Controlling
  the microstructure and phase behavior of confined soft colloids by active
  interaction switching},}\ }\href@noop {} {\bibfield  {journal} {\bibinfo
  {journal} {Physical Review Letters}\ }\textbf {\bibinfo {volume} {125}},\
  \bibinfo {pages} {078001} (\bibinfo {year} {2020})}\BibitemShut {NoStop}%
\bibitem [{\citenamefont {Lutsko}\ and\ \citenamefont
  {Nicolis}(2016)}]{LutskoN2016}%
  \BibitemOpen
  \bibfield  {author} {\bibinfo {author} {\bibfnamefont {J.~F.}\ \bibnamefont
  {Lutsko}}\ and\ \bibinfo {author} {\bibfnamefont {G.}~\bibnamefont
  {Nicolis}},\ }\bibfield  {title} {\enquote {\bibinfo {title} {Mechanism for
  the stabilization of protein clusters above the solubility curve},}\
  }\href@noop {} {\bibfield  {journal} {\bibinfo  {journal} {Soft Matter}\
  }\textbf {\bibinfo {volume} {12}},\ \bibinfo {pages} {93--98} (\bibinfo
  {year} {2016})}\BibitemShut {NoStop}%
\bibitem [{\citenamefont {Espa{\~n}ol}\ and\ \citenamefont
  {Donev}(2015)}]{EspanolD2015}%
  \BibitemOpen
  \bibfield  {author} {\bibinfo {author} {\bibfnamefont {P.}~\bibnamefont
  {Espa{\~n}ol}}\ and\ \bibinfo {author} {\bibfnamefont {A.}~\bibnamefont
  {Donev}},\ }\bibfield  {title} {\enquote {\bibinfo {title} {Coupling a
  nano-particle with isothermal fluctuating hydrodynamics: coarse-graining from
  microscopic to mesoscopic dynamics},}\ }\href@noop {} {\bibfield  {journal}
  {\bibinfo  {journal} {Journal of Chemical Physics}\ }\textbf {\bibinfo
  {volume} {143}},\ \bibinfo {pages} {234104} (\bibinfo {year}
  {2015})}\BibitemShut {NoStop}%
\bibitem [{\citenamefont {Nakamura}\ and\ \citenamefont
  {Yoshimori}(2009)}]{NakamuraY2009}%
  \BibitemOpen
  \bibfield  {author} {\bibinfo {author} {\bibfnamefont {T.}~\bibnamefont
  {Nakamura}}\ and\ \bibinfo {author} {\bibfnamefont {A.}~\bibnamefont
  {Yoshimori}},\ }\bibfield  {title} {\enquote {\bibinfo {title} {Derivation of
  the nonlinear fluctuating hydrodynamic equation from the underdamped
  {L}angevin equation},}\ }\href@noop {} {\bibfield  {journal} {\bibinfo
  {journal} {Journal of Physics A: Mathematical and Theoretical}\ }\textbf
  {\bibinfo {volume} {42}},\ \bibinfo {pages} {065001} (\bibinfo {year}
  {2009})}\BibitemShut {NoStop}%
\bibitem [{\citenamefont {Emmerich}\ \emph {et~al.}(2012)\citenamefont
  {Emmerich}, \citenamefont {L{\"o}wen}, \citenamefont {Wittkowski},
  \citenamefont {Gruhn}, \citenamefont {T{\'o}th}, \citenamefont {Tegze},\ and\
  \citenamefont {Gr{\'a}n{\'a}sy}}]{EmmerichEtAl2012}%
  \BibitemOpen
  \bibfield  {author} {\bibinfo {author} {\bibfnamefont {H.}~\bibnamefont
  {Emmerich}}, \bibinfo {author} {\bibfnamefont {H.}~\bibnamefont {L{\"o}wen}},
  \bibinfo {author} {\bibfnamefont {R.}~\bibnamefont {Wittkowski}}, \bibinfo
  {author} {\bibfnamefont {T.}~\bibnamefont {Gruhn}}, \bibinfo {author}
  {\bibfnamefont {G.~I.}\ \bibnamefont {T{\'o}th}}, \bibinfo {author}
  {\bibfnamefont {G.}~\bibnamefont {Tegze}}, \ and\ \bibinfo {author}
  {\bibfnamefont {L.}~\bibnamefont {Gr{\'a}n{\'a}sy}},\ }\bibfield  {title}
  {\enquote {\bibinfo {title} {Phase-field-crystal models for condensed matter
  dynamics on atomic length and diffusive time scales: an overview},}\
  }\href@noop {} {\bibfield  {journal} {\bibinfo  {journal} {Advances in
  Physics}\ }\textbf {\bibinfo {volume} {61}},\ \bibinfo {pages} {665--743}
  (\bibinfo {year} {2012})}\BibitemShut {NoStop}%
\end{thebibliography}%
\end{document}